\documentclass[hyper,12pt,paper]{article}
\pdfoutput=1 
\usepackage{jheppub} 

\usepackage[T1]{fontenc} % if needed
\usepackage{tabu}
\usepackage{graphicx,amsmath,amssymb,multirow,array,bm,mathrsfs}
\usepackage{upgreek}
\usepackage{mathtools}
\usepackage{color}
\usepackage{tikz}

\usepackage{subcaption}
\usepackage{float}
\usepackage{afterpage}
\allowdisplaybreaks  % allow page breaking for align environment

\def\ket#1{\mid \! #1\rangle} 
\def\bra#1{\langle \, #1 \! \mid}

\def\ptr#1{{{#1}^\Gamma}}
\def\Eneg{{\mathscr E}}
\def\Neg{{\mathscr N}}
\def\tmi{{\sf I3}}
\def\qsl#1{[\sf Q{#1}]}
\def\regA{{\cal A}} 
\def\regAc{{\cal A}^c}

\definecolor{rust}{rgb}{0.8,0.2,0.2}
\definecolor{green}{rgb}{0.1,0.8,0.2}
\definecolor{myblue}{rgb}{0.12,0.51,0.88}

\subheader{}
\title{Entanglement structures in qubit systems}

\author[]{Mukund Rangamani,}
\author[]{\! Massimiliano Rota}
\affiliation[]{Centre for Particle Theory \& Department of Mathematical Sciences,\\
                     Durham University, South Road, Durham DH1 3LE, UK.}
\emailAdd{mukund.rangamani@durham.ac.uk}
\emailAdd{massimiliano.rota@durham.ac.uk}

\abstract{Using measures of entanglement such as negativity and tangles we provide a detailed analysis of entanglement structures in pure states of non-interacting qubits. The motivation for this exercise primarily comes from holographic considerations, where entanglement is inextricably linked with the emergence of geometry. We use the qubit systems as toy models to probe the internal structure, and introduce some useful measures involving entanglement negativity to quantify general features of  entanglement. In particular,  our analysis focuses on various constraints on the pattern of entanglement which are known to be satisfied by holographic sates, such as the saturation of Araki-Lieb inequality (in certain  circumstances), and the monogamy of mutual information. We argue that even systems as simple as few non-interacting qubits can be useful laboratories to explore how the emergence of the bulk geometry may be related to quantum information principles.
}

\preprint{DCPT-15/27}

\begin{document}

\maketitle
\flushbottom

%~~~~~~~~~~~~~~~~~~~~~~~~~~~~~~~~~~~~~~~~~~~~~~~
\section{Introduction}
\label{sec:intro}
%~~~~~~~~~~~~~~~~~~~~~~~~~~~~~~~~~~~~~~~~~~~~~~

One of the key features distinguishing quantum mechanics is the presence of entanglement which is  a natural consequence of the superposition principle. Usually this is characterized  simply by the inability to separate a composite system into its constituent
parts without losing some information about the whole. The lack of knowledge of how the individual parts comprise the entire  
system is encoded by entanglement.

While the presence or absence of entanglement elicits a binary response, one often would like to know more and in particular be able to quantify the precise nature of entanglement in a quantum system. In simple bipartite systems, e.g., two qubits, this is easily done using the von Neumann entropy of the reduced density matrix for one of the components. This quantity which is referred to as  the entanglement entropy provides a complete characterization of the entanglement  inherent in the state.\footnote{  To be sure this only captures the entanglement under the obvious bipartitioning; we will later be careful to  distinguish this from the entanglement contained in further subdivisions of the each system.} However, this ceases to be the case in more general scenarios: density matrices of bipartite systems or equivalently pure states of multipartite systems. 

To quantify the amount of entanglement in  more general cases various measures of entanglement have been proposed in the quantum information literature. Some of these which we shall review in the sequel are easy to compute, while others have restricted applicability.  Nevertheless, given that the structure of entanglement in multi-component systems can get rather intricate (if only due to the rapid growth of potential permutations involved), it is interesting to contrast the different observables against each other.

While this is an interesting exercise in its own right in quantum mechanics, part of our motivation in attempting to understand such detailed structure of entanglement stems from potential insights it can offer in the context of holography.  One of the amazing facts about the holographic AdS/CFT correspondence is the observation that the entanglement in a class of  strongly coupled planar field theories is geometrized in terms of a gravitational background in higher dimensions. This statement is manifest in the holographic entanglement entropy proposals of Ryu and Takayanagi \cite{Ryu:2006bv,Ryu:2006ef} and its covariant generalization \cite{Hubeny:2007xt}. A more intricate and intriguing picture arises when we ask whether the structure of entanglement in the field theory is itself responsible for the emergence of geometry as was first suggested a few years ago in \cite{Swingle:2009bg,VanRaamsdonk:2009ar,VanRaamsdonk:2010pw}.  These ideas have been central to the recent thesis that ``entanglement builds geometry'' codified succinctly in the statement ER = EPR \cite{Maldacena:2013xja}.
 
One obvious question in this context is the following: is the emergence of geometry simply reliant on the presence or absence of entanglement, or does it depend more crucially on the structure of entanglement? Most discussions in the holographic context presuppose a  pure state of a bipartite system whence entanglement entropy suffices. However, we should be able to ask for the emergence of geometry in situations where the configuration in question is more complicated and admits no simple bipartite description. Typical scenarios we have in mind are multipartite systems exemplified by the multi-boundary wormhole geometries of 
 \cite{Brill:1995jv,Aminneborg:1997pz,Brill:1998pr} in three dimensions. Here the precise manner in which the individual parts are entangled does play a role in the emergence of some sort of semi-classical geometry and indeed previous investigations \cite{Gharibyan:2013aha,Balasubramanian:2014hda} indicate this to be the case.  

The prototype scenario for this discussion is a $N$-partite system wherein integrating out $(N-3)$ components leads to a density matrix for the residual three components. In this case, it is known following an interesting analysis of \cite{Hayden:2011ag} that the density matrix of the resulting tripartite system has to have non-positive definite tripartite information (see below) in order to admit a semi-classical geometry as a holographic dual.\footnote{ As far as we know this condition is necessary but not sufficient to guarantee a semi-classical geometric dual. The constraint was derived in \cite{Hayden:2011ag} by examining the properties of holographic entanglement entropy, essentially adapting the arguments leading to the proof of strong-subadditivity of holographic entanglement entropy. It appears not to follow simply from strong subadditivity, making it in an independent statement about systems with large numbers of degrees of freedom as is relevant for holography. We thank Matthew Headrick and Veronika Hubeny for an extremely illuminating discussion on this point.}
On the other hand for simple quantum systems the tripartite information can have either sign, so not all states of a tripartite system can a-priori admit  a semi-classical gravity dual. This point was already made in \cite{Gharibyan:2013aha}, cautioning that the ER=EPR statement should be accompanied by some riders. We will take this as sufficient motivation to examine the nature of entanglement in simple systems.

To set the stage for our discussion, let us recall that in the holographic context we are interested in studying continuum quantum field theories in the large central charge limit. While the central charges can be formally defined in terms of conformal anomalies, it  is operationally useful to think them as measuring the curvature scale of the holographic dual geometry $\ell_\text{AdS}$ in units of the Planck constant $\ell_P$, viz., $c_\text{eff} \sim \ell_\text{AdS}/\ell_P$. Thus $c_\text{eff} \gg 1$ corresponds to the regime where semi-classical geometry is trustworthy.\footnote{ Strictly speaking we also need $\ell_\text{AdS} \gg \ell_s$ where $\ell_s$ is the string scale, which requires the field theory coupling being large. If not we end up with a classical theory, but one which involves stringy excitations as well. } Heuristically this means that we are interested in considering systems with a large number of degrees of freedom. Given such a theory we want to understand what structures of entanglement are possible. 

The canonical route of exploration is to consider various measures of entanglement in continuum QFTs, such as entanglement entropy, Renyi entropies, negativity etc.. However, these quantities are rather difficult to compute generically in interacting systems. If we start with a pure state in the QFT and demarcate various (disjoint) regions $\regA_i$ $i=1,\cdots, M$, then while it is possible to compute the entanglement entropy for $\cup_i\, \regA_i$ in holographic systems,\footnote{ Given a collection of boundary regions as above, the \cite{Ryu:2006ef,Hubeny:2007xt} require one to solve a classical gravitational problem to find an extremal surface in the geometry dual to the state in question. Even in complicated geometries, this is a problem of solving classical partial differential equations, which whilst involved, is nevertheless a lot simpler than the quantum problem (the simplification is made possible by the $c_\text{eff} \gg 1$ limit). } it is harder to compute the R\'enyi entropies  and negativities.\footnote{ Similar statements apply for entangled states in tensor product of CFTs, e.g., \cite{Maldacena:2001kr,Balasubramanian:2014hda}.} 

Entanglement negativity, introduced in \cite{Vidal:2002zz}, is a  clean measure of the quantum entanglement even for mixed states, while the usual von Neumann entropy is contaminated by classical correlations. This is particularly pertinent, if we are interested in understanding the entanglement between two of our regions, say $\regA_j$ and $\regA_k$, after tracing out the state. Entanglement properties of the density matrix $\rho_{\regA_j\cup \regA_k}$ are more cleanly encoded in the negativity, which bounds the amount of distillable entanglement, so it tells us directly how many Bell pairs are common to this disjoint region.   It is however quite hard to compute it in continuum systems.\footnote{ In recent years negativity has been explored quite extensively in pure and thermal states of two dimensional CFTs in a series of works, cf., \cite{Calabrese:2012ew,Calabrese:2012nk,Calabrese:2014yza,Kulaxizi:2014nma,Coser:2014gsa,Hoogeveen:2014bqa,Wen:2015qwa}. We should also note that negativity in spin chains has been studied in \cite{Wichterich:2009fk,Calabrese:2013uq,Eisler:2015kx,Coser:2015rt}.} 

In a previous work \cite{Rangamani:2014ywa}, we examined the (logarithmic) negativity in the vacuum state of a CFT (for bipartitionings given by connected regions of spherical topology). We conjectured that the ratio of the universal part of the  logarithmic negativity of a pure state with respect to a given bipartitioning was bounded from below by the  entanglement  entropy of the reduced density matrix obtained by integrating out one of the components.\footnote{ The universal part here refers to the renormalization scheme independent term; in even dimensional CFTs it is the coefficient of the logarithmically divergent term, while in odd dimensional CFTs it corresponds to the finite part.} Further analysis of this result for more complicated regions (building on the analysis of  \cite{Lewkowycz:2014jia}) reveals a rather interesting interplay between the central charges of the CFT and the geometry and topology of the entangling surface \cite{Perlmutter:2015zr}.  Motivated by these observations in the continuum, we introduce a new measure involving negativity, called specific robustness, which is sensitive to the pattern of internal entanglement in  simple systems (see below).

To get further insight into features of quantum entanglement we look at a toy problem of non-interacting qubits.\footnote{ Strictly speaking we pick random pure states of a few qubits and are agnostic about the actual Hamiltonian (which may well be the identity operator); hopefully our terminology does not cause confusion.}  Our motivation here is to understand how the measures of entanglement that have been proposed in the quantum information literature serve to help us delineate the entanglement structure of the state. The advantage of working with qubit systems is that we can explicitly compute (at least for small numbers of qubits) various measures of entanglement. Starting with a pure state of $N$-qubits, we can consider tracing out $k<N$ qubits and examining the entanglement inherent in the remainder $(N-k)$-qubits. Furthermore we can consider different bipartitions (or multi-partitions) of the remaining qubits and investigate how entanglement is distributed among them and what are its properties. We perform some simple numerical experiments starting with randomly chosen pure states of $N$-qubits (with $N\leq 8$ for computational reasons) and argue that in general the combination of information emerging from different measures of entanglement can give useful insights about the properties of the state. Even for a fixed bipartition of a pure state, where entanglement entropy determines the amount of entanglement, other measures give additional information about the nature of this entanglement.  While the holographic systems we are really after are not as simple as non-interacting qubits, it is useful to use this toy model to build some intuition about the nature of entanglement inherent in many-body wavefunctions.\footnote{ The entanglement we explore is more closely related to the notion of particle partition entanglement used in certain contexts to gain information complementary to that contained in spatial cuts of the systems, cf., \cite{Eckert:2002rw,Haque:2007gf,Zozulya:2007ul,Herdman:2014qr}. }

Following our interest for the ratio between the logarithmic negativity and entanglement entropy motivated by field theory arguments \cite{Rangamani:2014ywa,Perlmutter:2015zr}, we consider a similar (and strictly related) quantity for qubit system, i.e., the ratio ${\cal R}$ between the negativity (not logarithmic) and the entropy. Using a known operational interpretation for the negativity in terms of robustness of entanglement against noise, we introduce the concept of {\em specific robustness} which is measured by the ratio and investigate how this property is related to the pattern of entanglement inside the state. In the case of 4 qubits, where a classification of the possible states under a particular class of operations (called stochastic LOCC, aka, SLOCC) exists, we show how this additional information allows a partial resolution of the classes and investigate the detailed structure of entanglement within this classification scheme.

We also undertake an analysis of mutual information and the monogamy constraint, both for generic states and also for the different classes of 4 qubits. We find that generically the monogamy constraint is not particularly restrictive. However, if we restrict attention to SLOCC classes of states, we find a unique class that respects monogamy.

In \cite{Rangamani:2014ywa} part of the motivation for considering negativities as a measure of  entanglement in holographic systems was its ability to provide clear  distinction between classicality and quantumness. For bipartitions of mixed states the von Neumann entropy mixes classical and quantum correlations, while negativities can distinguish between them. In a multipartite setting the same problem arises for the tripartite information. We use a measure of multipartite entanglement known as tangle in the quantum information literature, as a witness of intrinsically quantum multipartite correlations. Even though its interpretation is somewhat murky and its definition restricted to qubit systems alone,  we show that it provides useful information when compared against the tripartite information (and flesh out some connections to the monogamy of mutual information). Armed with these result for qubit systems, we attempt to draw some general lessons for continuum field theories.

The outline of the paper is as follows: in \S\ref{sec:measures} we introduce the measures of bipartite and multipartite entanglement as well as the notation that we will use for our investigations. We then start in \S\ref{sec:3qubits} with the simplest case of 3 qubits, which serves as an introduction to the kind of investigations that we will later conduct on larger systems. \S\ref{sec:4q} is the core of the paper, we first investigate generic states of 4 qubits and then present a detailed analysis of the structure of entanglement for the known equivalence classes. We extend the analysis to generic states of larger systems (6 and 8 qubits) in \S\ref{sec:largersys} and discuss the main results and potential implication for holography in \S\ref{sec:discuss}. Appendix~\ref{sec:appendix}, contains additional plots that complete the main results presented in the other sections.

%~~~~~~~~~~~~~~~~~~~~~~~~~~~~~~~~~~~~~~~~~~~~~~~
\section{Measures of entanglement}
\label{sec:measures}
%~~~~~~~~~~~~~~~~~~~~~~~~~~~~~~~~~~~~~~~~~~~~~~

The problem of quantifying entanglement has been at the center of research in quantum information theory for the last 15 years, nevertheless no conclusive measure has been found so far that enables us to fully capture the structure of entanglement of generic states. Several measures have been proposed in the literature but they are usually strongly dependent on the specificities of the application for which they have been developed, and often very difficult to compute. Alternatively, quantities with the correct mathematical properties to be good candidates for entanglement witness, often lack a clear physical interpretation. In this section we review some properties of entanglement and some measures that can be efficiently used to investigate its structure. In the following we will be careful in distinguishing classical from quantum correlations and reserve the term \textit{entanglement} for the latter.

%~~~~~~~~~~~~~~~~~~~~~~~~~~~~~~~~~~~~~~~~~~~~~~~
\subsection{Bipartite entanglement}
\label{subsec:bip}
%~~~~~~~~~~~~~~~~~~~~~~~~~~~~~~~~~~~~~~~~~~~~~~

In its original formulation entanglement is a form of correlation that is not compatible with local physics. Bell's inequalities impose a bound on the strength of correlations achievable by local physics, the violation of these constraints is a witness of entanglement, which actually serves as its definition. For pure states this is well understood --  states which are not products are entangled and always violate some (generalized) Bell's inequality. 

This is not always the case for mixed states.  A mixed state $\rho$ in a Hilbert space $\mathcal{H}_A\otimes\mathcal{H}_B$ is said to be \textit{separable} if it is a convex combination of product states
\begin{equation}
\rho=\sum_{i}\,p_i\,\rho_i^A\otimes\rho_i^B\,,
 \hspace{1cm} \sum_ip_i=1 \,,\hspace{1cm} p_i\geq0
\,,
\label{}
\end{equation}	
and \textit{entangled} otherwise. Product states contain no correlation, separable states have only classical correlation (i.e., correlations that can be produced by LOCC\footnote{ LOCC stands for local operations and classical communication, which includes action by unitaries, measurement and information exchange on classical channels.}) and entangled states contain some sort of quantum correlation.\footnote{ The question of whether these correlations may or may not be compatible with local physics is still open, see \citep{Rangamani:2014ywa} for additional comments and further references.} It is important to realize that the bipartition of the system is a crucial part of the definition.

From a more recent perspective entanglement, can also be interpreted as a powerful resource for specific protocols that are not possible when only classical resources are available. In this context one often would like to be able to manipulate entanglement, convert it into different forms, extract it from a system and transfer it to another and so on. In this context entanglement can be quantified depending on a specific task such as the preparation of a state (\textit{entanglement of formation}), or the extraction of Bell's pairs from a given state (\textit{entanglement of distillation}) (see \cite{Horodecki:2009aa}).

For pure states of a system where a bipartition is specified, it is well known that entanglement entropy is a measure of the amount of entanglement between the subsystems. In a practical situation where two parties only have limited access to the subsystems this is the best measure known so far, as it quantifies both non-locality and the value of entanglement as a resource for specific tasks such as teleportation.\footnote{ In this case entanglement entropy is known to be equal to both entanglement of formation and distillable entanglement \cite{Horodecki:2009aa}.} Na\"ively it is the number of Bell pairs available.\footnote{ This is precisely the meaning of distillable entanglement, but it is important to realize that the definition is an asymptotic statement in the limit where an infinite number of copies of the original state is available. In the context of holography one would prefer an interpretation in terms of a single system.}

We want instead to investigate how different subsystems are correlated among each other in a state of a given global system, being particularly careful in distinguishing classical from quantum correlations. One natural way to proceed is to consider different bipartitions of the entire system: this would certainly give additional information about the distribution of correlations. If we restrict this analysis to pure states, entanglement entropy is a reasonable way to capture the quantum entanglement inherent in the state.

The problem becomes much more intricate for mixed states, where the relation between non-locality and the ``task dependent'' formulation of entanglement is in general not clear at present.\footnote{ There exist \textit{bound entangled} states which are entangled (i.e., not separable), but at the same time they are of very limited value as resource for typical quantum information (QI) protocols; specifically Bell pairs cannot be distilled from them.} In this case the von Neumann entropy is not a good measure of entanglement any more.

This is also a problem one faces in the attempt to completely characterize the internal pattern of entanglement in a given state, even if the state is pure. One can use entanglement entropy as long as only bipartitions of the entire state are considered, but this is not enough to describe the state entirely. Given a pure state of a system $ABC$, if we want to study the entanglement among internal subsystems, for example $A$ and $B$ alone, we first need to trace out the degrees of freedom in $C$, but the result of this operation is a mixed state. In order to characterize the entanglement between $A$ and $B$ we need some other measure.

In a previous paper \cite{Rangamani:2014ywa} we focused on negativities \cite{Vidal:2002zz} as the measures of interest. Let us  recall their definitions and salient properties for convenience. Given a density matrix $\rho$ and a bipartition $\mathcal{H}_A\otimes\mathcal{H}_B$ one defines the \textit{partial transpose} $\ptr{\rho}$ as 
\begin{equation}
\bra{{\mathfrak r}^\text{\tiny(A)}_i\, {\mathfrak l}^\text{\tiny (B)}_n}\ptr{\rho}\ket{{\mathfrak r}^\text{\tiny (A)}_j 
{\mathfrak l}^\text{\tiny(B)}_m}=
\bra{{\mathfrak r}^\text{\tiny(A)}_i\, {\mathfrak l}^\text{\tiny (B)}_m} \rho \ket{{\mathfrak r}^\text{\tiny (A)}_j 
{\mathfrak l}^\text{\tiny(B)}_n}
\label{}
\end{equation}  
The \textit{logarithmic negativity} then is defined as 
\begin{equation}
\Eneg=\log{\|\ptr{\rho}\|} \,,
\end{equation}
where $\|...\|$ denotes the trace norm.\footnote{ The trace norm is defined as $\| {\cal O} \| = {\rm Tr}\left(\sqrt{{\cal O}^\dagger {\cal O}}\right)$ for any Hermitian operator ${\cal O}$.} This is known to be an upper bound to distillable entanglement. It is in general greater or equal to entanglement entropy (for pure bipartite states), with the  equality holding for maximally entangled states.  It is somewhat natural in continuum systems as one can give a suitable path integral representation \cite{Calabrese:2012ew}. One can also define the \textit{negativity} as 
\begin{equation}
\Neg=\frac{\|\ptr{\rho}\|-1}{2} \,.
\end{equation}
While simply related to the logarithmic negativity, it is more convenient to consider in simple discrete systems; hence we will focus for the most part on the negativity itself. 

Armed with the tools of entanglement entropy for bipartition of pure states and negativity for bipartition of pure or mixed states, one can ask how much information about the structure of entanglement of a given state can be extracted considering different partitionings and comparing the two measures. More specifically \cite{Rangamani:2014ywa,Lewkowycz:2014jia,Perlmutter:2015zr} considered the ratio between negativity and entanglement entropy for bipartitions of pure states. One of the motivations of the present work is to flesh out a possible interpretation of this quantity.

To this end it turns out it will be useful to interpret the negativity in terms of another measure of entanglement called \textit{robustness} \cite{Vidal:1999aa}. Given a state $\rho$ and a separable state $\rho_s$, one can consider mixtures of the two states and ask how much of $\rho_s$ is necessary to completely disentangle $\rho$. Formally
\begin{equation}
\tilde{\rho}=\frac{1}{1+s}\left(\rho+s\rho_s\right) .
\end{equation}
The minimal value of $s$ such that $\tilde{\rho}$ is separable is called the robustness of $\rho$ \textit{relative} to $\rho_s$. One can then ask what is the minimal value of $s$ for all possible choices of $\rho_s$, this is the robustness of $\rho$. Intuitively this corresponds to the robustness of $\rho$ against ``intelligent jamming''; it is the minimal amount of noise needed to disrupt the entanglement when we have full knowledge about the structure of the state. 

For finite dimensional systems the negativity is known to be equal to a half of the robustness
\begin{equation}
\Neg=\frac{1}{2}\min_{\forall \rho_s} s\,.
\label{eq:sNrel}
\end{equation}
 This then provides a potential interpretation of negativity. Note however that this way to quantify entanglement is operational and is quite different from the intuition we have for the entropy in terms of non-locality and separability. It is interesting to ask if and how the robustness is related  to the internal structure of entanglement, namely to the way entanglement is distributed among subsystems. 

Inspired by this concept, to get a quantitative handle, we introduce a measure, which we call {\em specific robustness}  ${\cal R}$. It is  defined as the ratio of the negativity of a given state (and bipartitioning), to the entanglement entropy of the reduced density matrix under the same bipartitioning. Schematically we can write\footnote{ In \cite{Rangamani:2014ywa,Perlmutter:2015zr} we considered the  ratio ${\cal X} = \Eneg/S$ which was convenient in continuum systems. For qubit systems the ratio $\mathcal{R}$ seems more appropriate, and one can anyway translate to ${\cal X}$ if necessary.}
\begin{equation}
{\cal R}= \frac{\Neg}{S}
\label{eq:Rdef}
\end{equation}	
Heuristically we want to think of it as a measure of the minimal amount of noise sufficient  to disentangle Bell pairs in a given state. More specifically, given two states with the same entropy but different negativity, we will then interpret the entanglement as more or less robust, depending on the ratio ${\mathcal R}$; higher values of ${\mathcal R}$ would correspond to greater robustness of the entanglement pattern. It is worth reiterating that such a notion of robustness  captures the operational sense of the concept, as it relies on a procedure that mixes the state with some noise. We want to ask whether (and how) this quantity depends on the internal pattern of entanglement of the state. In particular, given an entangling surface corresponding to a fixed bipartition, one can distinguish entanglement inside the two subsystems or entanglement ``across'' the entangling surface. In the rest of the paper we will examine this quantity in simple qubit systems.

Finally, as a measure of bipartite correlations we recall the definition of mutual information
\begin{equation}
I(A|B)=S(A)+S(B)-S(AB)
\label{}
\end{equation}
This has been argued to capture the total amount of correlations, both classical and quantum \cite{Groisman:aa}.

In the theory of quantum information it is often useful to consider inequalities that constrain the values of the measure of interest among different subsystems. One such example is a relation called \textit{monogamy}, which is defined for a quantum information theoretic function $f$ as 
\begin{equation}
f(A|B)+f(A|C)\leq f(A|BC)
\label{eq:mono}
\end{equation}
Monogamy is known to be a general feature of quantum entanglement. One can interpret Eq.~\eqref{eq:mono} as  follows: if $f$ is some entanglement measure, and  subsystem $A$ is almost maximally entangled both with subsystem $B$ and a larger one $BC$, then there is almost no entanglement between $A$ and $C$, i.e., $f(A|C)=0$. This corresponds to the common intuition for the concept of monogamy. Alternatively, the monogamy relation is the precise statement of the fact that the ``union is more than the sum of its parts''. Specifically, there is some subtle correlation between $A$ and the pair $BC$ which is lost if one only looks at the correlations $A|B$ and $A|C$. The latter interpretation of monogamy will be crucial in the definition of some measures of multipartite entanglement in the following.
 
We will be interested in exploring monogamy relations as a potential way to constrain the allowed entanglement structures.  Specifically, we will have occasion to explore the monogamy relation for the square of the negativity that was proved in \cite{He:2014aa}. Another monogamy relation involves the  mutual information. While $I(A|B)$ is not monogamous in general, it happens to be so for holographic theories, as proved in \cite{Hayden:2011ag} (this holds asymptotically at large $c_\text{eff}$). We note that recently \cite{Lin:2014hva,Lashkari:2014kda,Bhattacharya:2014vja} used a similar philosophy to derive a set of (inequality) constraints on holographic theories, using strong-subadditivity and relative entropy.

%~~~~~~~~~~~~~~~~~~~~~~~~~~~~~~~~~~~~~~~~~~~~~~~
\subsection{Multipartite entanglement}
\label{subsec:multip}
%~~~~~~~~~~~~~~~~~~~~~~~~~~~~~~~~~~~~~~~~~~~~~~

The definition of entangled and separable states introduced previously can be extended to a multipartite setting. For a system of $N$ parties a state is said to be \textit{fully separable} if it can be written as\footnote{ For more details about multipartite entanglement see \cite{Horodecki:2009aa} and \cite{Guhne:2009aa}.}
\begin{equation}
\rho=\sum_{i}\,p_i\,\rho_i^{A_1}\otimes\rho_i^{A_2}...\otimes\rho_i^{A_N}\,,
 \hspace{1cm} \sum_ip_i=1 \,,\hspace{1cm} p_i\geq 0
\,,
\label{}
\end{equation}	
where the states $\rho_i$ contain no entanglement. If some of the parties contain some entanglement the state is called \textit{m-separable} if it can be similarly decomposed into a convex linear combination of products of $m$ parts only. A state is said to contain genuine $N$-partite entanglement if it is neither fully separable nor $m$-separable for any $m>1$. 

We will focus on pure states as in the mixed case the properties of multipartite entanglement are much less understood. The prototype of multipartite entangled states is the well known GHZ state (a.k.a. cat state), whose general expression for $N$ qubits is given by 
\begin{equation}
\ket{\text{GHZ}_N}=\frac{1}{\sqrt{2}}(\ket{\underbrace{0\cdots0}_N}+\ket{\underbrace{1\cdots 1}_N}) 
\label{}
\end{equation}
This is sometimes called a \textit{maximally entangled} state (in an $N$-partite sense) as it is the state that violates a $N$-partite generalization of Bell's inequality maximally \citep{Gisin:1998aa}. 

The interpretation and quantification of multipartite entanglement is in general difficult and much less understood than in the bipartite case. Some measures exist but their physical meaning is usually not known. In the following our measure of interest for multipartite entanglement of pure states will be the $N$-\textit{tangle} ($\tau_N$), which was introduced for three qubits by \cite{Coffman:2000aa}. In the three qubits case $\tau_3$ is known to quantify the residual multipartite entanglement in the state, which is not captured by its bipartite counterpart\footnote{ For pure states the $2$-tangle is simply $\tau_2=4\det \rho_A$, where $\rho_A$ is the usual reduced density matrix of the qubit $A$.}
\begin{align}
\tau_2(A|B)+\tau_2(A|C)+\tau_3(ABC)=\tau_2(A|BC)
\end{align}
This equation precisely corresponds to the intuition we have from the previous discussion about monogamy and actually serves as the definition of the tripartite entanglement measure $\tau_3$.

A formal generalization of $\tau_3$ to any even number $N$ of qubits was given in \cite{Wong:aa}
\begin{align}
\tau_{N}=&2\left|\sum a_{\alpha_1...\alpha_N}a_{\beta_1...\beta_N}a_{\gamma_1...\gamma_N}a_{\delta_1...\delta_N}\times\right.
\nonumber\\
&\left.\epsilon_{\alpha_1\beta_1}\epsilon_{\alpha_2\beta_2}...\epsilon_{\alpha_{N-1}\beta_{N-1}}
\epsilon_{\gamma_1\delta_1}\epsilon_{\gamma_2\delta_2}...\epsilon_{\gamma_{N-1}\delta_{N-1}}
\epsilon_{\alpha_N\gamma_N}\epsilon_{\beta_N\delta_N}\right|
\label{}
\end{align}
where the coefficients correspond to the components of the state vector $\ket{\!\psi}$ in the conventional computational basis ($\ket{0\cdots00}$,$\ket{0\cdots01}$, \ldots). This quantity is known to be an entanglement monotone\footnote{ A measure of entanglement is called an \textit{entanglement monotone} if its value on a given state cannot increase under the effect of LOCC operations performed on the state.} and invariant under qubits permutation, nevertheless its interpretation for generic $N$ is not fully understood. For $N=4$ an interpretation in terms of residual entanglement analogous to the three qubits case was given in \cite{Gour:2010aa}.  Nevertheless it is worth noting that this interpretation has some peculiarities \cite{Gour:2010aa}: for example, 
$\tau_4=1$ for a product state of two Bell pairs suggesting that it cannot be interpreted as a measure of genuine $4$-partite entanglement. We emphasize that these measures are only defined for qubits --  there is no generalization to a continuous setting.

Another measure of mutipartite correlation that we will use is the so called \textit{tripartite} or \textit{interaction} information ($\tmi$). This is measure of tripartite correlation which is defined as:
\begin{align}
\tmi(A|B|C)&=S(A)+S(B)+S(C)
\nonumber\\
&-S(AB)-S(BC)-S(AC)+S(ABC)
\label{}
\end{align}
It is important to notice that this is a combination of mutual informations and hence it mixes classical and quantum correlation. For generic quantum states the tripartite information can be either negative or positive, but one can rephrase the condition for the monogamy of mutual information in terms of $\tmi$ simply as \cite{Hayden:2011ag}:
\begin{equation}
\tmi(A|B|C)\leq 0
\label{}
\end{equation}
In the following we will investigate the relation between monogamy of mutual information and the structure of internal entanglement for simple qubits systems. 

%~~~~~~~~~~~~~~~~~~~~~~~~~~~~~~~~~~~~~~~~~~~~~~~
\subsection{Notation}
\label{subsec:notation}
%~~~~~~~~~~~~~~~~~~~~~~~~~~~~~~~~~~~~~~~~~~~~~~

\begin{figure}[t!]
\centering
\begin{subfigure}{0.4\textwidth}
\centering
\begin{tikzpicture}
\filldraw [gray] (0,0) circle (2pt)
(1,0) circle (2pt)
(2,0) circle (2pt)
(0,1) circle (2pt)
(1,1) circle (2pt)
(2,1) circle (2pt);
\draw (-0.3,0.5) -- (2.3,0.5);
\draw (0,1) circle (6pt);
\draw (2,1) circle (6pt);
\node at (2.5,0.5) {\tiny{$\Sigma$}};
\end{tikzpicture}
\subcaption{}
\label{}
\end{subfigure}
\hfill
\begin{subfigure}{0.49\textwidth}
\centering
\begin{tikzpicture}
\filldraw [gray] (0,0) circle (2pt)
(1,0) circle (2pt)
(2,0) circle (2pt)
(0,1) circle (2pt)
(1,1) circle (2pt)
(2,1) circle (2pt);
\draw (-0.3,0.5) -- (2.3,0.5);
\draw (0,0) circle (6pt);
\draw (1.5,1) ellipse (22pt and 6pt);
\node at (2.5,0.5) {\tiny{$\Sigma$}};
\end{tikzpicture}
\subcaption{}
\label{}
\end{subfigure}
\caption{Example of our notation for $6$ qubits, the horizontal line represents the entangling surface $\Sigma$. (a) Local entanglement, $\Neg^{\text{loc}}_{1|1}$. (b) Entanglement across the entangling surface, $\Neg^\Sigma_{1|2}$ }
\label{fig:notation}
\end{figure}
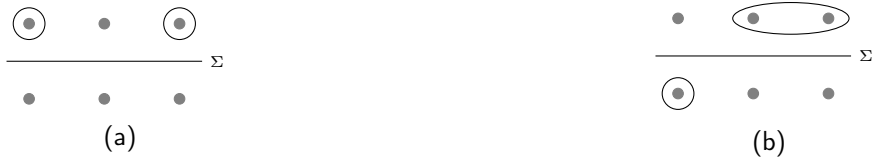

For the convenience of the reader we summarize here the notation that we will use in the sequel. Given a system of $N$ qubits we employ the following notation to denote various partitionings of interest: 
\begin{itemize}
\item single qubits will be labeled by small letters: $a,b,c, \ldots$.
\item capital letters will identify subsets of qubits: $A,B,C, \ldots$.
\item partitioning of $N$-qubits in groups of $m, n, k, \ldots$ etc., with $m+n +k + \cdots \leq N$ will be denoted simply as $m|n|k|\cdots$ when we don't need to specify the particularities of the grouping.
% \item numbers in expressions like $1|2$ will refer to the number of qubits in different subsets of qubits.
\end{itemize}
For example $\Neg_{a|bc}$ denotes the negativity between qubits $a$ and $bc$, $\Neg_{A|BC}$ is the negativity between a subset $A$ and a subset $BC$ (further decomposed into $B$ and $C$), $\Neg_{1|2}$ is the negativity between a generic single qubit and two other qubits in the system. Fig.~\ref{fig:notation} shows an example for $6$ qubits.

We will often consider averages of different quantities and use an overline to identify them. For example $\overline{\Neg_{1|2}}$ is the negativity between a single qubit and two qubits, averaged over all the possible choices of three qubits from the original set.
In the following we will mostly focus on systems made of an even number of qubits. For this systems it is obviously possible to consider bipartitions into two subsets with $N/2$ qubits each, we will refer to this particular bipartition as the \textit{maximal} one.

Using natural terminology from the context of holography we will call the fiducial surface that specifies a bipartition of a system the \textit{entangling surface} $(\Sigma)$. We will be interested in the entanglement among qubits that could lie in the two subsystems across the entangling surface $\Sigma$, or in the same one, see Fig.~\ref{fig:notation}. Expressions such as $\Neg^\Sigma_{1|1}$ refer to negativity between one qubit in a subset and one qubit in the other subsystem. In this case we use the expression \textit{entanglement across $\Sigma$}. We call instead \textit{local entanglement} the correlation between qubits in the same subsystem, and use expressions like $\Neg^{\text{loc}}_{1|1}$.

In the following sections we investigate the structure of entanglement for pure states of systems composed by few qubits. These are simple toy models where the structure of entanglement can be studied numerically, nevertheless the pattern of entanglement is highly non-trivial. Our goal is twofold, on the one hand we use qubits systems as simple laboratories to investigate the properties of different measures of entanglement and their interpretation. On the other hand we will look for entanglement structures that might be relevant for bulk reconstruction in holography. We start with the simplest case of three qubits, this serves as an introduction to the kinds of investigations which we will later apply to larger and more interesting systems. 

%~~~~~~~~~~~~~~~~~~~~~~~~~~~~~~~~~~~~~~~~~~~~~~~
\section{Warm up: Three qubits}
\label{sec:3qubits}
%~~~~~~~~~~~~~~~~~~~~~~~~~~~~~~~~~~~~~~~~~~~~~~

We start by recalling the definition of two particular states, the GHZ and the W states of three qubits:\footnote{ The definition of the W state here is given for the case of 3 qubits, but similarly to the GHZ case the generalization to a higher number of qubits is straightforward.}
\begin{equation}
\ket{\text{GHZ}_3}=\frac{1}{\sqrt{2}}(\ket{000}+\ket{111}) \qquad \ket{\text{W}_3}=\frac{1}{\sqrt{3}}(\ket{100}+\ket{010}+\ket{001})
\label{}
\end{equation}
These states are well known in the literature. The GHZ state only carries genuine tripartite entanglement, this is captured by the fact that the value of the 3-tangle is maximal, $\tau_3=1$. In particular after tracing out one of the qubits one is left with a system of two qubits in a separable (mixed) state, i.e., not entangled. On the other hand the W state contains the maximal possible amount of bipartite entanglement, i.e., the correlation between any pair of qubits is maximal. It does not contain any tripartite entanglement and $\tau_3=0$. This distribution of internal bipartite entanglement corresponds to the known fragility and robustness against qubits removal of the GHZ and W states respectively. An alternative motivation for considering the GHZ state as the maximally entangled state of three qubits (cf., the definition in terms of Bell inequalities in \S\ref{subsec:multip}) is precisely this notion of fragility \citep{Gisin:1998aa}.

\begin{figure}[t]
\centering
\begin{subfigure}{0.49\textwidth}
\includegraphics[width=\textwidth]{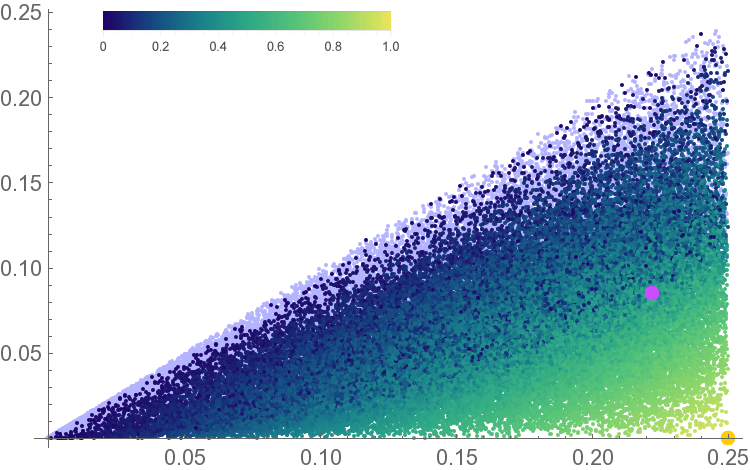}
\put(-12,-6){\makebox(0,0){{\tiny $\Neg^2_{a|bc}$}}}
\put(-204.5,142){\makebox(0,0){{\tiny $\Neg^2_{a|b}+\Neg^2_{a|c}$}}}
\put(-147,137){\makebox(0,0){{\tiny $\tau_3$}}}
\put (-12,16) {\makebox(0,0){
	\begin{tikzpicture}
	\draw[->] (0,0)--(7.07pt,-7.07pt);
	\end{tikzpicture}
}}
\put (-35,58) {\makebox(0,0){
	\begin{tikzpicture}
	\draw[->][white] (0,0)--(7.07pt,-7.07pt);
	\end{tikzpicture}
}}
\put(-20,24){\makebox(0,0){{\tiny{GHZ}}}}
\put(-42,65){\makebox(0,0){{\tiny{\textcolor{white}{W}}}}}
\subcaption{}
\label{subfig:monogamy3a}
\end{subfigure}
\hfill
\begin{subfigure}{0.49\textwidth}
\includegraphics[width=\textwidth]{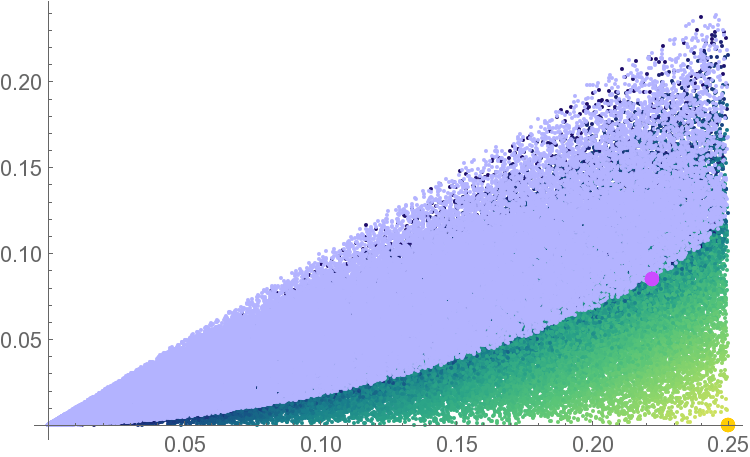}
\put(-12,-6){\makebox(0,0){{\tiny $\Neg^2_{a|bc}$}}}
\put(-204.5,140){\makebox(0,0){{\tiny $\Neg^2_{a|b}+\Neg^2_{a|c}$}}}
\put(-20,24){\makebox(0,0){{\tiny{GHZ}}}}
\put(-42,65){\makebox(0,0){{\tiny{W}}}}
\put (-12,16) {\makebox(0,0){
	\begin{tikzpicture}
	\draw[->] (0,0)--(7.07pt,-7.07pt);
	\end{tikzpicture}
}}
\put (-35,58) {\makebox(0,0){
	\begin{tikzpicture}
	\draw[->] (0,0)--(7.07pt,-7.07pt);
	\end{tikzpicture}
}}
\subcaption{}
\label{subfig:monogamy3b}
\end{subfigure}

\begin{subfigure}{0.49\textwidth}
\includegraphics[width=\textwidth]{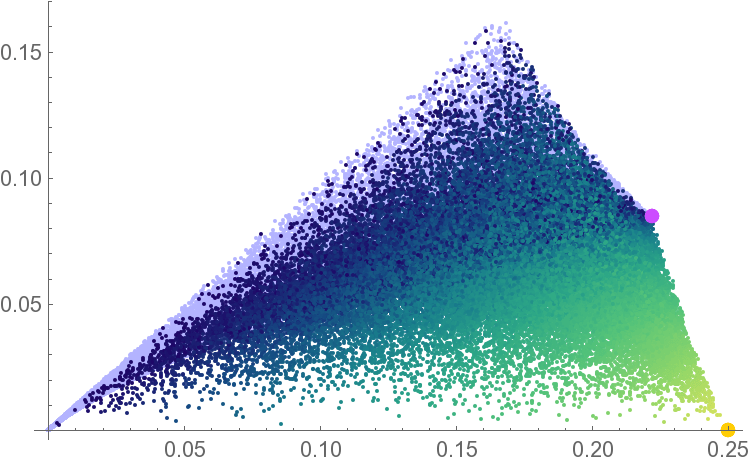}
\put(-204.5,142){\makebox(0,0){{\tiny $\overline{\Neg^2_{1|1}}$}}}
\put(-9,-6){\makebox(0,0){{\tiny $\overline{\Neg^2_{1|2}}$}}}
\put (-6,20) {\makebox(0,0){
	\begin{tikzpicture}
	\draw[->] (0,0)--(0,-10pt);
	\end{tikzpicture}
}}
\put(-5,30){\makebox(0,0){{\tiny{GHZ}}}}
\put (-28,82) {\makebox(0,0){
	\begin{tikzpicture}
	\draw[->] (0,0)--(0,-10pt);
	\end{tikzpicture}
}}
\put(-28,92){\makebox(0,0){{\tiny{W}}}}
\subcaption{}
\label{subfig:Amonogamy3a}
\end{subfigure}
\hfill
\begin{subfigure}{0.49\textwidth}
\includegraphics[width=\textwidth]{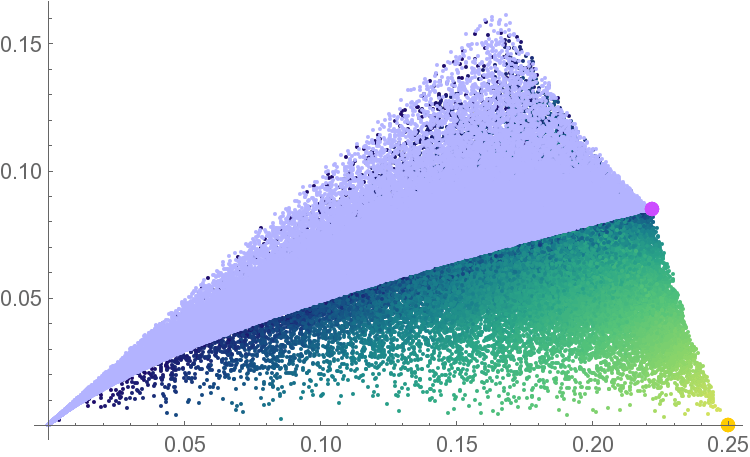}
\put(-204.5,140){\makebox(0,0){{\tiny $\overline{\Neg^2_{1|1}}$}}}
\put(-9,-6){\makebox(0,0){{\tiny $\overline{\Neg^2_{1|2}}$}}}
\put (-28,82) {\makebox(0,0){
	\begin{tikzpicture}
	\draw[->] (0,0)--(0,-10pt);
	\end{tikzpicture}
}}
\put (-6,20) {\makebox(0,0){
	\begin{tikzpicture}
	\draw[->] (0,0)--(0,-10pt);
	\end{tikzpicture}
}}
\put(-5,30){\makebox(0,0){{\tiny{GHZ}}}}
\put(-28,92){\makebox(0,0){{\tiny{W}}}}
\subcaption{}
\label{subfig:Amonogamy3b}
\end{subfigure}

\begin{subfigure}{0.49\textwidth}
\includegraphics[width=\textwidth]{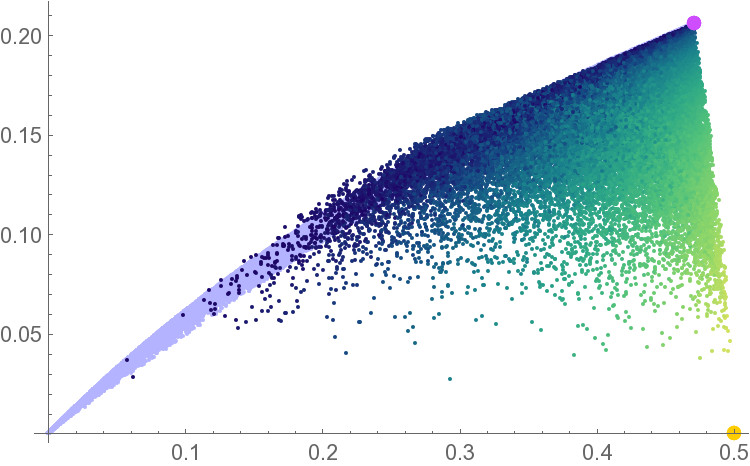}
\put(-204.5,142){\makebox(0,0){{\tiny $\overline{\Neg_{1|1}}$}}}
\put(-8,-6){\makebox(0,0){{\tiny $\overline{\Neg_{1|2}}$}}}
\put(-20,24){\makebox(0,0){{\tiny{GHZ}}}}
\put(-40,128){\makebox(0,0){{\tiny{W}}}}
\put (-30,128) {\makebox(0,0){
	\begin{tikzpicture}
	\draw[->] (0,0)--(10pt,0);
	\end{tikzpicture}
}}
\put (-12,16) {\makebox(0,0){
	\begin{tikzpicture}
	\draw[->] (0,0)--(7.07pt,-7.07pt);
	\end{tikzpicture}
}}
\subcaption{}
\label{subfig:internalN3a}
\end{subfigure}
\hfill
\begin{subfigure}{0.49\textwidth}
\includegraphics[width=\textwidth]{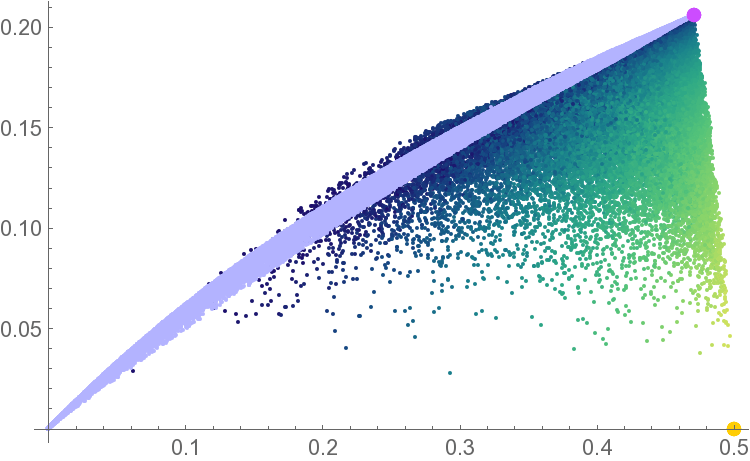}
\put(-204.5,140){\makebox(0,0){{\tiny $\overline{\Neg_{1|1}}$}}}
\put(-8,-6){\makebox(0,0){{\tiny $\overline{\Neg_{1|2}}$}}}
\put(-20,24){\makebox(0,0){{\tiny{GHZ}}}}
\put(-40,128.5){\makebox(0,0){{\tiny{W}}}}
\put (-30,128.5) {\makebox(0,0){
	\begin{tikzpicture}
	\draw[->] (0,0)--(10pt,0);
	\end{tikzpicture}
}}
\put (-12,16) {\makebox(0,0){
	\begin{tikzpicture}
	\draw[->] (0,0)--(7.07pt,-7.07pt);
	\end{tikzpicture}
}}
\subcaption{}
\label{subfig:internalN3b}
\end{subfigure} 
\caption{$50000$ random states in the W (light violet) and the GHZ (color map) classes. The color map corresponds to the value of $\tau_3$ in the range $(0,1)$ as shown in panel (a). The GHZ and W states correspond to the orange and purple large dots respectively. Left and right panels show the same plots with different overlay. \\
(a)-(b): Monogamy of the square of the negativity for a specific bipartition of the global system. (c)-(d): Average of the squared negativity over all possible bipartitions. (e)-(f): Average negativity between single qubits compared to the average negativity for bipartitions of the entire system.}
\label{fig:3plot2}
\end{figure}
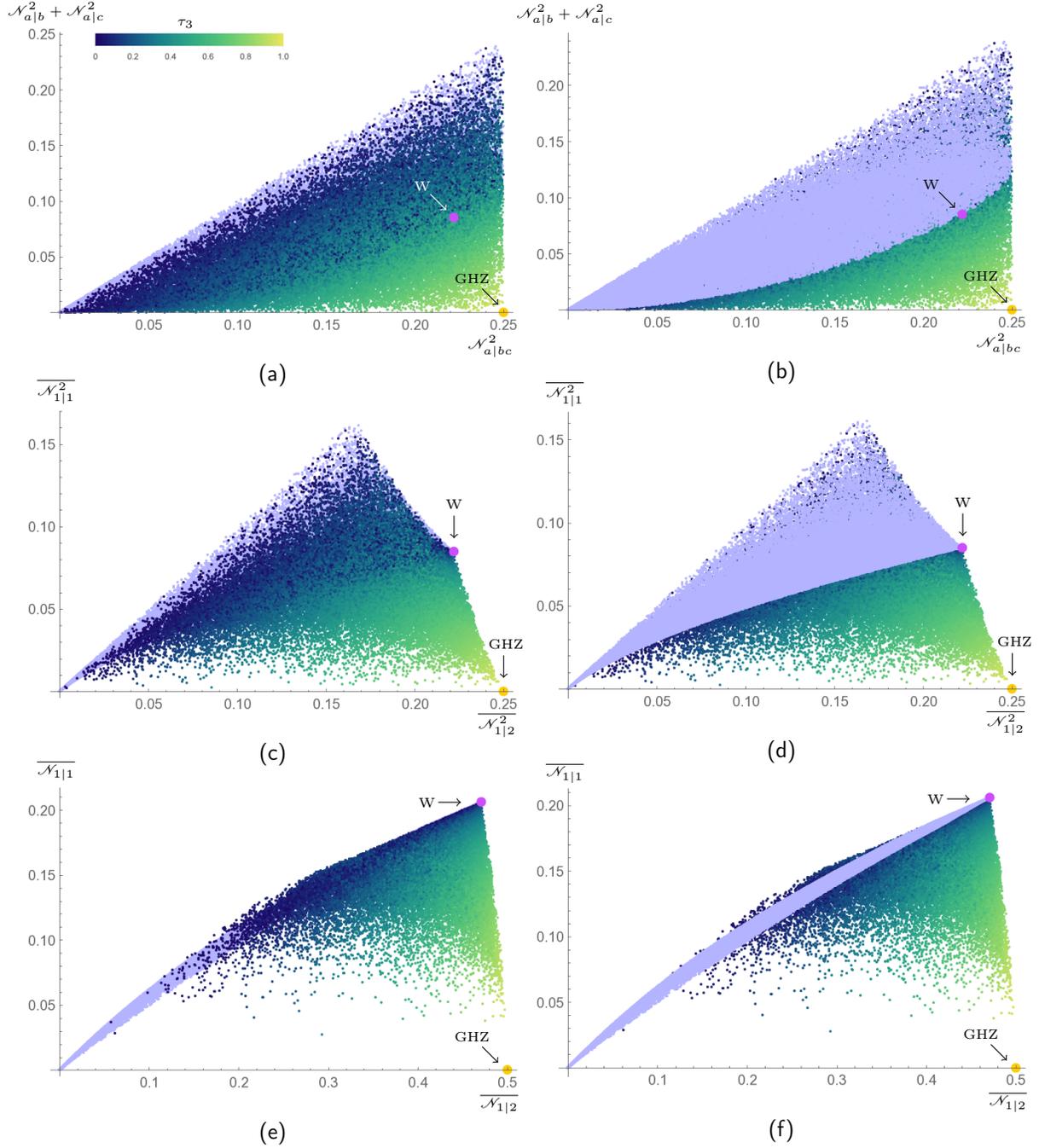
\afterpage{\clearpage}

General pure states of three qubits were classified in \cite{Dur:2000aa}. The classification relies on an equivalence relation under a class of operations called SLOCC (stochastic local operation and classical communication). Two states are considered equivalent when there is a non vanishing probability to convert one state into the other using LOCC and a single copy of the state.\footnote{ Equivalence under LOCC instead would require the ability to convert one state into the other with certainty. It is known that when only a single copy of a state is available two states are equivalent under LOCC if and only if they are related by local unitaries (LU) \cite{Vidal:2000aa}. In this case a classification under LOCC would result in an infinite number of inequivalent classes even for a system of only three qubits.} There is a total number of six classes: the class of product states, three classes of entangled pairs where the third qubit is not entangled, and two classes of states which contain entanglement involving all three qubits. These are the classes we will focus on, they are called GHZ and W classes from the names of their representatives. 

Given any pure state it is possible to decide with certainty which class it belongs to using $\tau_3$. When $\tau_3=0$ there is no truly 3-partite entanglement in the state, which is then in the W class. We will reverse the process and use instead $\tau_3$ to generate random states in the two classes for which we study the pattern of bipartite entanglement. It should be noted however that the W class has measure zero with respect to the generic class (GHZ), this means that our numerical investigation will not respect the statistics of the two classes.

We label the qubits by $abc$. There are three possible bipartitions of the entire system which are generally inequivalent, we will denote them by: $a|bc$, $b|ac$, $c|ab$. For such bipartitions one could in principle consider both entanglement entropy and negativity. Nevertheless since one of the parties only contains a single qubit, its entanglement is completely determined by a single number.\footnote{ Formally a single qubit density matrix has only one non-trivial eigenvalue, the other being determined by the trace normalization.} This means that if we only look at bipartitions of the entire system there is no additional information carried by a second measure and we can choose to equivalently use either the entropy or the negativity. The negativity being a well defined measure of entanglement, which works well for mixed states, we will prefer it to the entropy and  use it to quantify the entanglement between single qubits. 

We start by choosing a specific bipartition ($a|bc$) and computing the following negativities: $\Neg_{a|bc}$, $\Neg_{a|b}$, $\Neg_{a|c}$. As a first exercise we can check the monogamy relation for the square of the negativity which was proved in \cite{He:2014aa}
\begin{equation}
\Neg_{a|b}^2+\Neg^2_{a|c}\leq\Neg^2_{a|bc}
\label{}
\end{equation}
this is shown in Fig.~\ref{subfig:monogamy3a}-\ref{subfig:monogamy3b}. We notice that monogamy seems to be saturated more easily by W states. More interestingly we find that there is a lower bound on the internal negativity for W states. Some non-vanishing amount of tripartite correlation is necessary to disentangle pair of qubits while at the same time strongly entangling each qubit with the other pair. An analogous result holds for the average over the three possible bipartitions of the global state, Fig.~\ref{subfig:Amonogamy3a}-\ref{subfig:Amonogamy3b}.
\begin{equation}
2\overline{\Neg^2}_{1|1}\equiv\frac{2}{3}\left(\Neg^2_{a|b}+\Neg^2_{a|c}+\Neg^2_{b|c}\right)\leq\frac{1}{3}\left(\Neg^2_{a|bc}+\Neg^2_{b|ac}+\Neg^2_{c|ab}\right)\equiv\overline{\Neg^2}_{1|2}
\label{}
\end{equation}
One can notice how the distribution of the states in the W class is reproduced by the states in the GHZ class with small value of $\tau_3$ (dark blue). It is useful to contrast this behaviour with the saturation of the Arkai-Lieb inequality, but we postpone that discussion till we discuss the situation with more qubits.

As a measure of the strength of correlations in the state it is also interesting to look at the average of the internal negativity between single qubits
\begin{equation}
\overline{\Neg}_{1|1}\equiv\frac{1}{3}\left(\Neg_{a|b}+\Neg_{a|c}+\Neg_{b|c}\right)
\label{}
\end{equation}
and compare it to the average negativity for bipartitions of the entire state $\overline{\Neg}_{1|2}$. One clearly sees that the internal negativity is always close to the maximum for states in the W class, Fig.~\ref{subfig:internalN3a}-\ref{subfig:internalN3b}. This corresponds to the common intuition for W-like entanglement as being more robust when a qubit is removed from the system. 

Note however that this concept of robustness is different from the definition we gave in the previous section. In that case the robustness is proportional to the negativity for a bipartition of the global state ($\Neg_{1|2}$), which on average is actually maximized by the GHZ state, and not the W state. The relation between this last concept of robustness and the pattern of internal entanglement will be discussed extensively in the next sections for larger systems. In order to keep this distinction clear we will reserve the expression \textit{robustness} to the noise-related quantity and refer to the robustness against qubits removal simply as \textit{internal entanglement}.

Finally one can ask for the monogamy of mutual information. It is straightforward to check that for pure states of only three qubits \tmi\ is identically zero; consequently the mutual information is always monogamous. 

%~~~~~~~~~~~~~~~~~~~~~~~~~~~~~~~~~~~~~~~~~~~~~~~~~~~~~~~~
\section{Four qubits}
\label{sec:4q}
%~~~~~~~~~~~~~~~~~~~~~~~~~~~~~~~~~~~~~~~~~~~~~~~~~~~~~~~~~

In this section we consider the more interesting case of four qubits, we will see that the addition of a single qubits introduces much more structure and correspondingly several new investigations are possible.

For a four qubits system two kinds of bipartitions of the global state are possible, using the notation of the previous section we will refer to them as $(1|3)$ and $(2|2)$. As before, the first case is less interesting as the entropy essentially carries the same information as the negativity. The second case instead is more interesting, for such bipartition we can now compute both the negativity and entanglement entropy and ask what kind of information about the state one can gain by comparing them. More specifically we will consider the specific robustness  \eqref{eq:Rdef} and show that it conforms to the  interpretation we wish to give it.

A second novelty of a four qubits system is the possibility to investigate the properties of different states with respect to the disentangling theorem for the negativity. In particular, we explore the relation to the saturation of Araki-Lieb inequality for entanglement entropy. This is again inspired from holography owing the occurrence of the entanglement plateaux phenomena \cite{Hubeny:2013gta} there.

Finally, it is now possible to obtain a mixed state of three qubits by simply tracing out a single qubit, the value of $\tmi$ is then non-trivial and one can investigate the monogamy of mutual information in relation to the structure of entanglement of the state.

In \S\ref{subsec:4qgeneric} we will start these investigation for random generic states. Following this in \S\ref{subsec:4qclasses} we will introduce a SLOCC classification for four qubits systems and apply the extend the consideration to the different classes.

%~~~~~~~~~~~~~~~~~~~~~~~~~~~~~~~~~~~~~~~~~~~~~~~~~~~
\subsection{Generic states of 4 qubits}
\label{subsec:4qgeneric}
%~~~~~~~~~~~~~~~~~~~~~~~~~~~~~~~~~~~~~~~~~~~~~~~~~~~~

We begin our discussion here by picking out random pure states of 4-qubits. We will subject these to various examinations, testing for the saturation of the AL inequality, the disentangling theorem, and finally explain how the specific robustness can play a useful role in delineating internal entanglement structure.

% MONOGAMY AND ARAKI-LIEB 
%
\begin{figure}[tb]
\centering
\begin{subfigure}{0.49\textwidth}
\includegraphics[width=\textwidth]{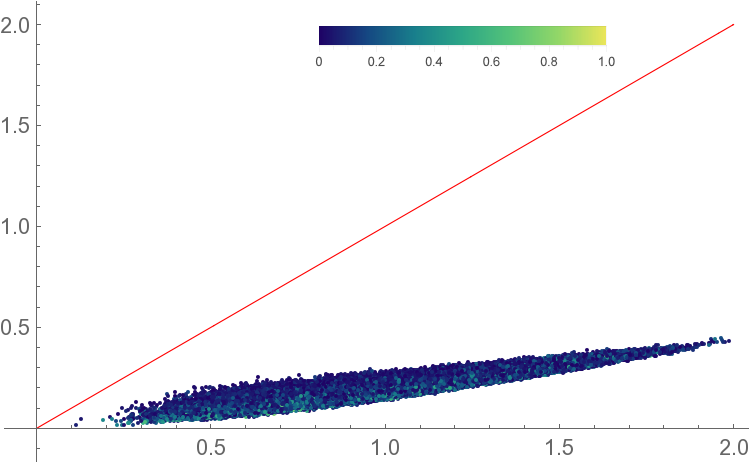}
\put(-204.5,143){\makebox(0,0){{\tiny $\Neg^2_{ab|c}+\Neg^2_{ab|d}$}}}
\put(-12,-6){\makebox(0,0){{\tiny $\Neg^2_{ab|cd}$}}}
\put(-82,136){\makebox(0,0){{\tiny $\tau_4$}}}
\caption{}
\label{fig:4qmonog_a}
\end{subfigure}
\hfill
\begin{subfigure}{0.49\textwidth}
\includegraphics[width=\textwidth]{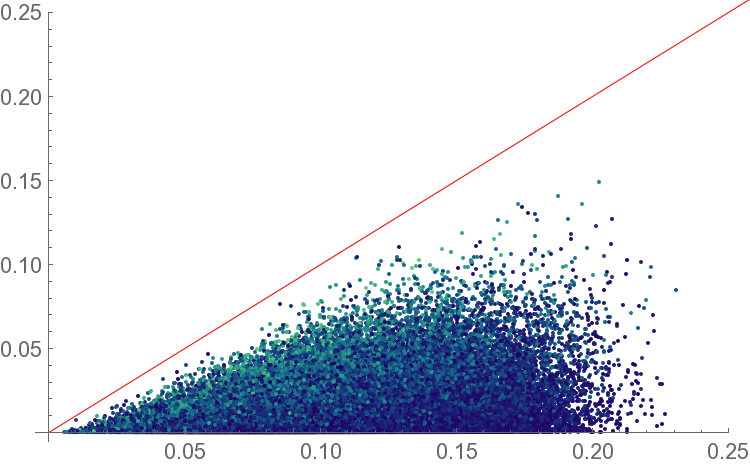}
\put(-204.5,140){\makebox(0,0){{\tiny $\Neg^2_{a|b}+\Neg^2_{a|c}$}}}
\put(-10,-6){\makebox(0,0){{\tiny $\Neg^2_{a|bc}$}}}
\put (-40,102) {\makebox(0,0){
	\begin{tikzpicture}
	\draw[->] (0,0)--(5.26pt,-8.5pt);
	\end{tikzpicture}
}}
\put(-28,105){\makebox(0,0){{\tiny{$f^{(3)}$}}}}
\caption{}
\label{fig:4qmonog_b}
\end{subfigure}

\begin{subfigure}{0.49\textwidth}
\includegraphics[width=\textwidth]{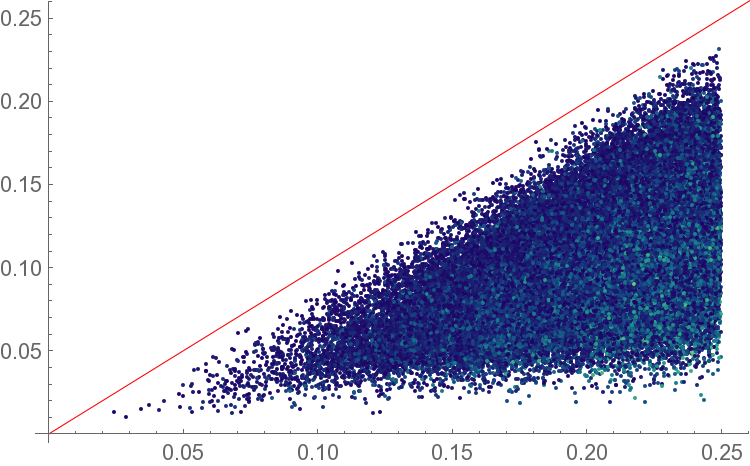}
\put(-204.5,141){\makebox(0,0){{\tiny $\Neg^2_{a|bc}+\Neg^2_{a|d}$}}}
\put(-12,-6){\makebox(0,0){{\tiny $\Neg^2_{a|bcd}$}}}
\put (-5,124) {\makebox(0,0){
	\begin{tikzpicture}
	\draw[->] (0,0)--(5.26pt,-8.5pt);
	\end{tikzpicture}
}}
\put(-10,135){\makebox(0,0){{\tiny{$f^{(4)}$}}}}
\caption{}
\label{fig:4qmonog_c}
\end{subfigure}
\hfill
\begin{subfigure}{0.49\textwidth}
\includegraphics[width=\textwidth]{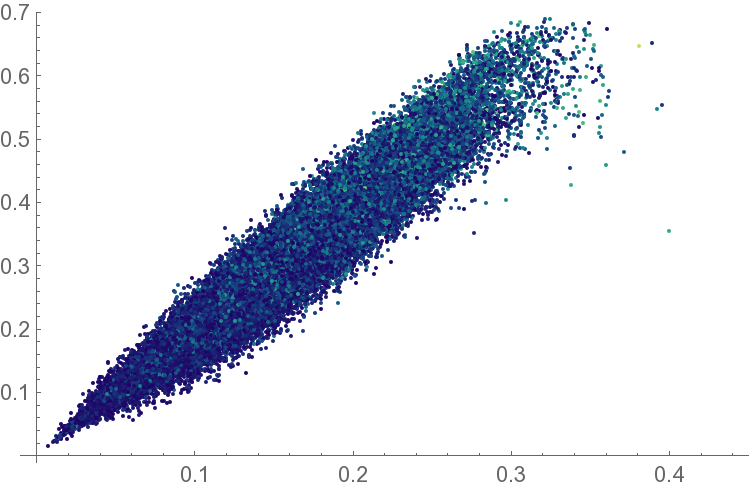}
\put(-16,-6){\makebox(0,0){{\tiny $\Delta\Neg_{ABC}$}}}
\put(-204.5,145){\makebox(0,0){{\tiny $\Delta S_{AB}$}}}
\caption{}
\label{fig:4qmonog_d}
\end{subfigure}
\caption{Monogamy of the squared negativity for $100000$ random states. (a), (b), (c) show inequalities \eqref{eq:4qmonog1}, \eqref{eq:4qmonog2}, \eqref{eq:4qmonog3} respectively. (d) Contrasting Araki-Lieb saturation with the disentangling theorem for negativities.}
\label{fig:4qmonog}
\end{figure}
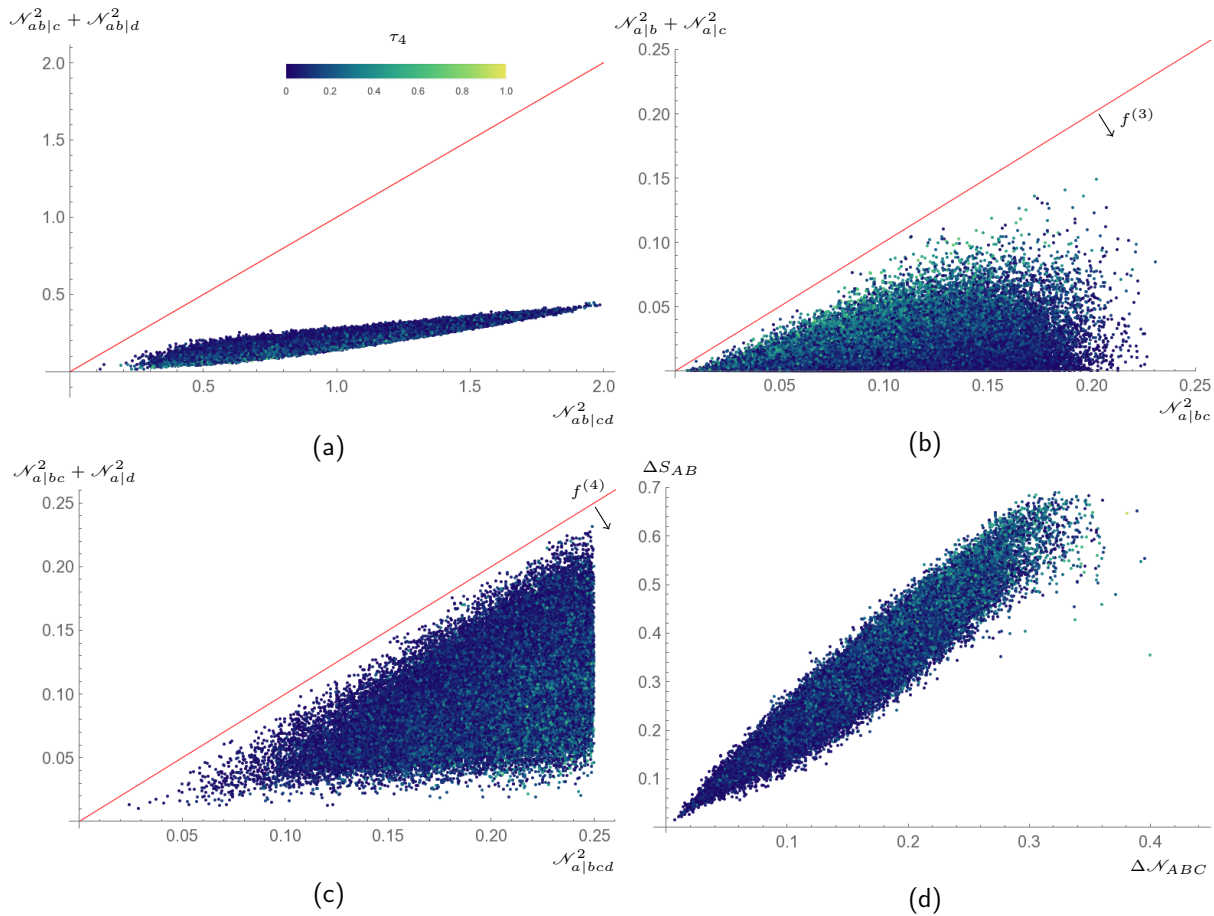
%\afterpage{\clearpage}

%~~~~~~~~~~~~~~~~~~~~~~~~~~~~~~~~~~~~~~~~~~~~~~~
\subsubsection{Monogamy of the negativity and disentangling theorem}
%~~~~~~~~~~~~~~~~~~~~~~~~~~~~~~~~~~~~~~~~~~~~~~

Similarly to the three qubits case, it is interesting to ask how the saturation of the monogamy is related to the four-partite entanglement measured by $\tau_4$. There are now three different kinds of monogamy relations (up to permutation of the qubits), depending on different ways to partition into subsystems:
\begin{subequations}
\begin{align}
\Neg^2_{ab|c}+\Neg^2_{ab|d}\leq\Neg^2_{ab|cd} \label{eq:4qmonog1}\\
\Neg^2_{a|b}+\Neg^2_{a|c}\leq\Neg^2_{a|bc}\label{eq:4qmonog2}\\
\Neg^2_{a|bc}+\Neg^2_{a|d}\leq\Neg^2_{a|bcd}\label{eq:4qmonog3}
\end{align}
\end{subequations}
The results are plotted in Fig.~\ref{fig:4qmonog_a}-\ref{fig:4qmonog_b}-\ref{fig:4qmonog_c}; note that the Hilbert space is now much larger and by random sampling one only covers a small portion of the space. Curiously, in the second case \eqref{eq:4qmonog2},  Fig.~\ref{fig:4qmonog_b}, the monogamy relation seems to be saturated more by states with a higher content of multipartite entanglement while in the third case \eqref{eq:4qmonog3}, Fig.~\ref{fig:4qmonog_c}, the states with low tangle saturate monogamy inequality.
 It is tempting to interpret this result in terms of residual multipartite entanglement  
 (cf., three qubit discussion in \S\ref{sec:3qubits}).

 Rewriting the saturation of \eqref{eq:4qmonog2} as
\begin{align}
\Neg^2_{a|b}+\Neg^2_{a|c}+f^{(3)}_{abc}=\Neg^2_{a|bc}
\end{align}
where $f^{(3)}$ is some measure of mixed residual tripartite correlation, it is natural to conjecture that high values of $\tau_4$ correspond to small values of bipartite ($\Neg^2$) and tripartite ($f^{(3)}$) correlation. $\tau_4\sim 1$ then implies $f^{(3)}\sim 0$ and the saturation of \eqref{eq:4qmonog2}. Similarly one could rewrite the saturation of \eqref{eq:4qmonog3} as 
\begin{align}
\Neg^2_{a|bc}+\Neg^2_{a|d}+f^{(4)}_{abcd}=\Neg^2_{a|bcd}
\end{align}
where now $f^{(4)}$ is some measure of $4$-partite correlation related to $\tau_4$. The saturation of \eqref{eq:4qmonog3} then corresponds to $\tau_4\sim 0$.

Let us now turn to the connection between the monogamy of the negativity and the saturation of Araki-Lieb inequality (AL) \cite{Araki:1970ba}  for entanglement entropy \cite{Zhang:2012fp}. Recall that the AL inequality reads:
\begin{equation}
|S(A)-S(B)|\leq S(AB)
\label{eq:AL1}
\end{equation}
For a joint system $A\cup B$ in a mixed state,  for reasons explained hitherto, it is difficult to interpret AL in terms of quantum correlations. In order to understand what kind of constraint AL implies for the internal structure of entanglement of the state, it is convenient to introduce the purification $C$ of the state $AB$. Thus for a system $U$ of $N$ qubits we then consider only tripartitions such that $A\cup B\cup C\equiv U$ and the global state is pure. With this choice one can then rewrite \eqref{eq:AL1} as
\begin{align}
|S_{A|BC}-S_{B|AC}|\leq S_{C|AB}
\label{eq:AL2}
\end{align}
where by expressions like $S_{A|BC}$ we mean the entropy of entanglement between $A$ and $BC$ (which of course is $S(A)=S(BC)$), stressing the interpretation of the von Neumann entropy as a measure of entanglement between a subsystem and its complement. 

In the case of four qubits then there is only one possible kind of tripartition, up to qubits permutation, i.e., $1|1|2$. We consider the set-up $A=\{a\}$, $B=\{bc\}$, $C=\{d\}$, \eqref{eq:AL2} then reads
\begin{align}
|S_{a|bcd}-S_{bc|ad}|\leq S_{d|abc}
\label{eq:AL3}
\end{align}
Permuting the qubits one obtains a set of constraints on the internal pattern of entanglement of the global state. We want to ask how this set of constraints is related to the one obtained from the monogamy of the negativity. In particular it is known that AL is in general difficult to saturate and it is certainly not saturated by generic states.\footnote{ For further discussion we refer the reader to \cite{Zhang:2012fp} for general analysis of AL saturation, \cite{Hubeny:2013gta} for explicit examples where the saturation occurs in holographic systems, and \cite{Headrick:2013zda} for a general analysis of AL saturation of holographic entanglement entropy. We should note that the saturation of AL in holography is not generic, but does happen in a large class of examples involving bulk spacetimes with horizons.} We want to investigate when this saturation actually happens and for which distribution of internal negativities. 

More specifically, we invoke the disentangling theorem for the negativity of \citep{He:2014aa} for this purpose. For a pure state of $U\equiv A\cup B\cup C$, if $\Neg_{A|BC}=\Neg_{A|B}$ then it is possible to partition $B=B_1\cup B_2$ such that the state factorizes $\ket{\Psi}=\ket{\psi_{AB_1}}\otimes\ket{\psi_{B_2C}}$. For our specific set-up the condition for the disentangling theorem is: $\Neg_{a|bcd}=\Neg_{a|bc}$. Eq.\eqref{eq:4qmonog3} then implies $\Neg^2_{a|d}\leq 0$, which is absurd; the only possible solution is saturation of the monogamy relation and in particular $\Neg_{a|d}=0$. The consequence of the disentangling theorem is even stronger, not only there is no distillable entanglement between $a|d$, but there is no entanglement at all and the global state factorizes either as $\ket{\psi_{ab}}\otimes\ket{\psi_{cd}}$ or as $\ket{\psi_{ac}}\otimes\ket{\psi_{bd}}$. 

The disentangling theorem implies saturation of AL in the following way: because of the factorization of the state one has $S(B)=S(B_1)+S(B_2)$, but since the individual states in the product are pure $S(B)=S(A)+S(C)$. Note that $C$ now is the purification of $AB$, hence $S(B)=S(A)+S(AB)$ i.e., AL is saturated. We measure the saturation for random states by 
$\Delta S_{AB}=S(AB)-|S(A)-S(B)|\geq 0$ and correspondingly the amount by which the states match the hypothesis of the disentangling theorem by $\Delta\Neg_{ABC}=\Neg_{A|BC}-\Neg_{A|B}\geq 0$. With our specific choice these quantities are
\begin{align}
\Delta S_{AB}=S_{d|abc}-|S_{a|bcd}-S_{bc|ad}|\,,\hspace{1cm} \Delta\Neg_{ABC}=\Neg_{a|bcd}-\Neg_{a|bc}
\label{eq:deltaSN}
\end{align}
The results are shown in Fig.~\ref{fig:4qmonog_d}. Note that as expected there are no states that saturate AL, while the states that get closer to the saturation correspond to the states that almost satisfy the conditions of the disentangling theorem. This in itself is not particularly surprising, as we discussed AL inequalities are statistically difficult to saturate. Furthermore the result shows that matching the hypothesis of the disentangling theorem is a sufficient conditions for a state to saturate AL. On the other hand it is interesting to ask if these conditions are necessary and what is the meaning of states that saturate AL without factorization. We will comment more on this issue as we investigate further aspects of the relation in greater detail in the following. 

% RATIO PLOTS (unconstrained entropy and average value)
%
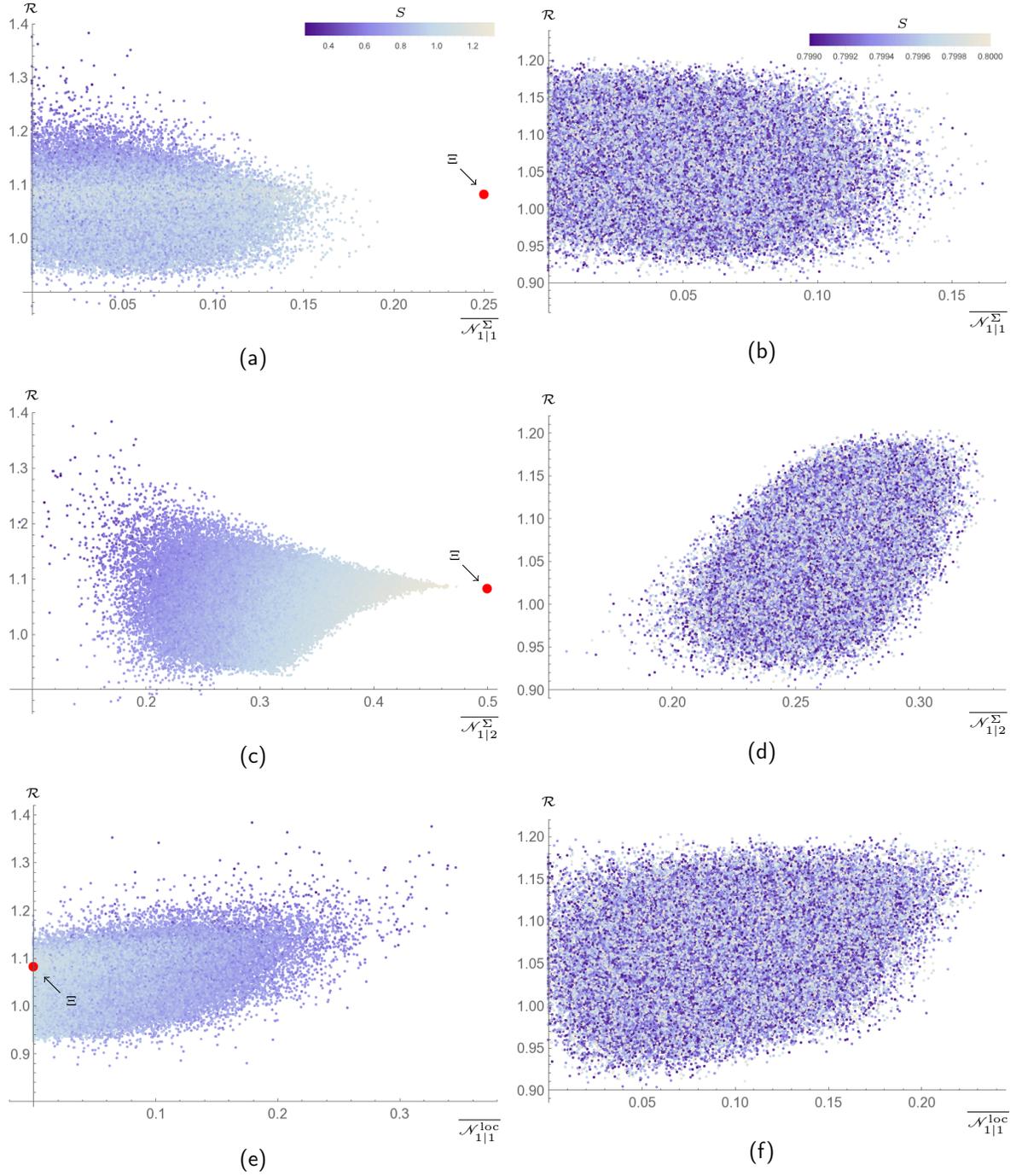
\begin{figure}[t]
\centering
\begin{subfigure}{0.49\textwidth}
\includegraphics[width=\textwidth]{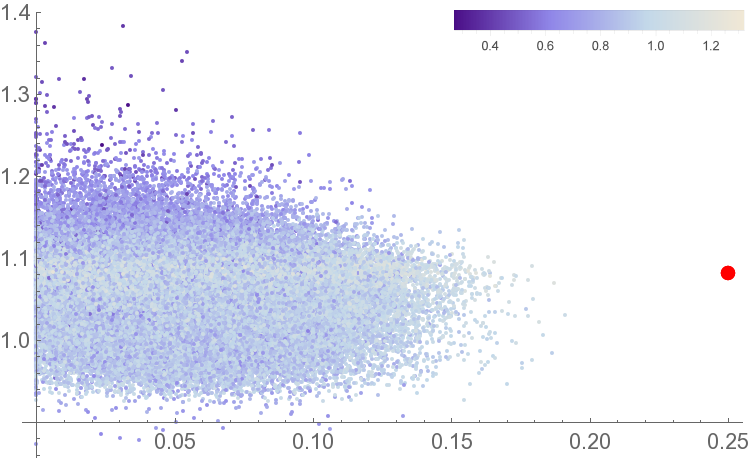}
\put(-8,-6){\makebox(0,0){{\tiny $\overline{\Neg^\Sigma_{1|1}}$}}}
\put(-207,137){\makebox(0,0){{\tiny $\mathcal{R}$}}}
\put(-43,135){\makebox(0,0){{\tiny $S$}}}
\put (-13,62) {\makebox(0,0){
	\begin{tikzpicture}
	\draw[->] (0,0)--(7.07pt,-7.07pt);
	\end{tikzpicture}
}}
\put(-20,70){\makebox(0,0){{\tiny{$\Xi$}}}}
\subcaption{}
\label{subfig:4qR1_1}
\end{subfigure}
\hfill
\begin{subfigure}{0.49\textwidth}
\includegraphics[width=\textwidth]{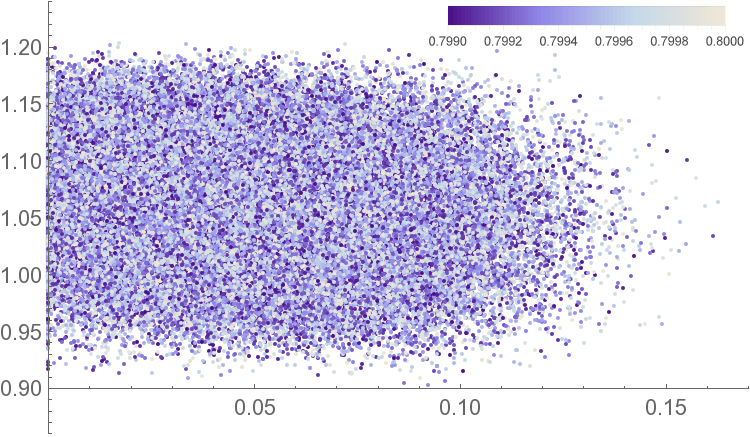}
\put(-8,-4){\makebox(0,0){{\tiny $\overline{\Neg^\Sigma_{1|1}}$}}}
\put(-203,133){\makebox(0,0){{\tiny $\mathcal{R}$}}}
\put(-46,129){\makebox(0,0){{\tiny $S$}}}
\subcaption{}
\label{subfig:4qR1_1Cons}
\end{subfigure}

\vspace{0.3cm}
\begin{subfigure}{0.49\textwidth}
\includegraphics[width=\textwidth]{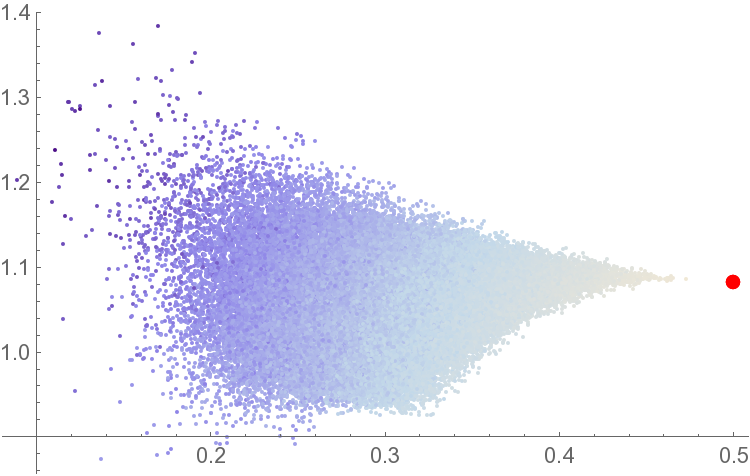}
\put(-8,-6){\makebox(0,0){{\tiny $\overline{\Neg^\Sigma_{1|2}}$}}}
\put(-206.5,142){\makebox(0,0){{\tiny $\mathcal{R}$}}}
\put (-12,63) {\makebox(0,0){
	\begin{tikzpicture}
	\draw[->] (0,0)--(7.07pt,-7.07pt);
	\end{tikzpicture}
}}
\put(-19,71){\makebox(0,0){{\tiny{$\Xi$}}}}
\subcaption{}
\label{subfig:4qR1_2}
\end{subfigure}
\hfill
\begin{subfigure}{0.49\textwidth}
\includegraphics[width=\textwidth]{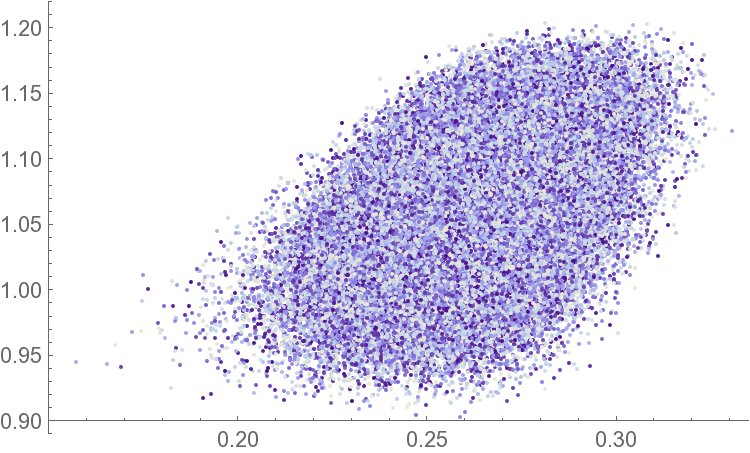}
\put(-8,-6){\makebox(0,0){{\tiny $\overline{\Neg^\Sigma_{1|2}}$}}}
\put(-203,138){\makebox(0,0){{\tiny $\mathcal{R}$}}}
\subcaption{}
\label{subfig:4qR1_2Cons}
\end{subfigure}

\vspace{0.3cm}
\begin{subfigure}{0.49\textwidth}
\includegraphics[width=\textwidth]{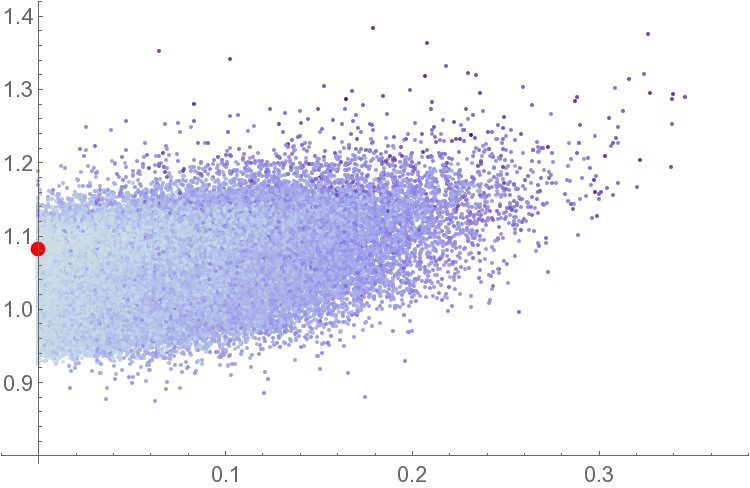}
\put(-8,-4){\makebox(0,0){{\tiny $\overline{\Neg^\text{loc}_{1|1}}$}}}
\put(-206,146){\makebox(0,0){{\tiny $\mathcal{R}$}}}
\put (-197.5,61) {\makebox(0,0){
	\begin{tikzpicture}
	\draw[->] (0,0)--(-7.07pt,7.07pt);
	\end{tikzpicture}
}}
\put(-189,54){\makebox(0,0){{\tiny{$\Xi$}}}}
\subcaption{}
\label{subfig:4qRInt}
\end{subfigure}
\hfill
\begin{subfigure}{0.49\textwidth}
\includegraphics[width=\textwidth]{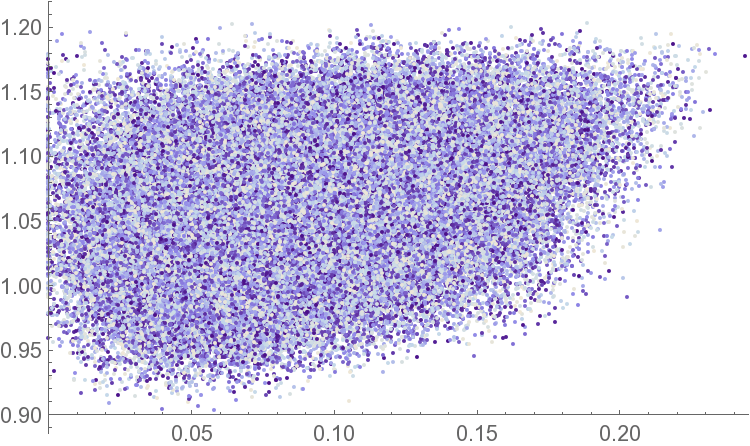}
\put(-8,-6){\makebox(0,0){{\tiny $\overline{\Neg^\text{loc}_{1|1}}$}}}
\put(-203,137){\makebox(0,0){{\tiny $\mathcal{R}$}}}
\subcaption{}
\label{subfig:4qRIntCons}
\end{subfigure} 
\caption{Average negativities across the entangling surface $\Sigma$ and inside the subsystems for a fixed bipartition ($100000$ states). The left panels show the results for random states with entropy in the range $(0.27292,1.32195)$, the large red dots show the maximally entangled state ($\Xi$). The right panels show states with a constrained value of the entropy $0.799\leq S(ab)\leq 0.8$.}
\label{}
\end{figure}
%\afterpage{\clearpage}

% RATIO PLOTS (low and high value of entropy)
%
\begin{figure}[t]
\centering
\begin{subfigure}{0.49\textwidth}
\includegraphics[width=\textwidth]{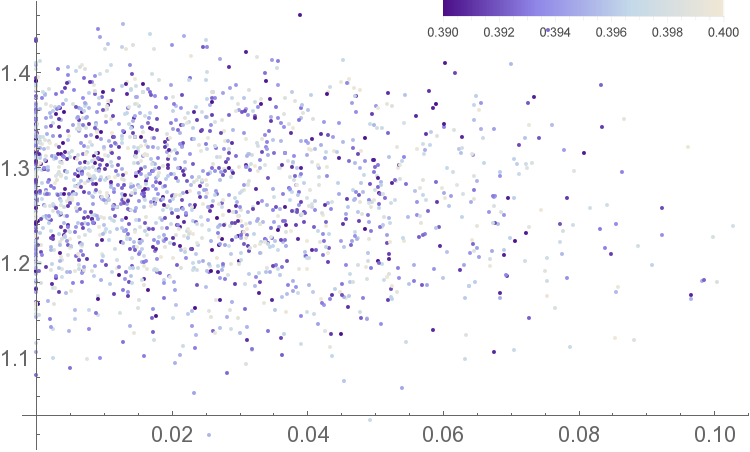}
\put(-8,-6){\makebox(0,0){{\tiny $\overline{\Neg^\Sigma_{1|1}}$}}}
\put(-207,137){\makebox(0,0){{\tiny $\mathcal{R}$}}}
\put(-45,135){\makebox(0,0){{\tiny $S$}}}
\subcaption{}
\label{subfig:4qR1_1Cons2}
\end{subfigure}
\hfill
\begin{subfigure}{0.49\textwidth}
\includegraphics[width=\textwidth]{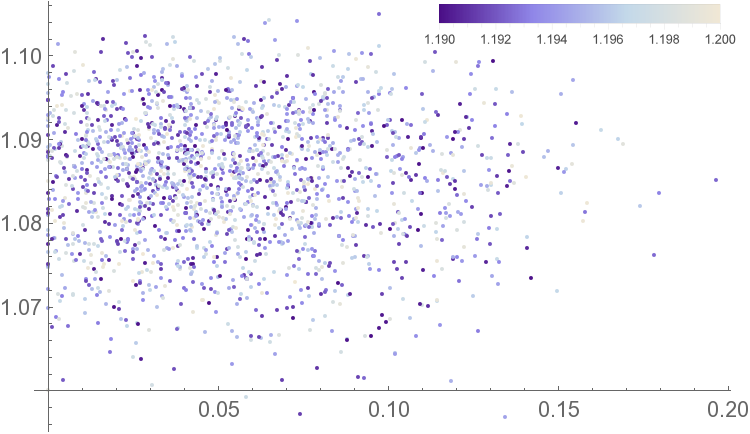}
\put(-8,-6){\makebox(0,0){{\tiny $\overline{\Neg^\Sigma_{1|1}}$}}}
\put(-203.5,132){\makebox(0,0){{\tiny $\mathcal{R}$}}}
\put(-48,129){\makebox(0,0){{\tiny $S$}}}
\subcaption{}
\label{subfig:4qR1_1Cons3}
\end{subfigure}

\vspace{0.4cm}
\begin{subfigure}{0.49\textwidth}
\includegraphics[width=\textwidth]{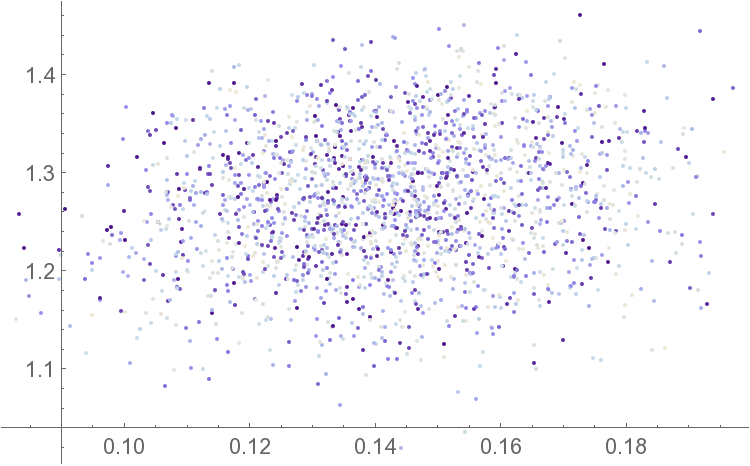}
\put(-8,-6){\makebox(0,0){{\tiny $\overline{\Neg^\Sigma_{1|2}}$}}}
\put(-200,141){\makebox(0,0){{\tiny $\mathcal{R}$}}}
\subcaption{}
\label{subfig:4qR1_2Cons2}
\end{subfigure}
\hfill
\begin{subfigure}{0.49\textwidth}
\includegraphics[width=\textwidth]{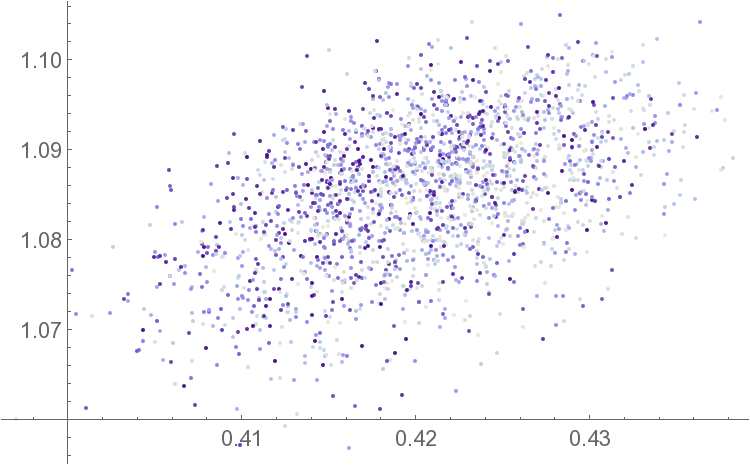}
\put(-8,-6){\makebox(0,0){{\tiny $\overline{\Neg^\Sigma_{1|2}}$}}}
\put(-198,141){\makebox(0,0){{\tiny $\mathcal{R}$}}}
\subcaption{}
\label{subfig:4qR1_2Cons3}
\end{subfigure}

\vspace{0.4cm}
\begin{subfigure}{0.49\textwidth}
\includegraphics[width=\textwidth]{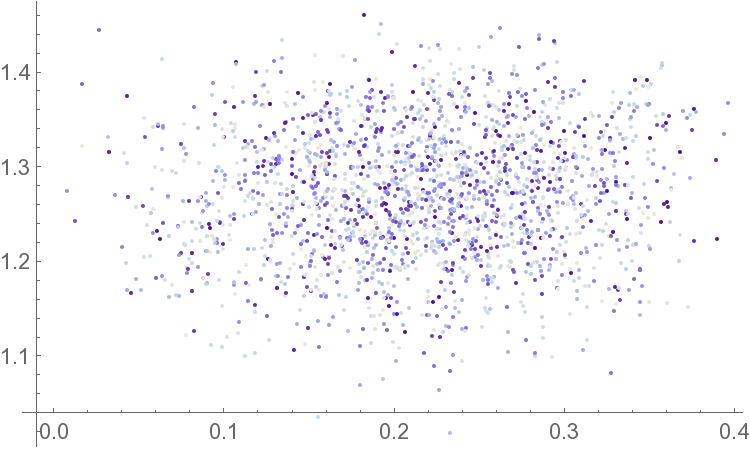}
\put(-8,-6){\makebox(0,0){{\tiny $\overline{\Neg^\text{loc}_{1|1}}$}}}
\put(-206.5,136){\makebox(0,0){{\tiny $\mathcal{R}$}}}
\subcaption{}
\label{subfig:4qRIntCons2}
\end{subfigure}
\hfill
\begin{subfigure}{0.49\textwidth}
\includegraphics[width=\textwidth]{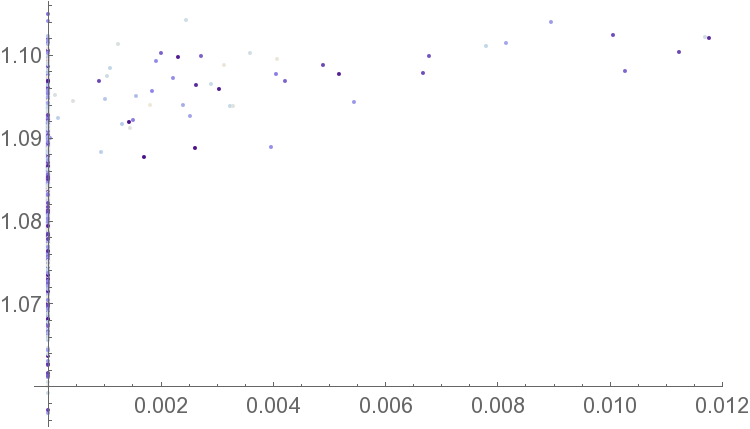}
\put(-8,-6){\makebox(0,0){{\tiny $\overline{\Neg^\text{loc}_{1|1}}$}}}
\put(-203.5,131){\makebox(0,0){{\tiny $\mathcal{R}$}}}
\subcaption{}
\label{subfig:4qRIntCons3}
\end{subfigure} 
\caption{Average negativities across the entangling surface $\Sigma$ and inside the subsystems for a fixed bipartition ($2000$ states). The left panels show the results for random states with entropy constrained in the range $0.39\leq S(ab)\leq 0.4$. The right panels show states with a constrained value of the entropy in $1.19\leq S(ab)\leq 1.2$.}
\label{fig:newslices}
\end{figure}
%\afterpage{\clearpage}

%~~~~~~~~~~~~~~~~~~~~~~~~~~~~~~~~~~~~~~~~~~~~~~~
\subsubsection{Negativity to entanglement ratio}
\label{sec:}
%~~~~~~~~~~~~~~~~~~~~~~~~~~~~~~~~~~~~~~~~~~~~~~
% \paragraph{2. Ratio $\Neg/S$:} 

We can now initiate the study of the ratio between entanglement entropy and negativity, which is one of the main motivations of the present work. In order for the entropy to be a sensible measure of quantum entanglement we only consider global bipartition of pure states. Furthermore, for the same reason discussed in the case of three qubits,  in the case of a 1|3 bipartition both the entanglement entropy and  negativity carry the same information. We will therefore focus on the $2|2$ bipartition only, where we can distinguish the negativity and entanglement.

Let us start with a particular partition ($ab|cd$) and refer to the fiducial separation of the two subsystems (i.e., the symbol ``$|$'') as the entangling surface ($\Sigma$). One can then compute both negativity and entropy for this particular bipartition. Since the state is pure the entropy is a measure of the amount of entanglement between the subsystems, intuitively the number of Bell's pairs that can be distilled.\footnote{ Note however that more precisely this would be an asymptotic statement.} On the other hand we interpret the negativity as the robustness of the entanglement between the subsystems. 

We stress again that this notion captures  the robustness against ``intelligent jamming'', which is in principle different from the common intuition about the robustness of the W states. In the latter case one actually refers to the amount of internal entanglement.  As described in \S\ref{sec:measures} we want to interpret the ratio ${\mathcal R}$ as capturing the specific robustness in a given state.  We explore the dependence of ${\mathcal R}$ on the entanglement structure of the state focusing on the internal pattern of entanglement.

For the bipartition $ab|cd$ we have $\Neg^{\text{loc}}_{a|b}$ and $\Neg^{\text{loc}}_{c|d}$ for local entanglement, while for entanglement across $\Sigma$ we consider all the possible negativities of the kind $\Neg^\Sigma_{1|1}$ and $\Neg^\Sigma_{1|2}$ where the sets of qubits are subsets of the original subsets of the bipartition. Finally we take the average of all the negativities of a particular kind, keeping the original bipartition fixed. The results are shown in Fig.~\ref{subfig:4qR1_1}-\ref{subfig:4qR1_2}-\ref{subfig:4qRInt} and show the dependence of the ratio on the internal entanglement with unconstrained value of the entropy across $\Sigma$. For states which are almost maximally entangled\footnote{ By maximally entangled state here we mean the usual state that maximizes entanglement for a given bipartition. For four qubits this is the state  which achieves this between two pairs, i.e., $\frac{1}{2}\sum_{i=1}^4 \ket{i}\otimes\ket{i}$ where $\ket{i}$ is the computational basis for a two-qubits system. We will reserve the symbol $\ket{\Xi_N}$ for such states.} the spread between possible values of entropy and negativity is very restricted. Interestingly the spread grows considerably for small values of the entropy, which also correspond to a higher value of the ratio. These are states that are less entangled but whose entanglement is particularly robust. In some sense it seems that if a state is highly entangled, its entanglement is more fragile against noise. Larger ${\cal R}$ corresponds to greater specific robustness, justifying our terminology.

One can also look for the distribution of the states when the entropy is almost fixed, the results are shown in Fig.~\ref{subfig:4qR1_1Cons}-\ref{subfig:4qR1_2Cons}-\ref{subfig:4qRIntCons}. The fixed value is picked to be somewhere in the middle-range of the entanglement spread for our sampling; other choices outside the edge regions lead to similar results.  Note that the dependence on the entropy now is completely random. Furthermore the ratio seems to be increasing when the ($1|2$) entanglement across the entangling surface is higher. On the other hand it seems not to depend on the average ($1|1$) entanglement. It will be useful to contrast this result against those for larger systems. For now we tentatively interpret the results as suggesting that states with a higher amount of ($1|2$) entanglement across the entangling surface, but the same value of the entropy, are more robust.

 Finally, we look at the average entanglement inside the subsystems specified by $\Sigma$. We find that high values of local negativity correspond to high value of the ratio although the converse is not always true. Fig.~\ref{fig:newslices} shows two other slices for different constraints on the entropy. For states which are almost maximally entangled (Fig.~\ref{subfig:4qR1_1Cons3}-\ref{subfig:4qR1_2Cons3}-\ref{subfig:4qRIntCons3}) the results agree with the previous analysis. On the other hand when the entropy is fixed but small (Fig.~\ref{subfig:4qR1_1Cons2}-\ref{subfig:4qR1_2Cons2}-\ref{subfig:4qRIntCons2}) the dependence of the ratio on $1|2$ entanglement becomes less evident. Furthermore a larger amount of local $1|1$ entanglement is not sufficient any more to produce higher values of the ratio. 

%~~~~~~~~~~~~~~~~~~~~~~~~~~~~~~~~~~~~~~~~~~~~~~~
\subsubsection{Monogamy of mutual information}
\label{sec:}
%~~~~~~~~~~~~~~~~~~~~~~~~~~~~~~~~~~~~~~~~~~~~~~
% \paragraph{3. Monogamy of mutual information:} 

For a system of four qubits there are in principle two different kinds of tripartite information one can look at, viz., either $\tmi(1|1|2)$, or $\tmi(1|1|1)$ after tracing out a single qubit. Since we are only working with pure states, in the first case we have $\tmi=0$, we will instead focus on the second case. There are in principle four different values of $\tmi$ depending on which qubit we choose to trace out. Nevertheless it is straightforward to check that they are all equal, there is actually a unique value of $\tmi$. 

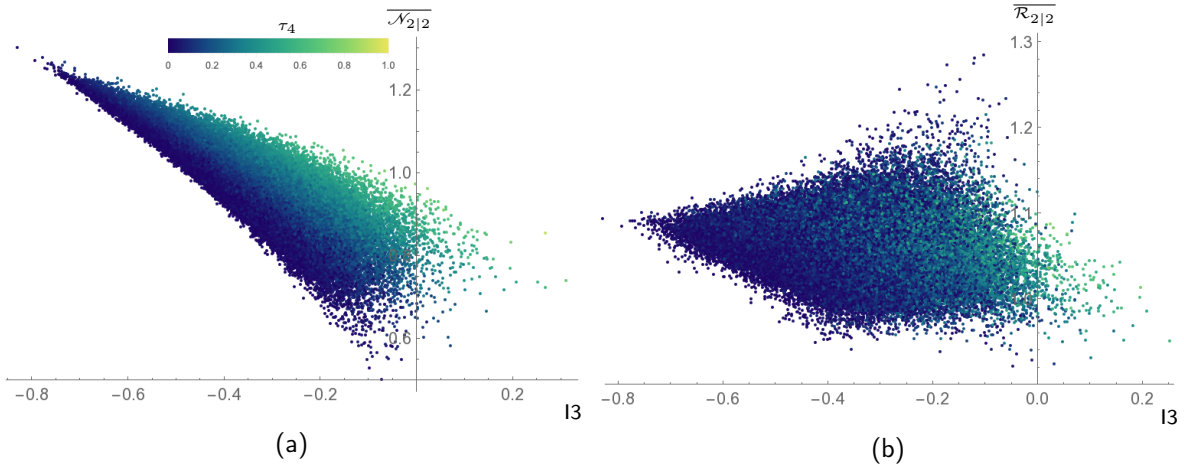
\begin{figure}[tb]
\centering
\begin{subfigure}{0.49\textwidth}
\includegraphics[width=\textwidth]{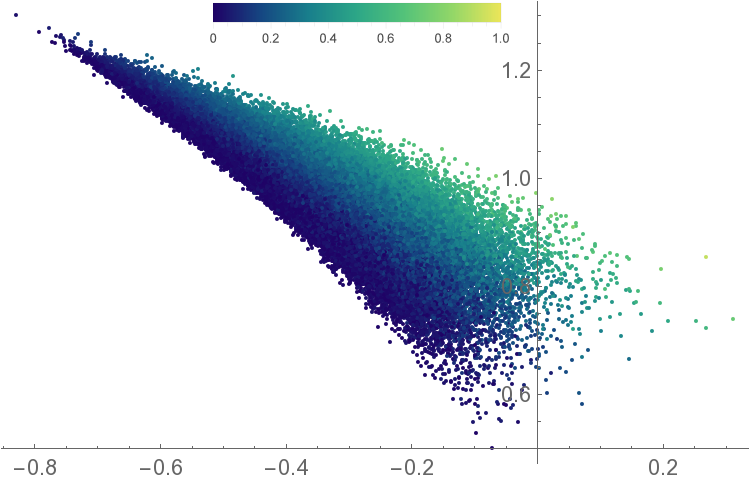}
\put(-2,-3){\makebox(0,0){{\scriptsize $\tmi$}}}
\put(-64,145){\makebox(0,0){{\tiny $\overline{\Neg_{2|2}}$}}}
\put(-110,142){\makebox(0,0){{\tiny $\tau_4$}}}
\caption{}
\label{subfig:4qI3N}
\end{subfigure}
\hfill
\begin{subfigure}{0.49\textwidth}
\includegraphics[width=\textwidth]{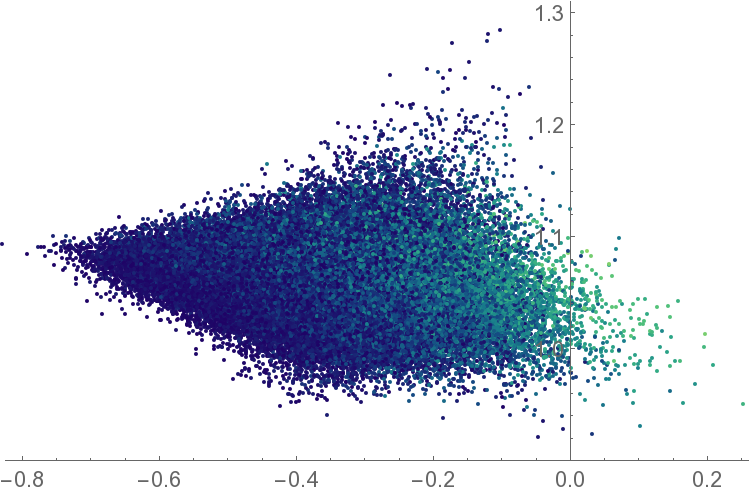}
\put(-2,-3){\makebox(0,0){{\scriptsize $\tmi$}}}
\put(-53,149){\makebox(0,0){{\tiny $\overline{\mathcal{R}_{2|2}}$}}}
\caption{}
\label{subfig:4qI3R}
\end{subfigure}
\caption{$100000$ random generic states to test monogamy of mutual information. (a) Average ratio between negativity and entropy compared to the values of I3 and $\tau_4$ (color map), (b) Average $2|2$ negativity compared to $\tmi$ and $\tau_4$ for the same states.}
\label{fig:4qI3}
\end{figure}

We compare this value to the amount of quantum multipartite entanglement measured by the tangle and the internal entanglement structure. As a measure of the robustness of the state we use the average negativity $\overline{\Neg_{2|2}}$, the result is shown in Fig.~\ref{subfig:4qI3N}. An evident result is the observation that statistically the monogamy of mutual information is not a very restrictive condition. States with low robustness and high values of $\tau_4$ seem to violate the monogamy of mutual information (positive $\tmi$) more easily. 

In Fig.~\ref{subfig:4qI3R} we show instead a similar plot for the average ratio $\overline{\mathcal R}$.  Now we note that the states that violate monogamy and have high value of $\tau_4$, also have  quite a small value of the ratio. This correlation is rather suggestive, and if true, implies  that the specific robustness measured by ${\mathcal R}$ could be a useful diagnostic vis a vis monogamy of mutual information. However, before we arrive at this conclusion, we should do some more sanity checks, which we now turn to, by considering a classification of four qubit states.

% MONOGAMY OF MI PLOTS
%
%\afterpage{\clearpage}

%~~~~~~~~~~~~~~~~~~~~~~~~~~~~~~~~~~~~~~~~~~~~~~~~~~~~~~~
\subsection{SLOCC classification of 4 qubit states}
\label{subsec:4qclasses}
%~~~~~~~~~~~~~~~~~~~~~~~~~~~~~~~~~~~~~~~~~~~~~~~~~~~~~~~~~~

For four or more qubits it is known that the number of inequivalent SLOCC classes is infinite \cite{Dur:2000aa}. Nevertheless, motivated by the result for three qubits, one can still look for special states that maximize mixed internal correlations. These are a higher dimensional generalization of the W states of three qubits. For states of four qubits \cite{Verstraete:aa} gave an SLOCC classification into eight special classes of this kind, plus an additional class ($\qsl{1}$ in the following) which contains infinitely many SLOCC classes and is related to generic states (see below). Using the standard computational basis for 4-qubits we can write down the classes explicitly as\footnote{ For simplicity we drop the normalization factor in the definition of the classes.}
\begin{align}
\ket{\qsl{1}}=\;&\frac{a+d}{2}(\ket{0000}+\ket{1111})+\frac{a-d}{2}(\ket{0011}+\ket{1100})+\frac{b+c}{2}(\ket{0101}+\ket{1010})\nonumber\\
	& +\; \frac{b-c}{2}(\ket{0110}+\ket{1001})\nonumber\\
\ket{\qsl{2}} = \;&\frac{a+b}{2}(\ket{0000}+\ket{1111})+\frac{a-b}{2}(\ket{0011}+\ket{1100})+c(\ket{0101}+\ket{1010})+\ket{0110}\nonumber\\
\ket{\qsl{3}}=\;&a(\ket{0000}+\ket{1111})+b(\ket{0101}+\ket{1010})+\ket{0110}+\ket{0011}\nonumber\\
\ket{\qsl{4}}=\;&a(\ket{0000}+\ket{1111})+\frac{a+b}{2}(\ket{0101}+\ket{1010})+\frac{a-b}{2}(\ket{0110}+\ket{1001})\nonumber\\
	&+\; \frac{i}{\sqrt{2}}(\ket{0001}+\ket{0010}+\ket{0111}+\ket{1011})\nonumber\\
\ket{\qsl{5}}=\;&a(\ket{0000}+\ket{0101}+\ket{1010}+\ket{1111})+i\ket{0001}+\ket{0110}-i\ket{1011}\nonumber\\
\ket{\qsl{6}}=\;&a(\ket{0000}+\ket{1111})+\ket{0011}+\ket{0101}+\ket{0110}\nonumber\\
\ket{\qsl{7}}=\;&\ket{0000}+\ket{0101}+\ket{1000}+\ket{1110}\nonumber\\
\ket{\qsl{8}}=\;&\ket{0000}+\ket{1011}+\ket{1101}+\ket{1110}\nonumber\\
\ket{\qsl{9}}=\;&\ket{0000}+\ket{0111}
\label{eq:qsl19}
\end{align}
Here $a,b,c,d$ are complex numbers which appear as eigenvalues of an operator used in constructing the classification scheme. The classification only includes states where all the qubits are entangled.\footnote{ The class $\qsl{9}$ is an exception as it can be written as $\ket{0}\otimes\ket{\text{GHZ}_3}$. This was indeed one of the motivation for \cite{Lamata:2007aa} to consider an alternative classification of the four qubits states into eight classes.} The first class is the ``generic class'' in the sense that any generic state of four qubits can be mapped to a state in $\qsl{1}$ by SLOCC. This class is not  unique under SLOCC; as clarified in \cite{Gour:2010aa} it is dense in the space of generic states, but it actually contains an infinite number of classes. The remaining classes are thus of measure zero, but contain the maximal amount of internal mixed bipartite or tripartite entanglement (there are some exceptions which we note below). In the following we will focus in particular on the classes $\qsl{1}$-$\qsl{6}$ -- the last three only contain exceptional states. The W state of four qubits belongs to $\qsl{4}$, while the GHZ 4-qubit state is in $\qsl{1}$ ($a=d$, $b=c=0$), as expected.

%DISTINGUISHING THE CLASSES (PLOTS)
%
\begin{figure}[tb]
\centering
\begin{subfigure}{0.49\textwidth}
\includegraphics[width=\textwidth]{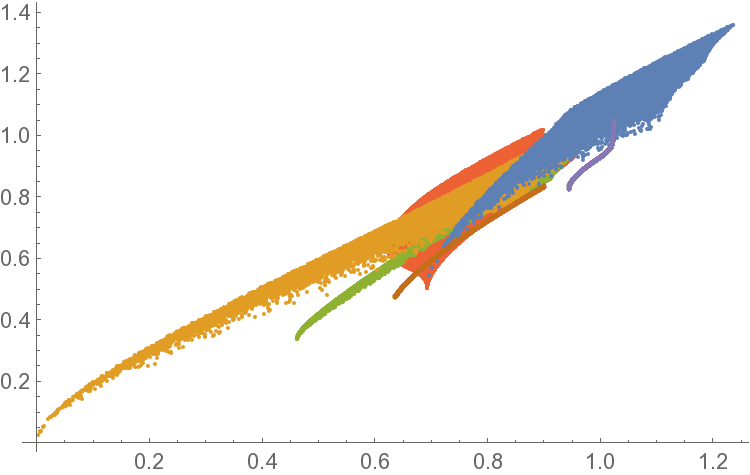}
\put(-7,-6){\makebox(0,0){{\tiny $\overline{S_{2|2}}$}}}
\put(-204.5,142){\makebox(0,0){{\tiny $\overline{\Neg_{2|2}}$}}}
\caption{}
\label{}
\end{subfigure}
\hfill
\begin{subfigure}{0.49\textwidth}
\includegraphics[width=\textwidth]{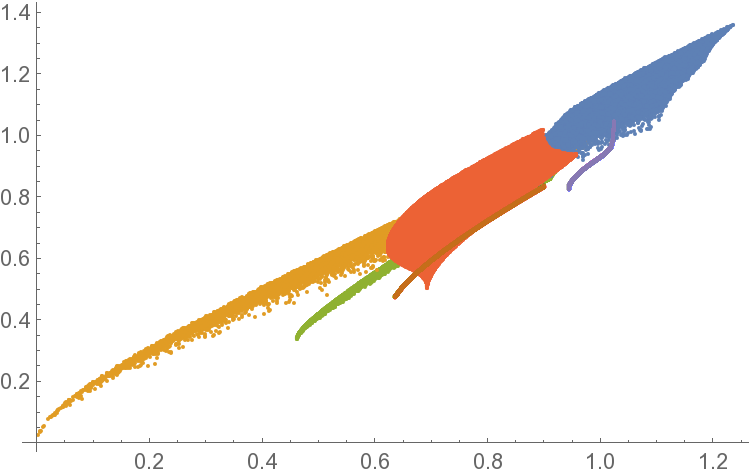}
\put(-7,-6){\makebox(0,0){{\tiny $\overline{S_{2|2}}$}}}
\put(-204.5,142){\makebox(0,0){{\tiny $\overline{\Neg_{2|2}}$}}}
\caption{}
\label{}
\end{subfigure}

\begin{subfigure}{0.49\textwidth}
\includegraphics[width=\textwidth]{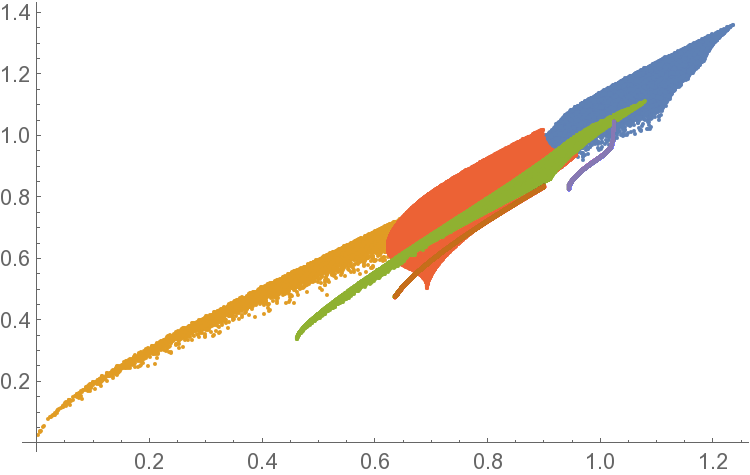}
\put(-7,-6){\makebox(0,0){{\tiny $\overline{S_{2|2}}$}}}
\put(-204.5,142){\makebox(0,0){{\tiny $\overline{\Neg_{2|2}}$}}}
\caption{}
\label{}
\end{subfigure}
\hfill
\begin{subfigure}{0.49\textwidth}
\includegraphics[width=\textwidth]{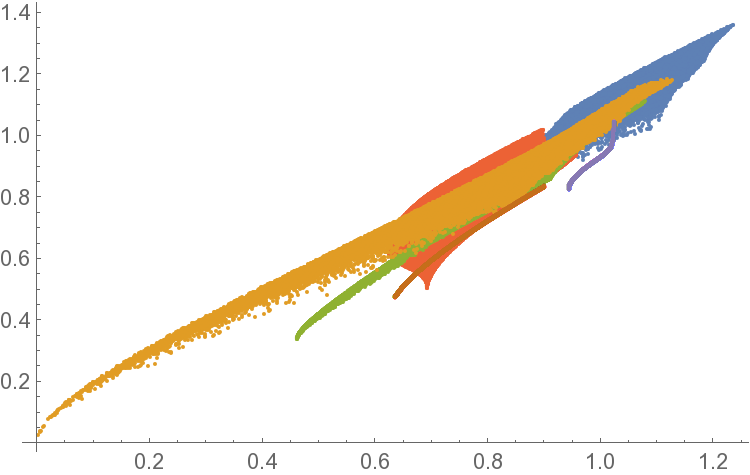}
\put(-7,-6){\makebox(0,0){{\tiny $\overline{S_{2|2}}$}}}
\put(-204.5,142){\makebox(0,0){{\tiny $\overline{\Neg_{2|2}}$}}}
\caption{}
\label{}
\end{subfigure}
\caption{Comparison of the averaged entropy and negativity for the maximal bipartition ($50000$ random states per class). The four panels show the same plot with different overlap. The color-class correspondence is as follows: $\textcolor[rgb]{0.368417, 0.506779, 0.709798}{\blacksquare}=\qsl{1}$, $\textcolor[rgb]{0.880722, 0.611041, 0.142051}{\blacksquare}=\qsl{2}$, $\textcolor[rgb]{0.560181, 0.691569, 0.194885}{\blacksquare}=\qsl{3}$, $\textcolor[rgb]{0.922526, 0.385626, 0.209179}{\blacksquare}=\qsl{4}$, $\textcolor[rgb]{0.528488, 0.470624, 0.701351}{\blacksquare}=\qsl{5}$, $\textcolor[rgb]{0.772079, 0.431554, 0.102387}{\blacksquare}=\qsl{6}$. Note that while generic states can be mapped into class $\qsl{1}$ by a SLOCC, picking states in $\qsl{1}$ according to the ansatz in \eqref{eq:qsl19} does not sample this genericity. Hence the region covered by $\textcolor[rgb]{0.368417, 0.506779, 0.709798}{\blacksquare}=\qsl{1}$ is not the entire domain of the plot above, but only a subregion thereof.  }
\label{fig:4qAEvAN}
\end{figure}
%\afterpage{\clearpage}

\paragraph{Distinguishing the classes:} Given a single state one could ask how it is possible to identify the corresponding class using different measures of entanglement. In principle one could compute the negativity for each bipartiton of the entire system and all bipartitions of each possible subsystem. It is natural to expect that the collection of this data allows some resolution of the classes. 

We want instead to ask a different question. As advertised earlier, we are concerned with knowing to what extent it is possible to distinguish states in different classes if one is restricted to use measures of entanglement for pure states. Our motivation comes from holography, where at present we only know how to compute the negativity for pure states. We can then combine information one extracts from both negativity and entropy, and see how much we can learn about the entanglement structure. 

For simplicity, we  focus on the maximal bipartitions, i.e., $2|2$.  One can easily compute the average negativity and average entropy for random states in the various classes, and average over the three possible inequivalent bipartitions. The results are shown in Fig.~\ref{fig:4qAEvAN}. One can see that even if the combined information extracted from negativity and entropy is not enough to completely resolve the classes, one can still discriminate them in some ranges of the values of the two measures. 

% MONOGAMY AND ARAKI-LIEB FOR CLASSES 
%
\begin{figure}[t]
\centering
\begin{subfigure}{0.49\textwidth}
\includegraphics[width=\textwidth]{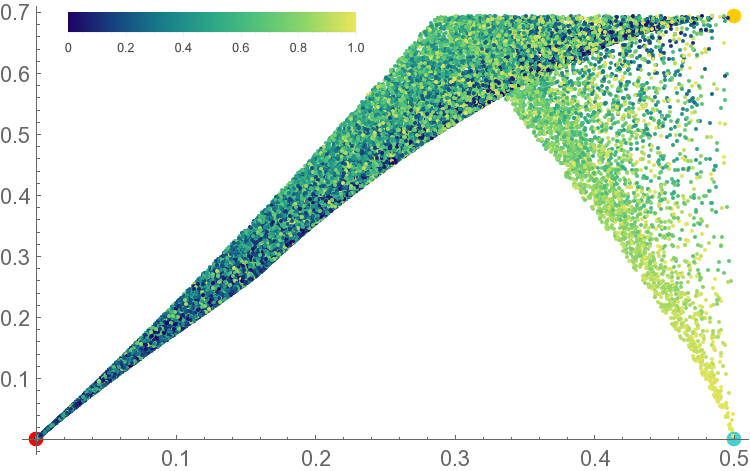}
\put(-12,-6){\makebox(0,0){{\tiny $\Delta\Neg_{ABC}$}}}
\put(-204.5,140){\makebox(0,0){{\tiny $\Delta S_{AB}$}}}
\put(-157,137){\makebox(0,0){{\tiny $\tau_4$}}}
\put (-13.5,15) {\makebox(0,0){
	\begin{tikzpicture}
	\draw[->] (0,0)--(8.5pt,-5.26pt);
	\end{tikzpicture}
}}
\put(-22,21){\makebox(0,0){{\tiny{$\Phi$}}}}
\put (-201,21) {\makebox(0,0){
	\begin{tikzpicture}
	\draw[->] (0,0)--(-3.71pt,-9.28pt);
	\end{tikzpicture}
}}
\put(-197,30){\makebox(0,0){{\tiny{$\Xi$}}}}
\put (-15,135) {\makebox(0,0){
	\begin{tikzpicture}
	\draw[->] (0,0)--(9.8pt,-1.96pt);
	\end{tikzpicture}
}}
\put(-29,139){\makebox(0,0){{\tiny{GHZ}}}}
\caption{\qsl{1}, $A=\{b\}$, $B=\{cd\}$, $C=\{a\}$}
\label{}
\end{subfigure}
\hfill
\begin{subfigure}{0.49\textwidth}
\includegraphics[width=\textwidth]{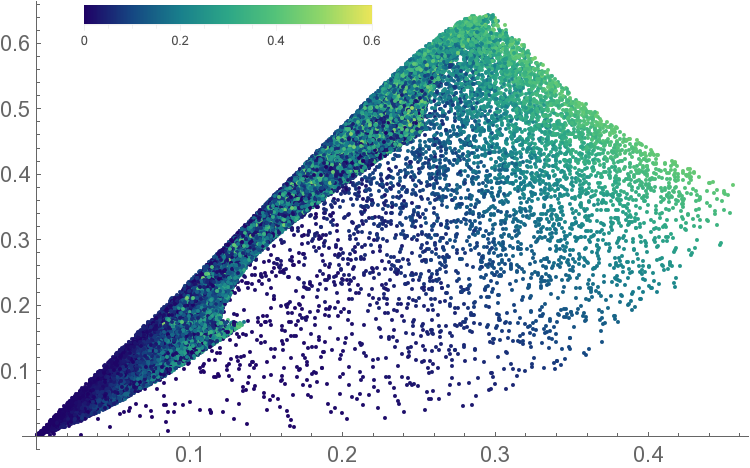}
\put(-12,-6){\makebox(0,0){{\tiny $\Delta\Neg_{ABC}$}}}
\put(-204.5,140){\makebox(0,0){{\tiny $\Delta S_{AB}$}}}
\put(-149,137){\makebox(0,0){{\tiny $\tau_4$}}}
\caption{\qsl{2}, $A=\{a\}$, $B=\{cd\}$, $C=\{b\}$}
\label{}
\end{subfigure}

\vspace{0.5cm}
\begin{subfigure}{0.49\textwidth}
\includegraphics[width=\textwidth]{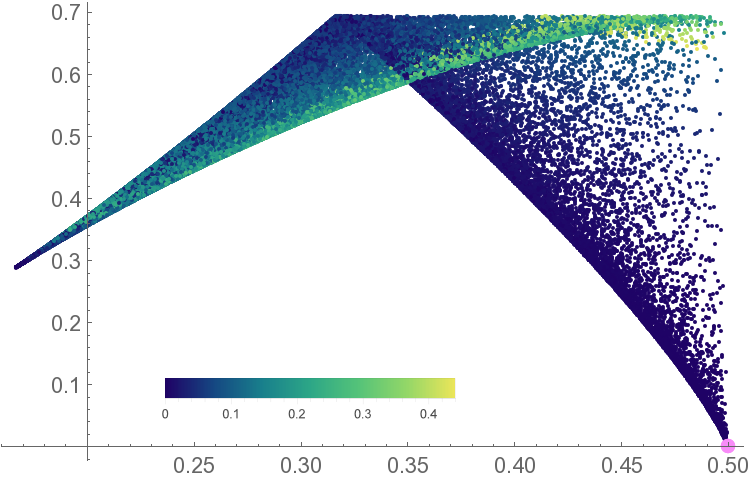}
\put(-12,-6){\makebox(0,0){{\tiny $\Delta\Neg_{ABC}$}}}
\put(-195,143){\makebox(0,0){{\tiny $\Delta S_{AB}$}}}
\put(-130,33){\makebox(0,0){{\tiny $\tau_4$}}}
\put (-16,15) {\makebox(0,0){
	\begin{tikzpicture}
	\draw[->] (0,0)--(8.5pt,-5.26pt);
	\end{tikzpicture}
}}
\put(-25,21){\makebox(0,0){{\tiny{$\Psi$}}}}
\caption{\qsl{3}, $A=\{b\}$, $B=\{ac\}$, $C=\{d\}$}
\label{}
\end{subfigure}
\hfill
\begin{subfigure}{0.49\textwidth}
\includegraphics[width=\textwidth]{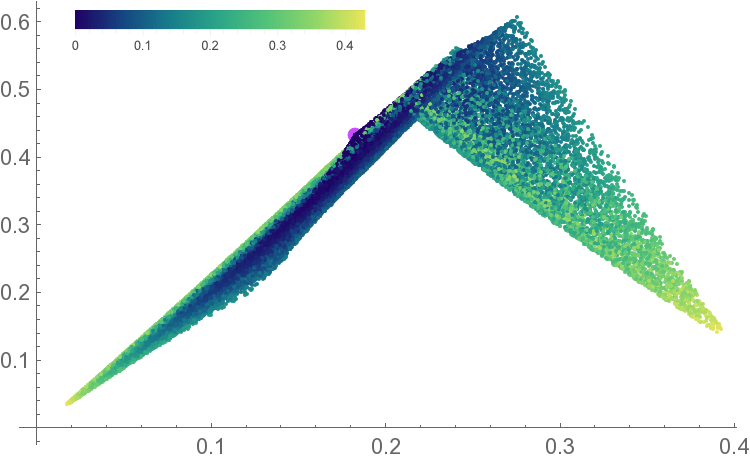}
\put(-12,-6){\makebox(0,0){{\tiny $\Delta\Neg_{ABC}$}}}
\put(-204.5,140){\makebox(0,0){{\tiny $\Delta S_{AB}$}}}
\put(-155,136){\makebox(0,0){{\tiny $\tau_4$}}}
\put (-125,94) {\makebox(0,0){
	\begin{tikzpicture}
	\draw[->] (0,0)--(10pt,0);
	\end{tikzpicture}
}}
\put(-136,94){\makebox(0,0){{\tiny{W}}}}
\caption{\qsl{4}, $A=\{a\}$, $B=\{bc\}$, $C=\{d\}$}
\label{}
\end{subfigure}

\vspace{0.5cm}
\begin{subfigure}{0.49\textwidth}
\includegraphics[width=\textwidth]{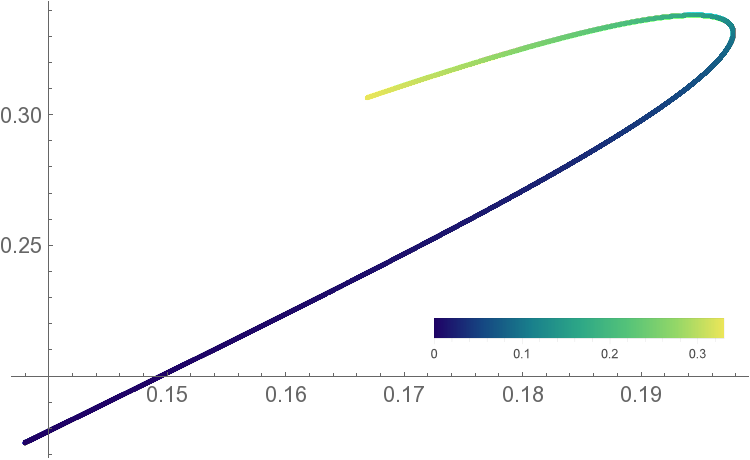}
\put(-12,2){\makebox(0,0){{\tiny $\Delta\Neg_{ABC}$}}}
\put(-204.5,140){\makebox(0,0){{\tiny $\Delta S_{AB}$}}}
\put(-49,44){\makebox(0,0){{\tiny $\tau_4$}}}
\caption{\qsl{5}, $A=\{a\}$, $B=\{bc\}$, $C=\{d\}$}
\label{}
\end{subfigure}
\hfill
\begin{subfigure}{0.49\textwidth}
\includegraphics[width=\textwidth]{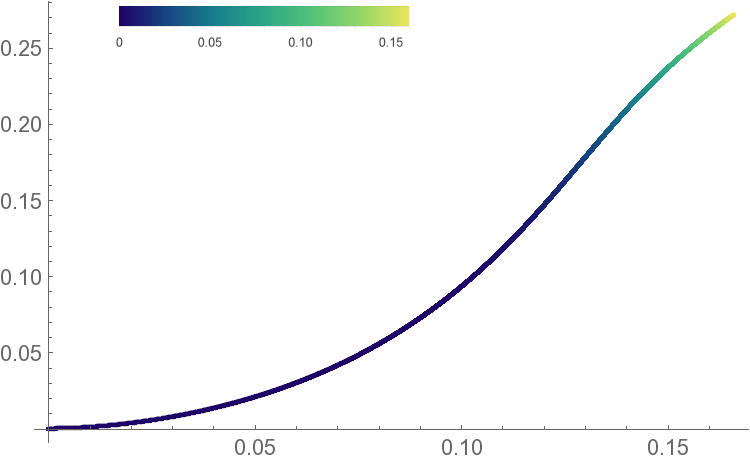}
\put(-12,-6){\makebox(0,0){{\tiny $\Delta\Neg_{ABC}$}}}
\put(-204.5,140){\makebox(0,0){{\tiny $\Delta S_{AB}$}}}
\put(-142,135){\makebox(0,0){{\tiny $\tau_4$}}}
\caption{\qsl{6}, $A=\{a\}$, $B=\{bc\}$, $C=\{d\}$}
\label{}
\end{subfigure}
\caption{Disentangling theorem and saturation of Araki-Lieb (see Eq.~\eqref{eq:deltaSN}) for the SLOCC classified 4-qubit states ($50000$ random states per class). The large dots show the GHZ (orange), $\ket{\!\Phi}$ (turquoise), $\ket{\!\Psi}$ (pink), W (purple) and maximally entangled 
$\ket{\Xi}$ (red) states.}
\label{fig:ALclasses}
\end{figure}
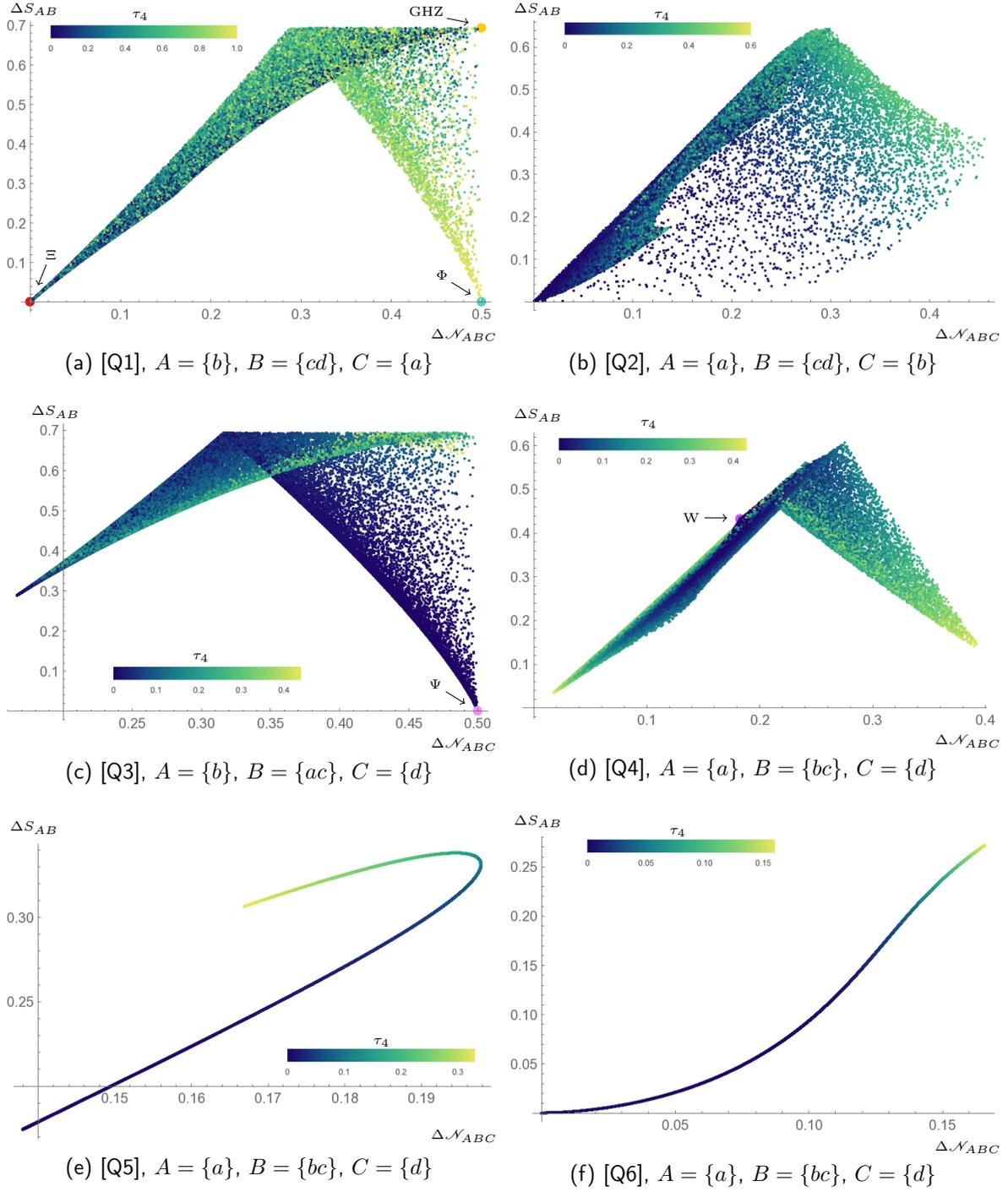
%\afterpage{\clearpage}

\paragraph{Disentangling theorem and Araki-Lieb:} We repeat the analysis of the previous subsection for the disentangling theorem of the negativity and the saturation of AL. In the previous discussion \S\ref{sec:3qubits} the states were generic and all the possible qubits permutations gave the same result. Here we expect the behaviour to be different depending on the specific set-up that we choose -- this is manifest in the fact that the classes are not completely symmetric under qubits permutations. In principle there are $12$ different situations to consider, but many choices give equivalent results. For each class we report the result for the most interesting partition in Fig.~\ref{fig:ALclasses}. The full list of possible results for each class and the corresponding plots can be found in Appendix~\ref{sec:appendix}. In each class there are states that get arbitrarily close to saturating AL and respect the conditions for the disentangling theorem, with the only exception of $\qsl{5}$. This is the class that will be of particular interest for us in the sequel. 

Interestingly, it is worth pointing out that in two classes there are states that saturate AL even without satisfying the conditions for the disentangling theorem.  These states have (a) high values of multipartite entanglement in $\qsl{1}$ and (b) a low value in $\qsl{3}$ (the color map in the various plots shows the values of $\tau_4$). The specific states (indicated by colored dots in Fig.~\ref{fig:ALclasses}) are respectively a product of two Bell pairs $\ket{\!\Phi}$, and the product between a Bell pair and two disentangled qubits $\ket{\!\Psi}$.\footnote{ The explicit expressions of these states are: $\ket{\!\Phi}=\ket{\phi^+}_{ab}\otimes\ket{\phi^+}_{cd}$ and $\ket{\!\Psi}=\ket{01}_{ac}\otimes\ket{\phi^+}_{bd}$ where $\ket{\phi^+}$ is the maximally entangled two qubits state $\ket{\phi^+}=\ket{01}+\ket{10}$, i.e., a Bell pair.} This means that the hypothesis of the disentangling theorem would be satisfied for a different permutation of qubits. Indeed the tangle is sensitive to the factorization inherent in one of the two subsystems. 

AL saturation without factorization ought to be a very restrictive condition on the pattern of internal entanglement. This may exaplain why it hard see the saturation for generic states. Looking at the results one might be tempted to conjecture that such a state could exist in $\qsl{2}$. A numerical search for such a state showed that states with $\Delta\Neg>0$ and $\Delta S\sim 0$ approach to the product $\ket{0110}$. The reason why these states may be highly shifted to the right of the origin is that their entanglement is particularly robust. An extremely small but non vanishing value of the entropy can produce much higher values of the negativity. We conclude then that in the case of four qubits the saturation of AL only happens if there exists a permutation such that the conditions for disentanglement are matched and factorization actually happens.

%RATIO AND LOCAL ENTANGLEMENT FOR THE CLASSES
%
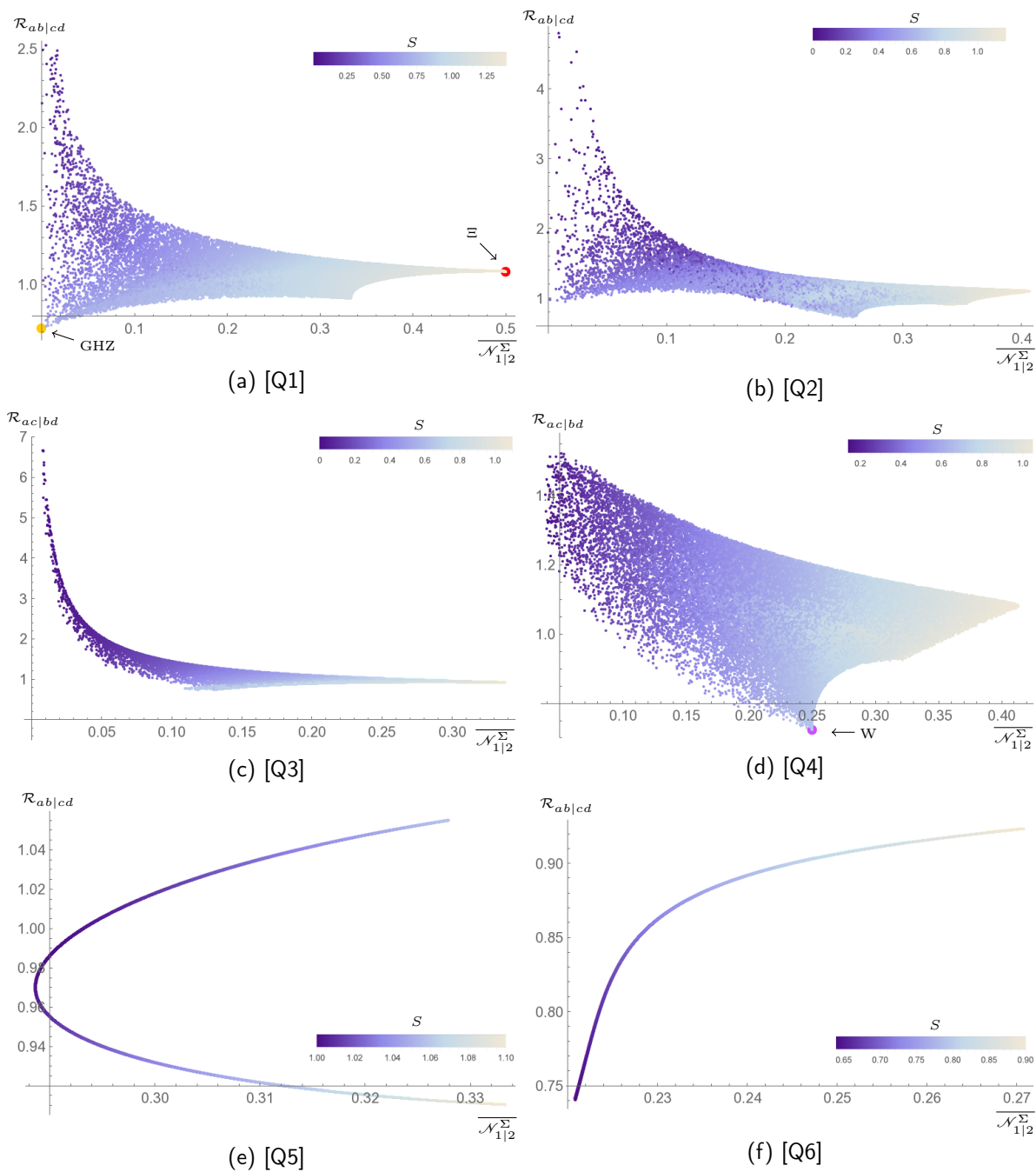
\begin{figure}[t]
\centering
\begin{subfigure}{0.49\textwidth}
\includegraphics[width=\textwidth]{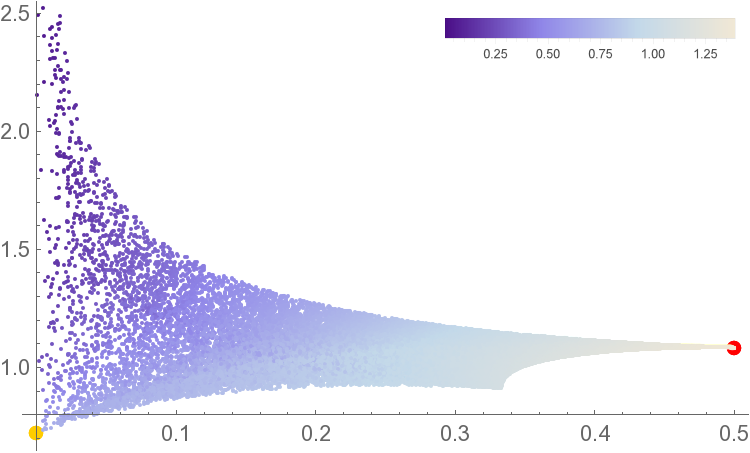}
\put(-207,137){\makebox(0,0){{\tiny $\mathcal{R}_{ab|cd}$}}}
\put(-8,-5){\makebox(0,0){{\tiny $\overline{\Neg^\Sigma_{1|2}}$}}}
\put(-45,130){\makebox(0,0){{\tiny $S$}}}
\put (-12,39) {\makebox(0,0){
	\begin{tikzpicture}
	\draw[->] (0,0)--(7.07pt,-7.07pt);
	\end{tikzpicture}
}}
\put (-197,2) {\makebox(0,0){
	\begin{tikzpicture}
	\draw[->] (0,0)--(-9.49pt,3.16pt);
	\end{tikzpicture}
}}
\put(-182,-2){\makebox(0,0){{\tiny{GHZ}}}}
\put(-19,47){\makebox(0,0){{\tiny{$\Xi$}}}}
\subcaption{\qsl{1}}
\label{subfig:4qRC1}
\end{subfigure}
\hfill
\begin{subfigure}{0.49\textwidth}
\includegraphics[width=\textwidth]{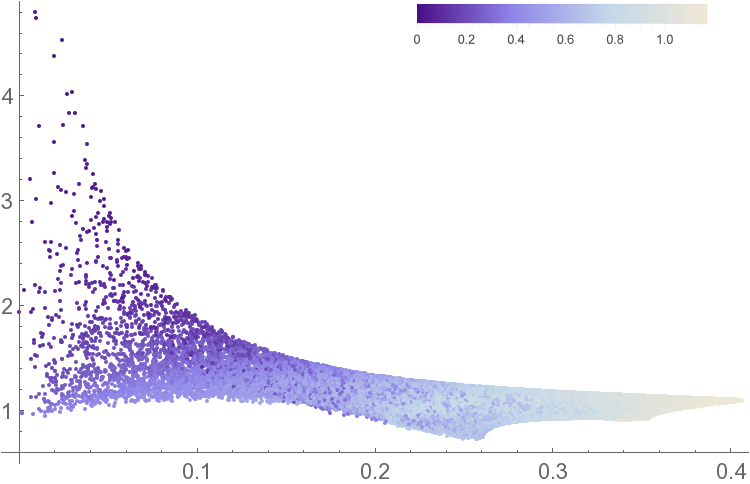}
\put(-211,143){\makebox(0,0){{\tiny $\mathcal{R}_{ab|cd}$}}}
\put(-8,-6){\makebox(0,0){{\tiny $\overline{\Neg^\Sigma_{1|2}}$}}}
\put(-52,143){\makebox(0,0){{\tiny $S$}}}
\subcaption{\qsl{2}}
\label{subfig:4qRC2}
\end{subfigure}

\vspace{0.2cm}
\begin{subfigure}{0.49\textwidth}
\includegraphics[width=\textwidth]{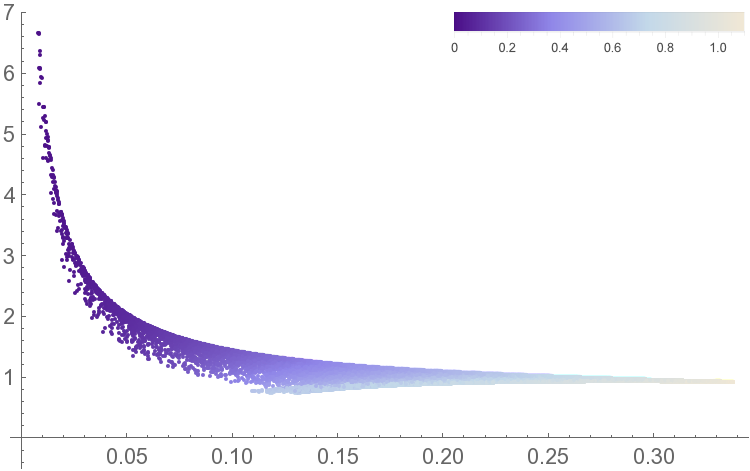}
\put(-210,140){\makebox(0,0){{\tiny $\mathcal{R}_{ac|bd}$}}}
\put(-9,0){\makebox(0,0){{\tiny $\overline{\Neg^\Sigma_{1|2}}$}}}
\put(-42,138){\makebox(0,0){{\tiny $S$}}}
\subcaption{\qsl{3}}
\label{subfig:4qRC3}
\end{subfigure}
\hfill
\begin{subfigure}{0.49\textwidth}
\includegraphics[width=\textwidth]{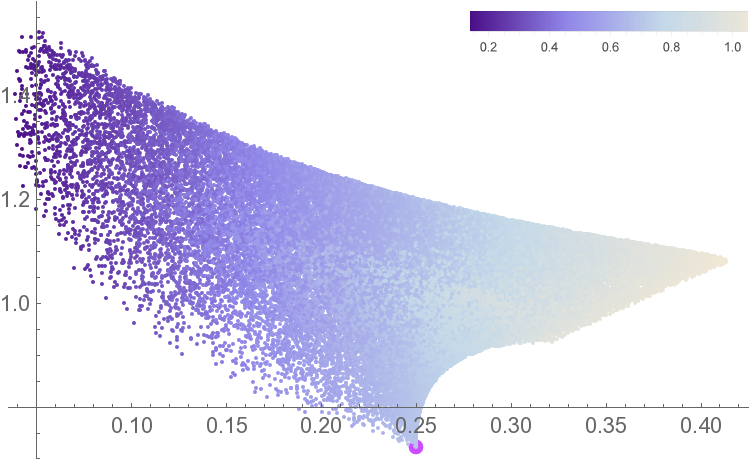}
\put(-207,138){\makebox(0,0){{\tiny $\mathcal{R}_{ac|bd}$}}}
\put(-9,0){\makebox(0,0){{\tiny $\overline{\Neg^\Sigma_{1|2}}$}}}
\put(-42,135){\makebox(0,0){{\tiny $S$}}}
\put (-83,3) {\makebox(0,0){
	\begin{tikzpicture}
	\draw[->] (0,0)--(-10pt,0);
	\end{tikzpicture}
}}
\put(-72,2.5){\makebox(0,0){{\tiny{W}}}}
\subcaption{\qsl{4}}
\label{subfig:4qRC4}
\end{subfigure}

\vspace{0.2cm}
\begin{subfigure}{0.49\textwidth}
\includegraphics[width=\textwidth]{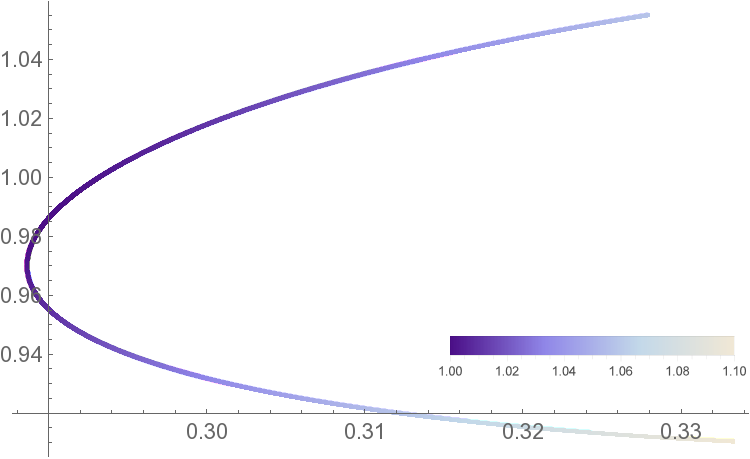}
\put(-203.5,137){\makebox(0,0){{\tiny $\mathcal{R}_{ab|cd}$}}}
\put(-8,-6){\makebox(0,0){{\tiny $\overline{\Neg^\Sigma_{1|2}}$}}}
\put(-43,41){\makebox(0,0){{\tiny $S$}}}
\subcaption{\qsl{5}}
\label{subfig:4qRC5}
\end{subfigure}
\hfill
\begin{subfigure}{0.49\textwidth}
\includegraphics[width=\textwidth]{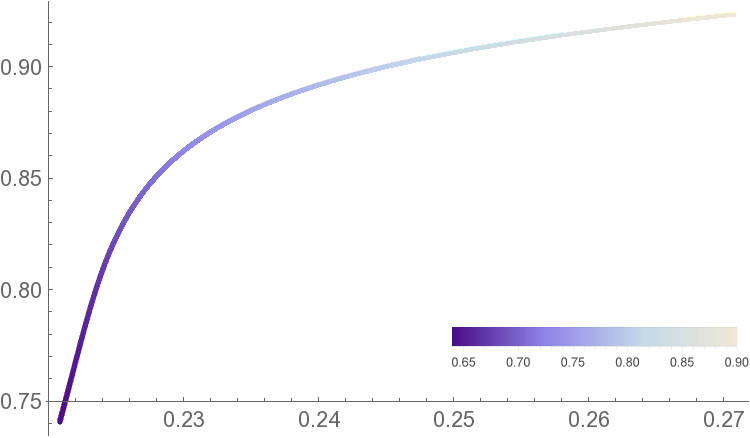}
\put(-203.5,132){\makebox(0,0){{\tiny $\mathcal{R}_{ab|cd}$}}}
\put(-8,-6){\makebox(0,0){{\tiny $\overline{\Neg^\Sigma_{1|2}}$}}}
\put(-43,37){\makebox(0,0){{\tiny $S$}}}
\subcaption{\qsl{6}}
\label{subfig:4qRC6}
\end{subfigure} 
\caption{The ratio for a given $\Sigma$ compared to the average $\Neg_{1|2}$ across $\Sigma$. $50000$ random states per class. Panels (a-b) are truncated, few states (not shown) approach the vertical axes with high values of the ratio and small values of the entropy. The large dots show the GHZ (orange), W (purple) and maximally entangled $\ket{\!\Xi}$ (red) states. (a)(b)(c) are truncated, few states with small values of $\Neg$ and ratio up to $\mathcal{R}\sim 7.4$ (a),  $\mathcal{R}\sim 16$ (b) and $\mathcal{R}\sim 32$ (c), are not shown.}
\label{fig:ratioclasses}
\end{figure}
%
%\afterpage{\clearpage}

\paragraph{Specific robustness $\mathcal{R}$:} For generic states we have seen that the most sensible quantity to the internal structure of entanglement is the negativity across $\Sigma$ between a qubit in a subsystem and the other pair, viz., $\overline{\Neg^\Sigma_{1|2}}$.  We now repeat the same analysis for the classes and compare specific robustness \eqref{eq:Rdef} to the negativity $\overline{\Neg^\Sigma_{1|2}}$. The results are shown in Fig.~\ref{fig:ratioclasses}; we keep track of the values of the entropy (color map). 

States in $\qsl{1}$ reproduce the behaviour of generic states; similar patterns are visible in $\qsl{2}$-$\qsl{4}$ -- see Fig.~\ref{subfig:4qRC1}-\ref{subfig:4qRC4}. On the other hand, for a different choice of $\Sigma$, a different pattern manifests itself in classes 
$\qsl{3}$ and $\qsl{4}$, see Appendix \ref{sec:appendix},  Figs.~\ref{subfig:R3a} and \ref{subfig:R4a}. In this case the highest values of the ratio in the class correspond to states with the highest entropy and negativity. There are no states with small negativity and high value of the ratio (and small entropy). $\qsl{6}$ has a similar behaviour as evidenced in Fig.~\ref{subfig:4qRC6}. In contrast,  $\qsl{5}$ instead is quite peculiar, states with high negativity and entropy correspond both to the highest and smallest values of the ratio, cf., Fig.~\ref{subfig:4qRC5}. 
Highly entangled states in $\qsl{5}$ are divided in two branches, one with fragile entanglement, the other whose entanglement is more robust. Note however that the three quantities that characterize the state are constrained in a small range of values.

While the behaviour of the various classes is more or less similar to the generic class, the curious feature is the exceptional behaviour in $\qsl{5}$. We have previously also seen that this class of 4-qubit states is also peculiar when we analyzed the consequences of the disentangling theorem, vis a vis, saturation of the AL inequality. We shall now compare the ratio to the value of multipartite entanglement and the monogamy of mutual information and find yet another distinguishing feature of this class.

% I3 FOR THE CLASSES (PLOT)
%
\begin{figure}[t]
\centering
\begin{subfigure}{0.49\textwidth}
\includegraphics[width=\textwidth]{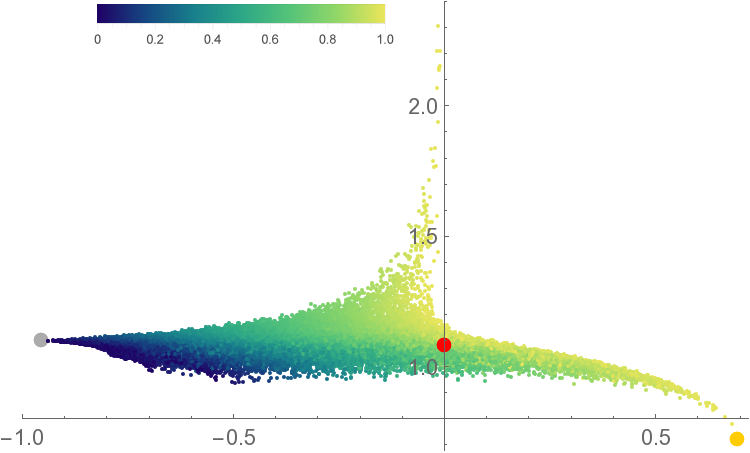}
\put(-87,138){\makebox(0,0){{\tiny $\overline{\mathcal{R}}$}}}
\put(-3,19){\makebox(0,0){{\scriptsize $\tmi$}}}
\put(-145,134){\makebox(0,0){{\tiny $\tau_4$}}}
\put (-205,42) {\makebox(0,0){
	\begin{tikzpicture}
	\draw[->] (0,0)--(0,-10pt);
	\end{tikzpicture}
}}
\put(-205,52){\makebox(0,0){{\tiny{$\mathcal{M}$}}}}
\put (-78,40) {\makebox(0,0){
	\begin{tikzpicture}
	\draw[->] (0,0)--(-7.07pt,-7.07pt);
	\end{tikzpicture}
}}
\put(-70,47){\makebox(0,0){{\tiny{$\Xi$}}}}
\put (-14,0) {\makebox(0,0){
	\begin{tikzpicture}
	\draw[->] (0,0)--(9.49pt,3.16pt);
	\end{tikzpicture}
}}
\put(-30,-4){\makebox(0,0){{\tiny{GHZ}}}}
\caption{\qsl{1}}
\label{subfig:4qI3C1}
\end{subfigure}
\hfill
\begin{subfigure}{0.49\textwidth}
\includegraphics[width=\textwidth]{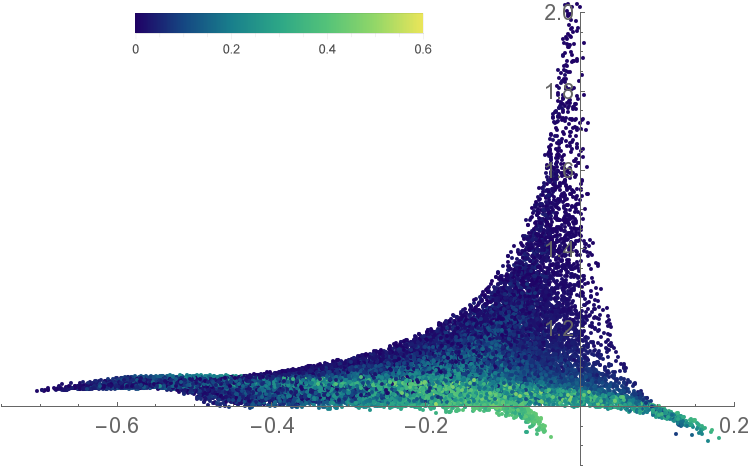}
\put(-49,141){\makebox(0,0){{\tiny $\overline{\mathcal{R}}$}}}
\put(-4,-2){\makebox(0,0){{\scriptsize $\tmi$}}}
\put(-135,136){\makebox(0,0){{\tiny $\tau_4$}}}
\caption{\qsl{2}}
\label{subfig:4qI3C2}
\end{subfigure}

\vspace{0.4cm}
\begin{subfigure}{0.49\textwidth}
\includegraphics[width=\textwidth]{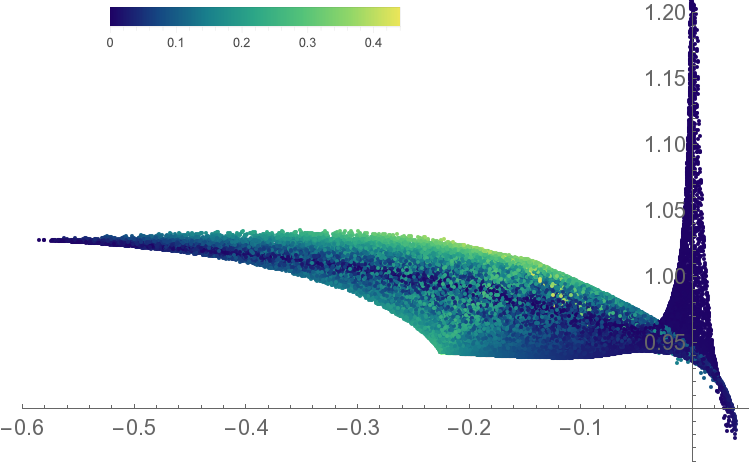}
\put(-16,141){\makebox(0,0){{\tiny $\overline{\mathcal{R}}$}}}
\put(-4,1){\makebox(0,0){{\scriptsize $\tmi$}}}
\put(-141,136){\makebox(0,0){{\tiny $\tau_4$}}}
\caption{\qsl{3}}
\label{subfig:4qI3C3}
\end{subfigure}
\hfill
\begin{subfigure}{0.49\textwidth}
\includegraphics[width=\textwidth]{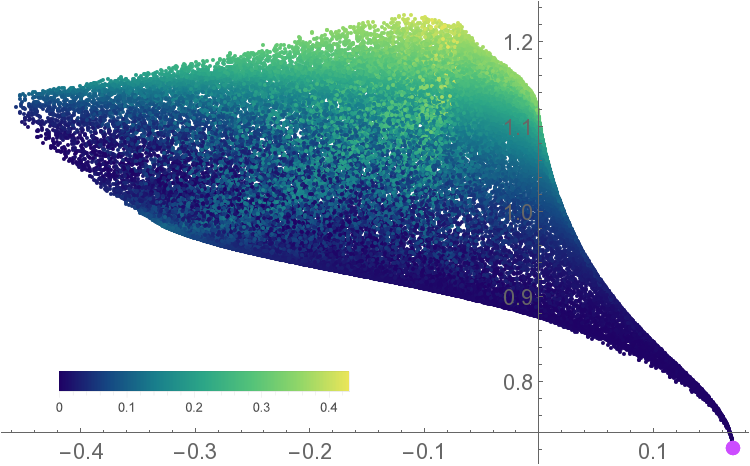}
\put(-61,141){\makebox(0,0){{\tiny $\overline{\mathcal{R}}$}}}
\put(-4,20){\makebox(0,0){{\scriptsize $\tmi$}}}
\put(-160,30){\makebox(0,0){{\tiny $\tau_4$}}}
\put (-14,1) {\makebox(0,0){
	\begin{tikzpicture}
	\draw[->] (0,0)--(9.49pt,3.16pt);
	\end{tikzpicture}
}}
\put(-27,-4){\makebox(0,0){{\tiny{W}}}}
\caption{\qsl{4}}
\label{subfig:4qI3C4}
\end{subfigure}

\vspace{0.4cm}
\begin{subfigure}{0.49\textwidth}
\includegraphics[width=\textwidth]{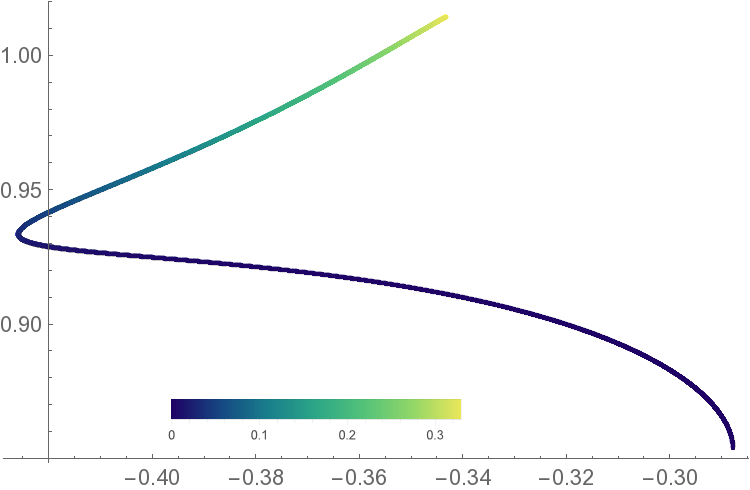}
\put(-203,148){\makebox(0,0){{\tiny $\overline{\mathcal{R}}$}}}
\put(-4,3){\makebox(0,0){{\scriptsize $\tmi$}}}
\put(-124,30){\makebox(0,0){{\tiny $\tau_4$}}}
\caption{\qsl{5}}
\label{subfig:4qI3C5}
\end{subfigure}
\hfill
\begin{subfigure}{0.49\textwidth}
\includegraphics[width=\textwidth]{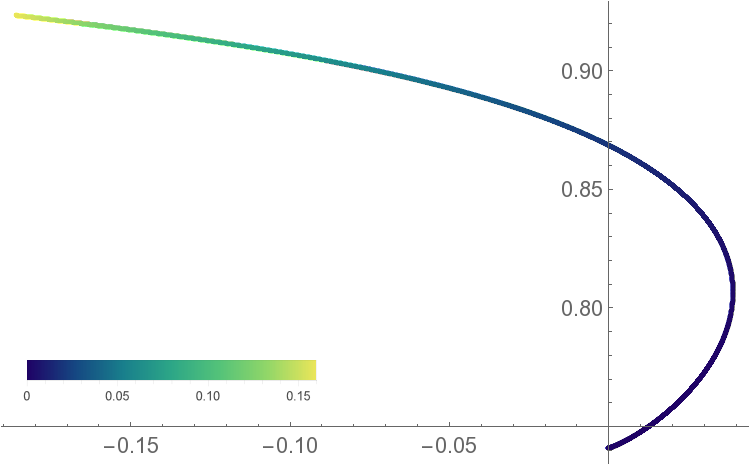}
\put(-41,141){\makebox(0,0){{\tiny $\overline{\mathcal{R}}$}}}
\put(-3,4){\makebox(0,0){{\scriptsize $\tmi$}}}
\put(-166,34){\makebox(0,0){{\tiny $\tau_4$}}}
\caption{\qsl{6}}
\label{subfig:4qI3C6}
\end{subfigure}
\caption{Monogamy of mutual information compared to the average ratio for the maximal bipartitions and mutipartite entanglement $\tau_4$ (color map). The large dots show the GHZ (orange), $\ket{\!\mathcal{M}}$ (gray), W (purple) and maximally entangled $\ket{\!\Xi}$ (red) states. $50000$ states per class. (b)(c) are truncated, few states with $\tmi\sim 0$ and values of the average ratio up to $\mathcal{R}\sim 8.5$ (b) and $\mathcal{R}\sim 11$ (c) are not shown.}
\label{fig:4qI3classes}
\end{figure}
%\afterpage{\clearpage}

\paragraph{Monogamy of mutual information:} 
Finally, we compare the value of $\tmi$, to the tangle, and the average ratio over the three maximal bipartitions. The results for the different classes are shown in Fig.~\ref{fig:4qI3classes}. Surprisingly, there is only a single class $\qsl{5}$ shown in Fig.~\ref{subfig:4qI3C5}, whose states always have a negative value of $\tmi$.

This the most interesting aspect of our analysis of the SLOCC classified 4-qubit states. What it suggests is the following: in all the other classes $\qsl{k}$ with $k\neq5$, a given state with a particular value of $\tmi$ can always be turned into a state with positive $\tmi$ by SLOCC. Even more strongly, since every generic state of four qubits can be mapped to the first class by SLOCC, the states in $\qsl{5}$ are the only one that can never violate the monogamy of mutual information. In effect, what this suggests is that the subset of 4-qubit states lying in Class $\qsl{5}$ are likely to be most holographic. Of course, this statement should be taken with a large grain of salt, for we are discussing here the structure of entanglement in qubit systems with no interactions.  Furthermore, it is unclear to us that the holographic map respects the SLOCC operations used to classify states herein.\footnote{ We thank Veronika Hubeny for a discussion on this issue.} Nevertheless we think the presence of a distinguished class of states suggests that certain patterns of entanglement are more likely than others to play a role in holographic systems (at least for the purposes of building semi-classical geometry).\footnote{ For completeness let us also record the values of $\tmi$ for the exceptional states 
$\qsl{7}, \qsl{8}, \qsl{9}$:\\
$$\tmi(\qsl{7}) = -0.356135\,, \qquad \tmi(\qsl{8}) = -0.477386 \,,\qquad \tmi(\qsl{9}) = 0 \,. $$
}

It is also interesting to look at the result for the generic class $\qsl{1}$, cf., Fig.~\ref{subfig:4qI3C1}. As expected from previous results for generic states the value of $\tmi$ is generally negative, in particular for small values of $\tau_4$. Nevertheless states which are strongly entangled in a multipartite sense are divided into two branches. One branch minimizes the average ratio and violates monogamy of mutual information, the other seems to respect monogamy and maximizes the ratio. The states in the latter case asymptotically approach the state $\Phi$ discussed before in the context of AL inequalities.\footnote{ This was checked numerically over $1000000$ states that maximize the ratio.} A numerical search for the states that minimizes the value of $\tmi$ gives instead the following state
\begin{align}
\ket{\mathcal{M}}=\ket{0011}+e^{-\frac{\pi}{3}i}\ket{0101}-e^{\frac{\pi}{3}i}\ket{0110}-e^{\frac{\pi}{3}i}\ket{1001}+e^{-\frac{\pi}{3}i}\ket{1010}+\ket{1100}
\end{align}
This state is interesting in its own right; it appears to be a highly scrambled state. It has maximal entanglement under all partitionings of the qubits \cite{Gour:2010aa}. Preliminary investigations indicate a similar pattern for higher qubit systems; it would be interesting to explore this class of states further. A behaviour similar to that of $\qsl{1}$ is manifest for $\qsl{4}$ (Fig.~\ref{subfig:4qI3C4}). On the other hand this should be contrasted with the behaviour of $\qsl{2}$ and $\qsl{3}$ (see Fig.~\ref{subfig:4qI3C2} and \ref{subfig:4qI3C3} respectively), where the states that maximize the average ratio contain a small amount of multipartite entanglement and can violate monogamy. 

% 6 QUBITS ENTANG. ACROSS SIGMA (NO CONSTRAINT ON THE ENTROPY)
%
\begin{figure}[tb]
\centering
\begin{subfigure}{0.49\textwidth}
\includegraphics[width=\textwidth]{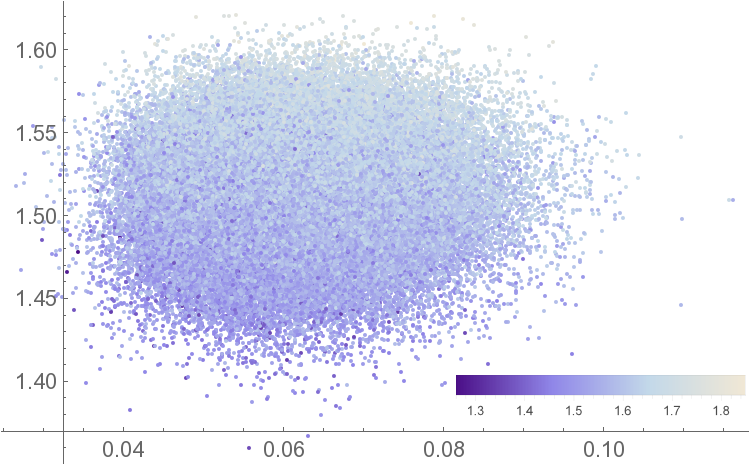}
\put(-8,-6){\makebox(0,0){{\tiny $\overline{\Neg^{\Sigma}_{1|2}}$}}}
\put(-198,139){\makebox(0,0){{\tiny $\mathcal{R}$}}}
\put(-43,31){\makebox(0,0){{\tiny $S$}}}
\caption{}
\label{}
\end{subfigure}
\hfill
\begin{subfigure}{0.49\textwidth}
\includegraphics[width=\textwidth]{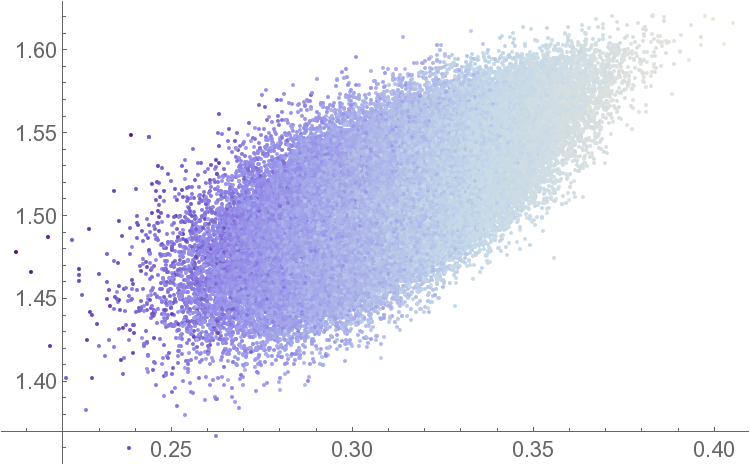}
\put(-8,-6){\makebox(0,0){{\tiny $\overline{\Neg^{\Sigma}_{1|3}}$}}}
\put(-198,139){\makebox(0,0){{\tiny $\mathcal{R}$}}}
\caption{}
\label{}
\end{subfigure}

\begin{subfigure}{0.49\textwidth}
\includegraphics[width=\textwidth]{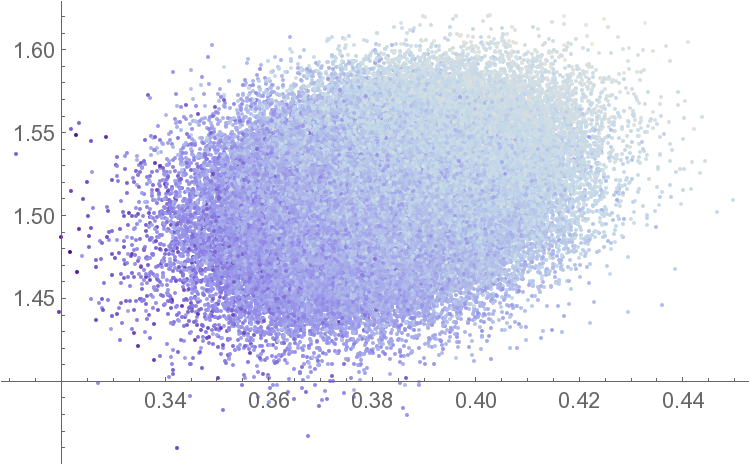}
\put(-8,8){\makebox(0,0){{\tiny $\overline{\Neg^{\Sigma}_{2|2}}$}}}
\put(-199,140){\makebox(0,0){{\tiny $\mathcal{R}$}}}
\caption{}
\label{}
\end{subfigure}
\hfill
\begin{subfigure}{0.49\textwidth}
\includegraphics[width=\textwidth]{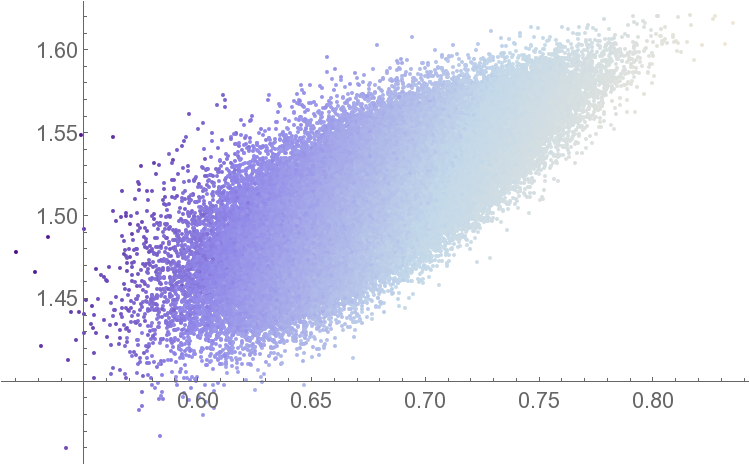}
\put(-8,8){\makebox(0,0){{\tiny $\overline{\Neg^{\Sigma}_{2|3}}$}}}
\put(-193,140){\makebox(0,0){{\tiny $\mathcal{R}$}}}
\caption{}
\label{}
\end{subfigure}
\caption{Entanglement across $\Sigma$ for a system of 6 qubits ($100000$ random states). The entangling surface $\Sigma$ is fixed and divides the states into two subsystems of three qubits each. The panels show the dependence of the ratio on the internal negativity. The color map shows the values of the entropy in the range ($1.26147, 1.84577$).}
\label{fig:6qratio_across}
\end{figure}
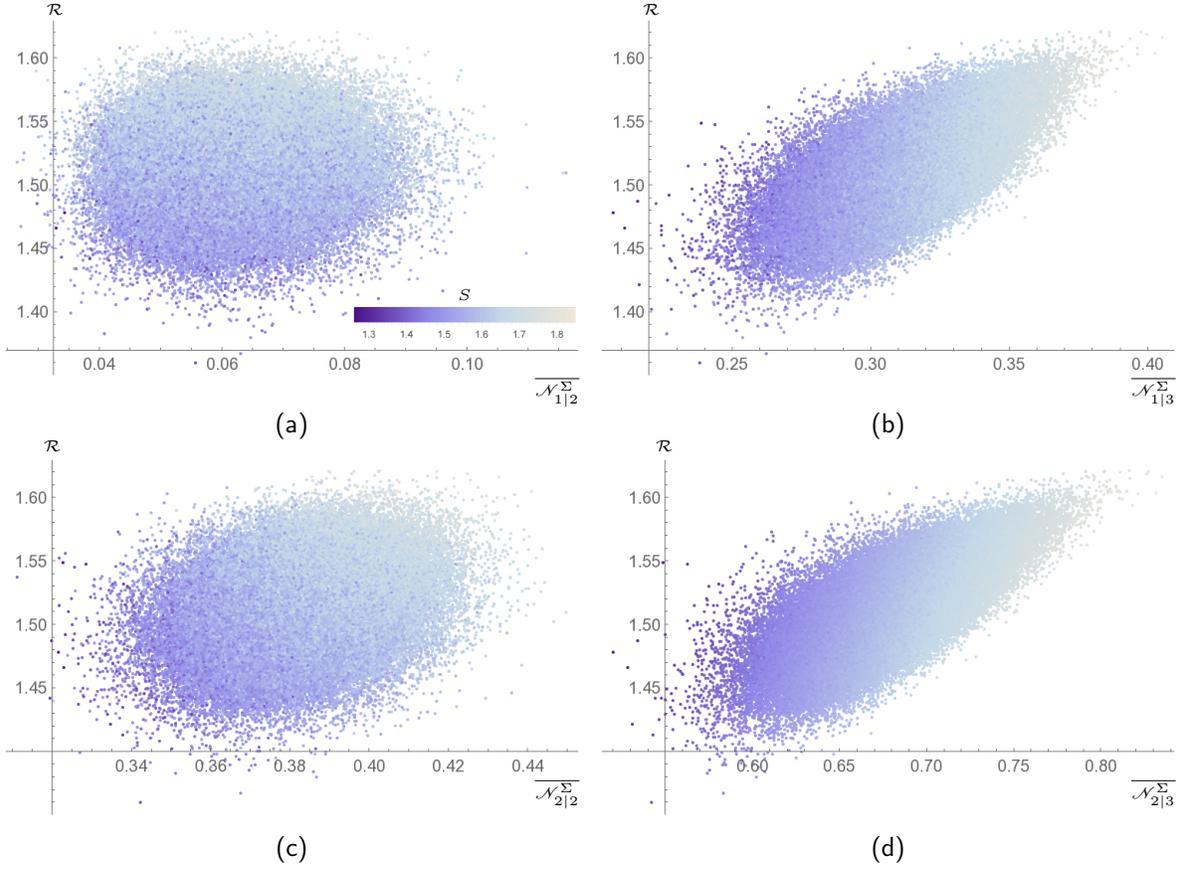
%\afterpage{\clearpage}

%~~~~~~~~~~~~~~~~~~~~~~~~~~~~~~~~~~~~~~~~~~~~~~~~~~~~~~~~~
\section{Large $N$ qubit systems}
\label{sec:largersys}
%~~~~~~~~~~~~~~~~~~~~~~~~~~~~~~~~~~~~~~~~~~~~~~~~~~~~~~~~~

At present no classification is known for pure states of five or more qubits. As a result any analysis of larger number of qubits must necessarily  be restricted to  generic states. We now explore  specific robustness characterized by the ratio ${\mathcal R}$, focusing on its dependence on internal entanglement. We also  look to examining  the relations amongst the ratio, multipartite entanglement and the monogamy of mutual information. We will specifically focus on states of 6 and 8 qubits, primarily because the tangle is only defined for an even number of qubits (for $N>3$). Much of the other results we derive ought not to change considerably for an odd number of qubits. 

%~~~~~~~~~~~~~~~~~~~~~~~~~~~~~~~~~~~~~~~~~~~~~~~
\subsection{Negativity versus entanglement}
\label{subsec:negent68}
%~~~~~~~~~~~~~~~~~~~~~~~~~~~~~~~~~~~~~~~~~~~~~~

% 6 QUBITS ENTANG. ACROSS SIGMA (WITH CONSTRAINT ON THE ENTROPY)
%
\begin{figure}[t]
\centering
\begin{subfigure}{0.49\textwidth}
\includegraphics[width=\textwidth]{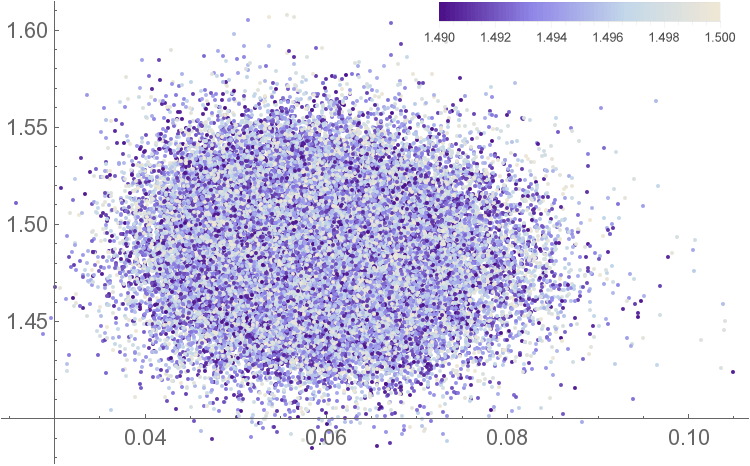}
\put(-8,-2){\makebox(0,0){{\tiny $\overline{\Neg^{\Sigma}_{1|2}}$}}}
\put(-200,139){\makebox(0,0){{\tiny $\mathcal{R}$}}}
\put(-47,138){\makebox(0,0){{\tiny $S$}}}
\caption{}
\label{}
\end{subfigure}
\hfill
\begin{subfigure}{0.49\textwidth}
\includegraphics[width=\textwidth]{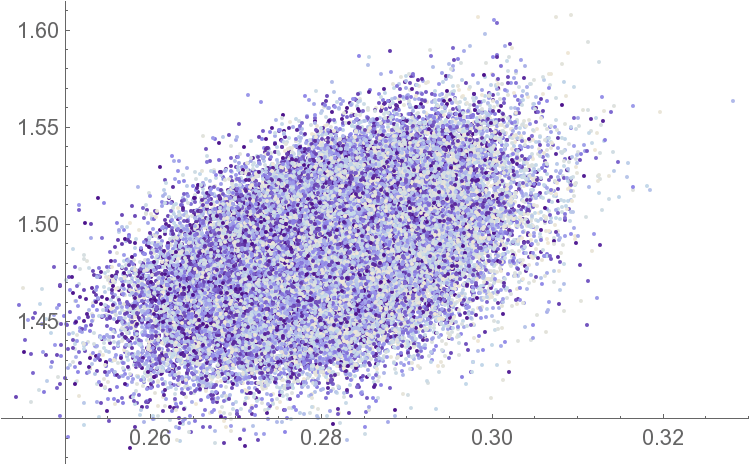}
\put(-8,-2){\makebox(0,0){{\tiny $\overline{\Neg^{\Sigma}_{1|3}}$}}}
\put(-197,139){\makebox(0,0){{\tiny $\mathcal{R}$}}}
\caption{}
\label{}
\end{subfigure}

\begin{subfigure}{0.49\textwidth}
\includegraphics[width=\textwidth]{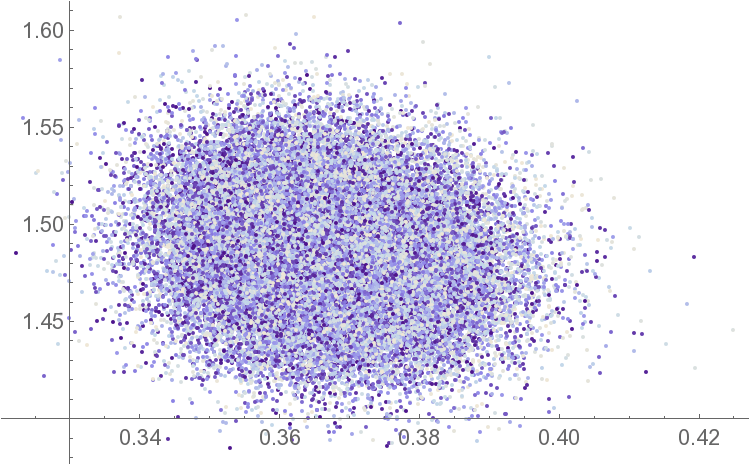}
\put(-8,-2){\makebox(0,0){{\tiny $\overline{\Neg^{\Sigma}_{2|2}}$}}}
\put(-197,139){\makebox(0,0){{\tiny $\mathcal{R}$}}}
\caption{}
\label{}
\end{subfigure}
\hfill
\begin{subfigure}{0.49\textwidth}
\includegraphics[width=\textwidth]{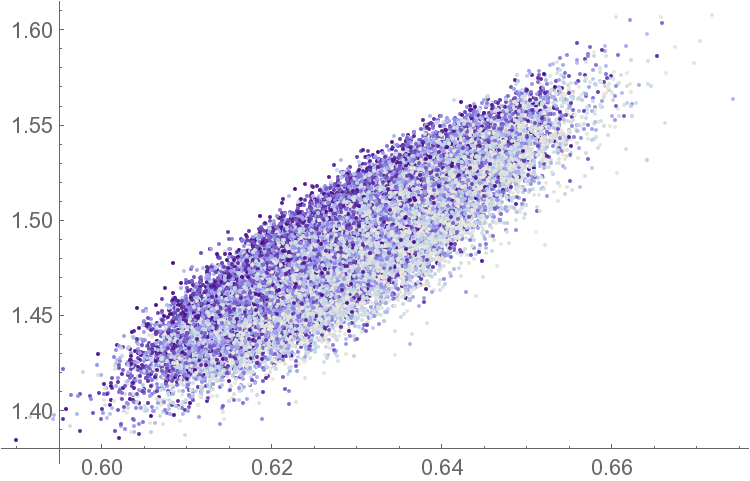}
\put(-8,-2){\makebox(0,0){{\tiny $\overline{\Neg^{\Sigma}_{2|3}}$}}}
\put(-199,143){\makebox(0,0){{\tiny $\mathcal{R}$}}}
\caption{}
\label{}
\end{subfigure}
\caption{A slice of Fig.~\ref{fig:6qratio_across}, for $50000$ states, now with a constraint on the entropy which takes values in the interval ($1.49, 1.5$) and is shown by the color map.}
\label{fig:6qratio_across_cons}
\end{figure}
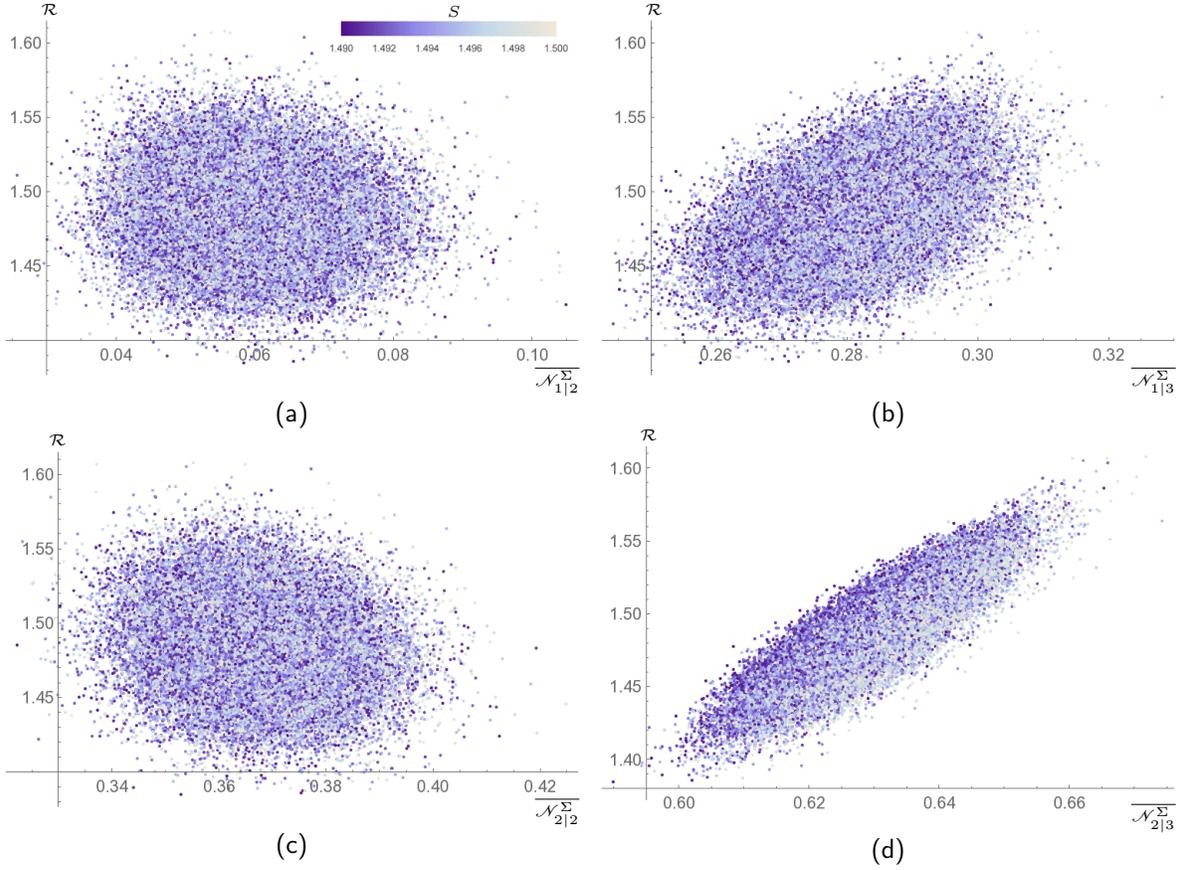
%\afterpage{\clearpage}

% 6 QUBITS LOCAL
%
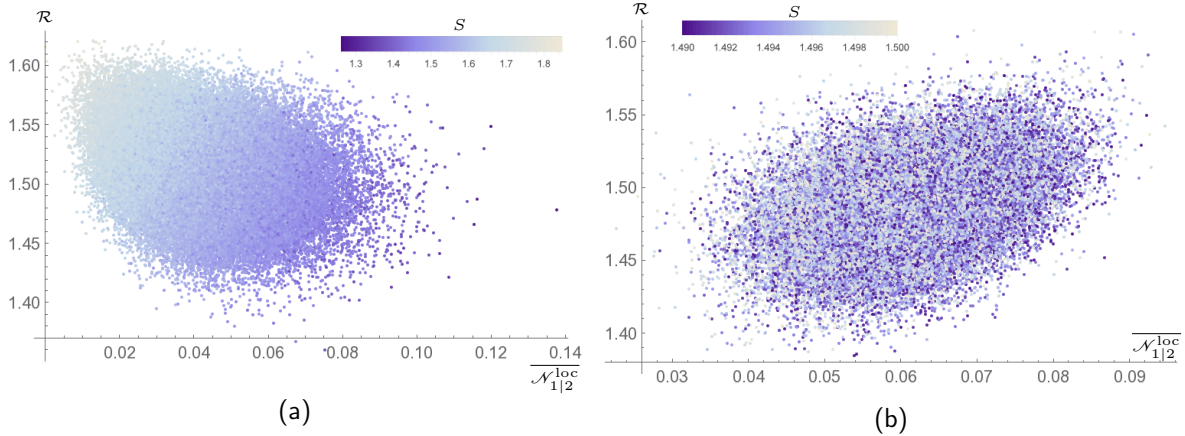
\begin{figure}[tb]
\centering
\begin{subfigure}{0.49\textwidth}
\includegraphics[width=\textwidth]{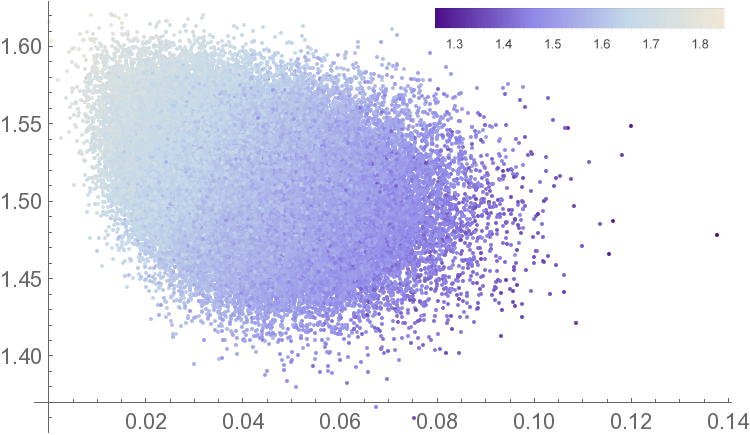}
\put(-10,-6){\makebox(0,0){{\tiny $\overline{\Neg^{\text{loc}}_{1|2}}$}}}
\put(-203,131){\makebox(0,0){{\tiny $\mathcal{R}$}}}
\put(-46,128){\makebox(0,0){{\tiny $S$}}}
\caption{}
\label{}
\end{subfigure}
\hfill
\begin{subfigure}{0.49\textwidth}
\includegraphics[width=\textwidth]{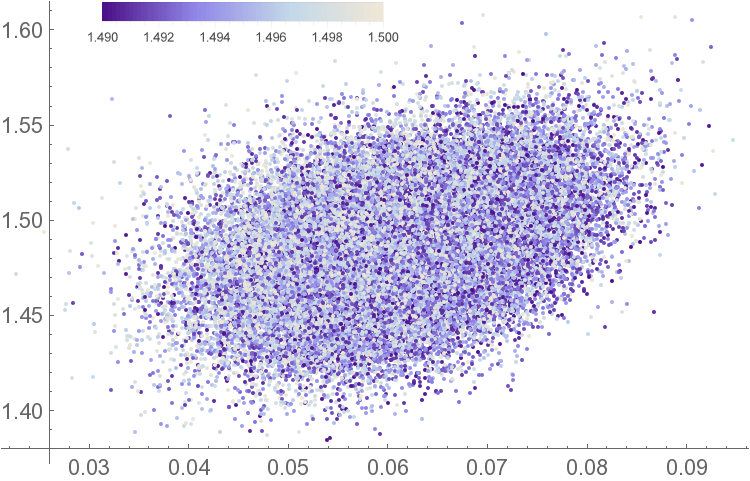}
\put(-8,15){\makebox(0,0){{\tiny $\overline{\Neg^{\text{loc}}_{1|2}}$}}}
\put(-203,143){\makebox(0,0){{\tiny $\mathcal{R}$}}}
\put(-145,142){\makebox(0,0){{\tiny $S$}}}
\caption{}
\label{}
\end{subfigure}
\caption{Dependence of the ratio on the average negativity between qubits inside the subsystems specified by $\Sigma$ for pure states of 6-qubits. (a) no constraint on the entropy (color map), for $100000$ states, (b) entropy constrained in the range ($1.49, 1.5$), for $50000$ states.}
\label{fig:6qratio_local}
\end{figure}
%\afterpage{\clearpage}

% 8Q (NO CONSTRAINT ON THE ENTROPY)
%
\begin{figure}[tb]
\centering
\begin{subfigure}{0.49\textwidth}
\includegraphics[width=\textwidth]{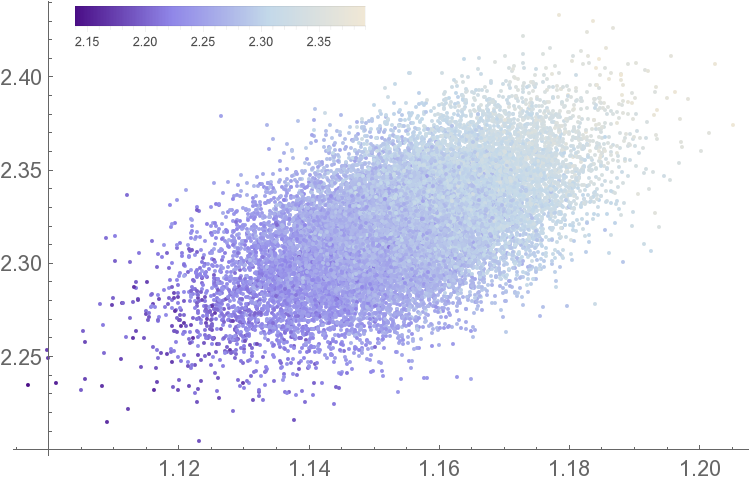}
\put(-8,-6){\makebox(0,0){{\tiny $\overline{\Neg^{\Sigma}_{3|3}}$}}}
\put(-203,145){\makebox(0,0){{\tiny $\mathcal{R}$}}}
\put(-152,142){\makebox(0,0){{\tiny $S$}}}
\caption{}
\label{}
\end{subfigure}
\hfill
\begin{subfigure}{0.49\textwidth}
\includegraphics[width=\textwidth]{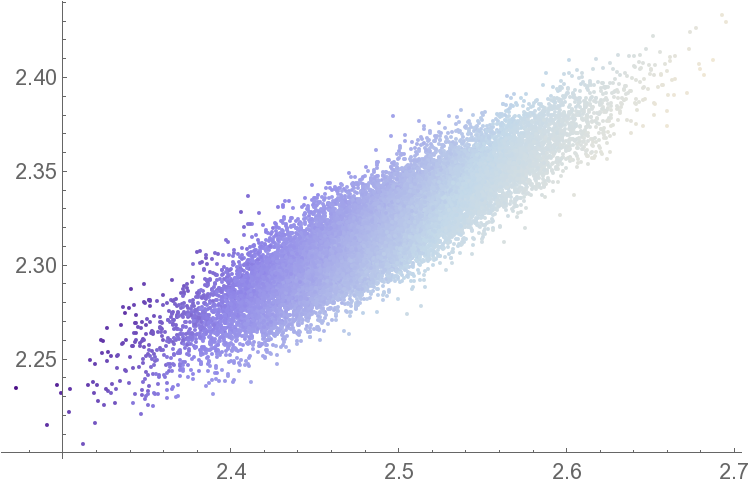}
\put(-8,19){\makebox(0,0){{\tiny $\overline{\Neg^{\Sigma}_{3|4}}$}}}
\put(-198,145){\makebox(0,0){{\tiny $\mathcal{R}$}}}
\caption{}
\label{}
\end{subfigure}

\begin{subfigure}{0.49\textwidth}
\includegraphics[width=\textwidth]{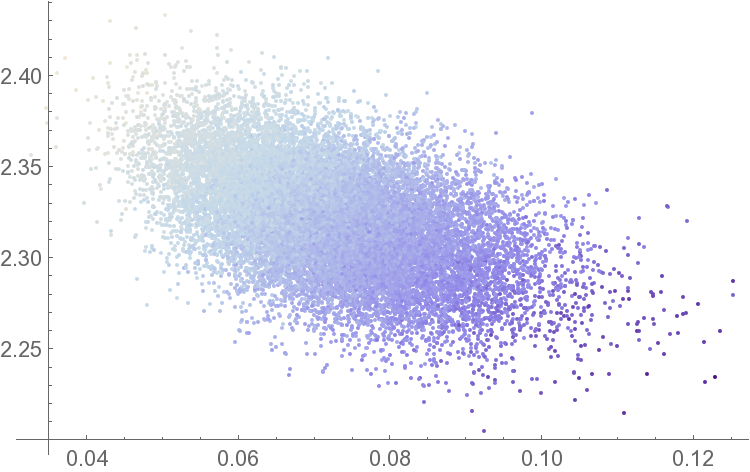}
\put(-8,-6){\makebox(0,0){{\tiny $\overline{\Neg^{\text{loc}}_{2|2}}$}}}
\put(-202,142){\makebox(0,0){{\tiny $\mathcal{R}$}}}
\caption{}
\label{}
\end{subfigure}
\hfill
\begin{subfigure}{0.49\textwidth}
\includegraphics[width=\textwidth]{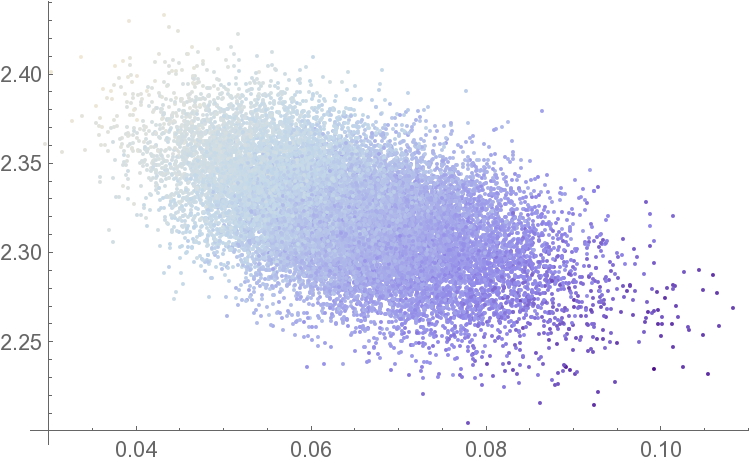}
\put(-8,0){\makebox(0,0){{\tiny $\overline{\Neg^{\text{loc}}_{1|3}}$}}}
\put(-202,140){\makebox(0,0){{\tiny $\mathcal{R}$}}}
\caption{}
\label{}
\end{subfigure}
\caption{Entanglement across $\Sigma$ for a system of 8-qubits ($20000$ random states). The entangling surface $\Sigma$ is fixed and divides the states into two subsystems of four qubits each. The panels show the dependence of the ratio on the internal negativity. The color map shows the values of the entropy in the range ($2.14035, 2.38651$). (a)-(b) entanglement across $\Sigma$, (c)-(d) entanglement inside subsystems.}
\label{fig:8qratio}
\end{figure}
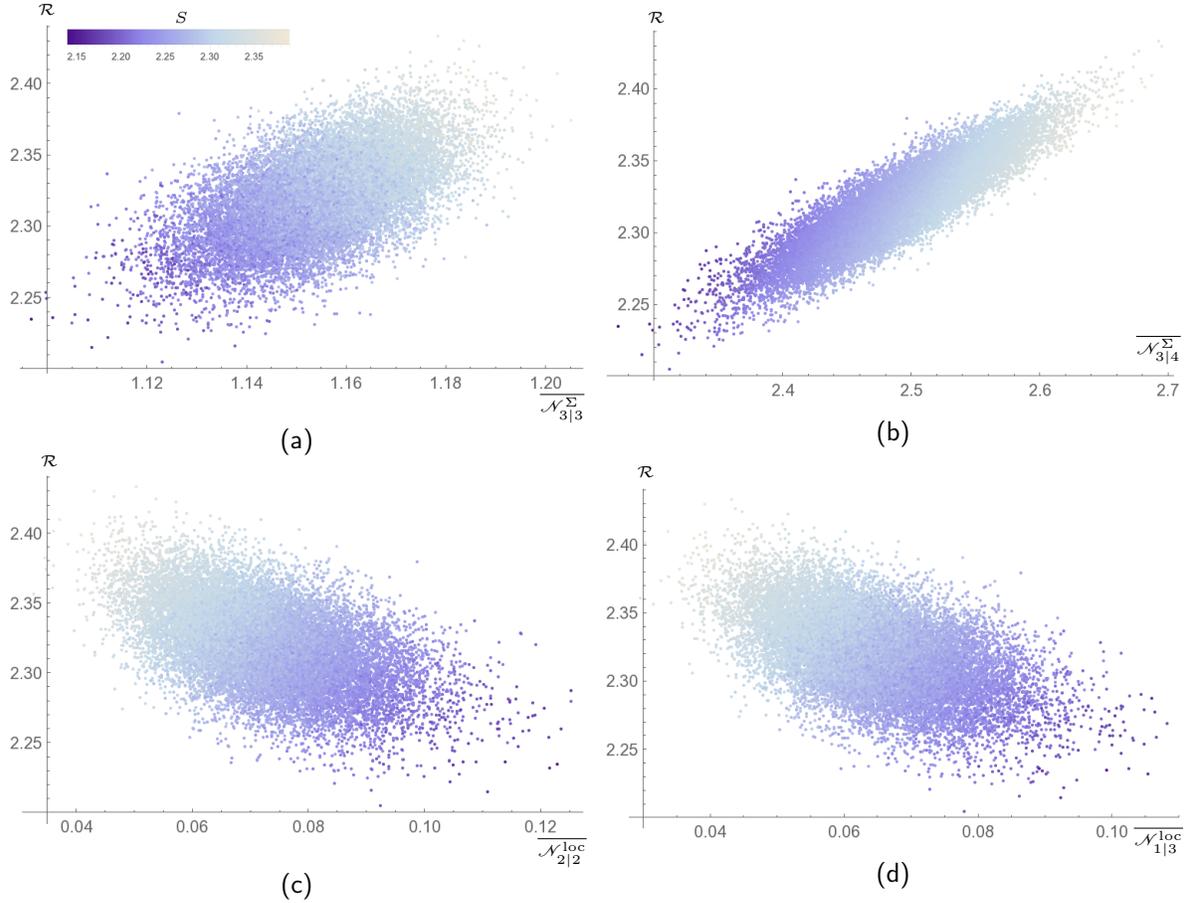
%\afterpage{\clearpage}

% 8 QUBITS WITH CONSTRAINT ON THE ENTROPY
%
\begin{figure}[tb]
\centering
\begin{subfigure}{0.49\textwidth}
\includegraphics[width=\textwidth]{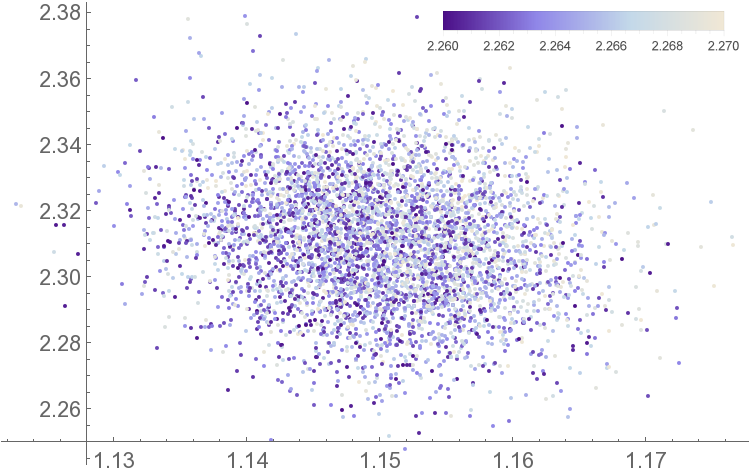}
\put(-8,0){\makebox(0,0){{\tiny $\overline{\Neg^{\Sigma}_{3|3}}$}}}
\put(-192,140){\makebox(0,0){{\tiny $\mathcal{R}$}}}
\put(-48,138){\makebox(0,0){{\tiny $S$}}}
\caption{}
\label{}
\end{subfigure}
\hfill
\begin{subfigure}{0.49\textwidth}
\includegraphics[width=\textwidth]{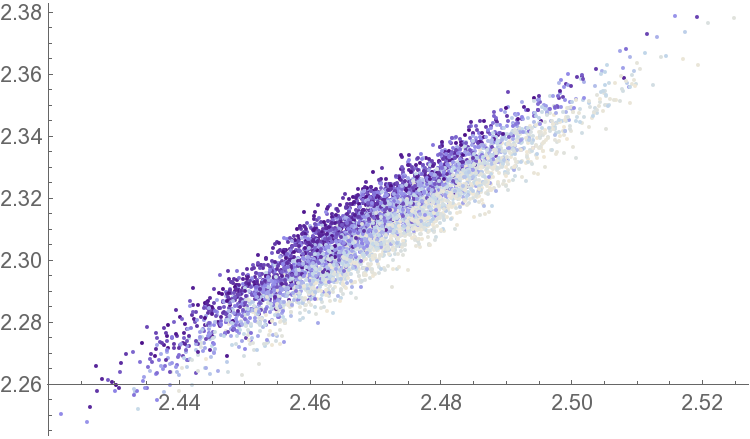}
\put(-8,0){\makebox(0,0){{\tiny $\overline{\Neg^{\Sigma}_{3|4}}$}}}
\put(-203,132){\makebox(0,0){{\tiny $\mathcal{R}$}}}
\caption{}
\label{}
\end{subfigure}

\begin{subfigure}{0.49\textwidth}
\includegraphics[width=\textwidth]{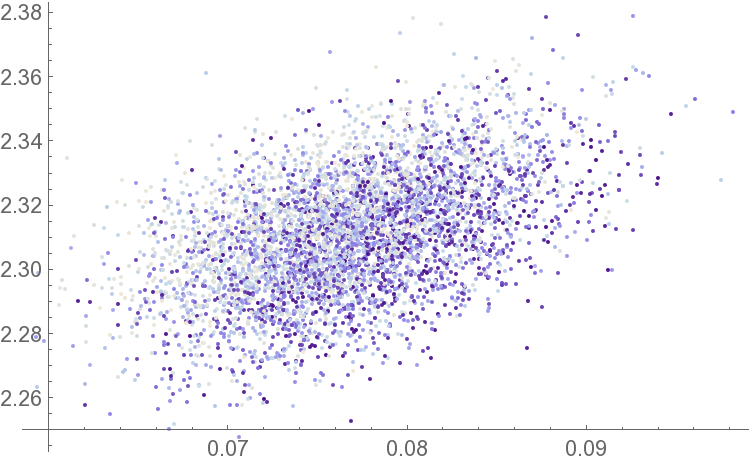}
\put(-8,0){\makebox(0,0){{\tiny $\overline{\Neg^{\text{loc}}_{2|2}}$}}}
\put(-203,138){\makebox(0,0){{\tiny $\mathcal{R}$}}}
\caption{}
\label{}
\end{subfigure}
\hfill
\begin{subfigure}{0.49\textwidth}
\includegraphics[width=\textwidth]{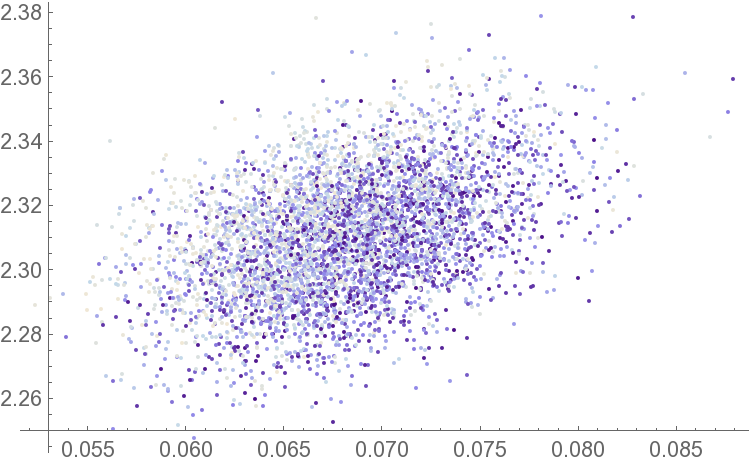}
\put(-8,15){\makebox(0,0){{\tiny $\overline{\Neg^{\text{loc}}_{1|3}}$}}}
\put(-202,139){\makebox(0,0){{\tiny $\mathcal{R}$}}}
\caption{}
\label{}
\end{subfigure}
\caption{A slice of Fig.~\ref{fig:8qratio}, now with a constraint on the entropy which takes values in the interval ($2.26, 2.27$) and is shown by the color map. Plot is for $5000$  random states.}
\label{fig:8qratio_cons}
\end{figure}
%\afterpage{\clearpage}

As in the case of four qubits we want to investigate the dependence of the specific robustness on the internal pattern of entanglement of the system. As before we are interested both in the entanglement between qubits inside a single subsystem and entanglement across the entangling surface $\Sigma$. We will choose $\Sigma$ such that the size of the subsystems is maximal, i.e., commit a 3|3 split for the 6 qubits case, which we keep fixed in what follows.

Let us start by listing all the possible internal negativities we can consider for a pure state of 6-qubits.
\begin{align}
&\text{across }\Sigma\text{:}&\overline{\Neg^{\Sigma}_{2|3}}\qquad\overline{\Neg^{\Sigma}_{1|3}}\qquad\overline{\Neg^{\Sigma}_{2|2}}\qquad\overline{\Neg^{\Sigma}_{1|2}}\qquad\overline{\Neg^{\Sigma}_{1|1}} \nonumber \\
&\text{local:}&\overline{\Neg^{\text{loc}}_{1|2}}\qquad\overline{\Neg^{\text{loc}}_{1|1}}
\label{eq:6qopts}
\end{align}
In the above, averages are computed by considering all possible permutations of qubits (with $\Sigma$ held fixed). 

Of the set of possibilities listed in \eqref{eq:6qopts}, $\overline{\Neg^{\Sigma}_{1|1}}$ is uninteresting,
and so we will ignore it from now on. In Fig.~\ref{fig:6qratio_across} we show the results for the other cases. The color map indicates values of the entropy across $\Sigma$ for the entire system. 

We see that the ratio ${\mathcal R}$ clearly increases when $\overline{\Neg^{\Sigma}_{2|3}}$ or $\overline{\Neg^{\Sigma}_{1|3}}$ increase. Some dependence is also manifest for $\overline{\Neg^{\Sigma}_{2|2}}$ while there is no clear dependence on $\overline{\Neg^{\Sigma}_{1|2}}$. In addition, note that the average negativity is well correlated with the entanglement entropy (they both increase in concert). This should  be compared to Fig.~\ref{subfig:4qR1_2} for the four qubits case. We just have to bear in mind  the obvious fact that  random sampling actually covers a larger portion of the space in the 4-qubit case. For system of six qubits the Hilbert space is much larger and random generation only gives access to a small portion of it. We are probably exploring only a region analogous to the one near the tip of  
Fig.~\ref{subfig:4qR1_2}. This is consistent with the fact that statistically we get high values of the entropy.\footnote{ A maximally entangled state of six qubits, in a bipartite 3|3 sense, has  entropy  $S=2.079$, specific robustness ${\mathcal R} = 1.683$.}  It is entirely possible that as in the four qubits case, ${\mathcal R}$ is maximized by states in another region, again with a much smaller value of the entropy. 

Bearing this caveat in mind we can still investigate how the specific robustness depends on internal entanglement for states in this region of the space. In order to understand whether ${\mathcal R}$ increases with the negativity independently of the entropy, we look for random states with the entropy constrained in some small range of values.  The results for the same bipartitions as above are shown in Fig.~\ref{fig:6qratio_across_cons}. One can see that the weak dependence  noticed above in the case 
$\overline{\Neg^{\Sigma}_{2|2}}$  actually disappears. On the other hand our suspicions are  vindicated for $\overline{\Neg^{\Sigma}_{1|3}}$ and in particular for $\overline{\Neg^{\Sigma}_{2|3}}$. In the last case we also notice that for states with a fixed value of the $\Neg$, the ratio ${\mathcal R}$ seems to be statistically maximized by states with a lower value of the entropy.

We can also look at the local entanglement (again ignoring $\overline{\Neg^{\text{loc}}_{1|1}}$).The results for $\overline{\Neg^{\text{loc}}_{1|2}}$ are shown in Fig.~\ref{fig:6qratio_local}. Quite curiously the ratio seems to slightly decrease as the internal negativity increases. At the same time the entropy seems to decrease as well. If we look instead at states with a constrained value of the entropy, the ratio seems to increase as the negativity increases. 

One lesson that we learn is that the specific robustness ${\mathcal R}$ is particularly sensible if one of subsystems coincides with one of the subsystems of the original bipartition. When we start to trace out qubits, the sensitivity of the ratio progressively fades. Similarly, for the internal entanglement the sensibility of the ratio seems to be higher when do not trace out any qubit.

We can check this result for an even larger system. For a system of eight qubits the list of all possible negativities we could look at is the following:
\begin{align}
&\text{across }\Sigma\text{:}&\overline{\Neg^{\Sigma}_{3|4}}\qquad\overline{\Neg^{\Sigma}_{2|4}}\qquad\overline{\Neg^{\Sigma}_{1|4}}\qquad\overline{\Neg^{\Sigma}_{3|3}}\qquad\overline{\Neg^{\Sigma}_{2|3}}\qquad\overline{\Neg^{\Sigma}_{1|3}}\qquad\overline{\Neg^{\Sigma}_{2|2}}\qquad\overline{\Neg^{\Sigma}_{1|2}}\nonumber\\
&\text{local:}&\overline{\Neg^{\text{loc}}_{1|3}}\qquad\overline{\Neg^{\text{loc}}_{2|2}}\qquad\overline{\Neg^{\text{loc}}_{1|2}}\qquad\overline{\Neg^{\text{loc}}_{1|1}}
\label{}
\end{align}
Fig.~\ref{fig:8qratio} shows the results without a constraint on the entropy, the results for states with constrained entropy are shown in Fig.~\ref{fig:8qratio_cons}.  Similar comments as for the 6-qubit case hold.\footnote{ A maximally entangled state of eight qubits, in a bipartite 4|4 sense, has  entropy  $S=2.773$, specific robustness ${\mathcal R} = 2.705$.}

%~~~~~~~~~~~~~~~~~~~~~~~~~~~~~~~~~~~~~~~~~~~~~~~
\subsection{Exploring multipartite entanglement}
\label{sec:mul68}
%~~~~~~~~~~~~~~~~~~~~~~~~~~~~~~~~~~~~~~~~~~~~~~

% 6 QUBITS I3
%
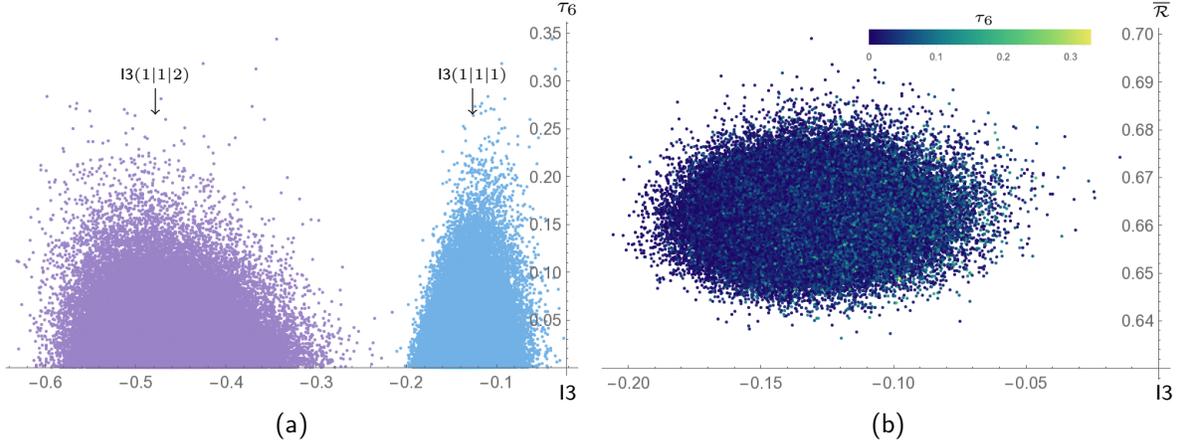
\begin{figure}[tb]
\centering
\begin{subfigure}{0.49\textwidth}
\includegraphics[width=\textwidth]{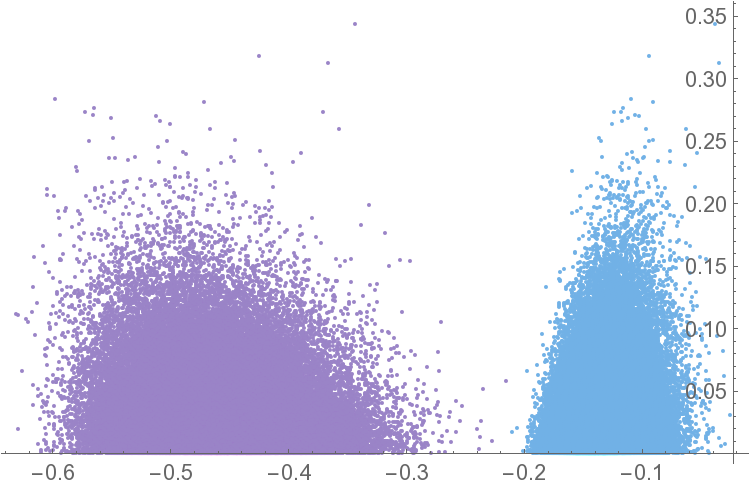}
\put(-4,0){\makebox(0,0){{\scriptsize $\tmi$}}}
\put(-4,145){\makebox(0,0){{\scriptsize $\tau_6$}}}
\put(-40,120){\makebox(0,0){{\tiny$\tmi(1|1|1)$}}}
\put (-40,110) {\makebox(0,0){
	\begin{tikzpicture}
	\draw[->] (0,0)--(0,-10pt);
	\end{tikzpicture}
}}
\put(-160,120){\makebox(0,0){{\tiny $\tmi(1|1|2)$}}}
\put (-160,110) {\makebox(0,0){
	\begin{tikzpicture}
	\draw[->] (0,0)--(0,-10pt);
	\end{tikzpicture}
}}
\caption{}
\label{subfig:6qI3kinds}
\end{subfigure}
\hfill
\begin{subfigure}{0.49\textwidth}
\includegraphics[width=\textwidth]{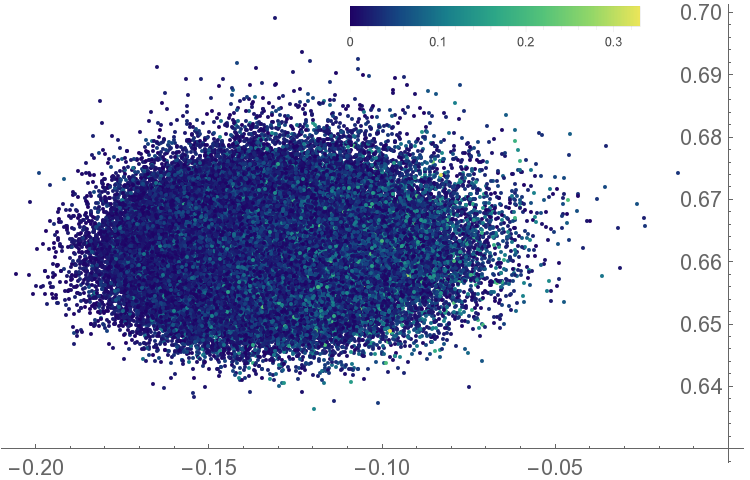}
\put(-4,0){\makebox(0,0){{\scriptsize $\tmi$}}}
\put(-5,145){\makebox(0,0){{\tiny $\overline{\mathcal{R}}$}}}
\put(-72,141){\makebox(0,0){{\tiny $\tau_6$}}}
\caption{}
\label{subfig:6qI3tangle}
\end{subfigure}
\caption{a) Maximal value of $\tmi$ for the two inequivalent kinds of partitionings of \ref{eq:6qI3classes}, from right to left. The values of $\tmi$ are compared to the tangle, $100000$ states per class, b) Max value of $\tmi(1|1|1)$ compared to tangle and the average ratio (color map), $100000$ states.}
\label{fig:6qI3}
\end{figure}
%\afterpage{\clearpage}

We now move to the analysis of multipartite correlations. Consider a pure state of a system $U$ of 
$N$-qubits. We can choose subsystems $A$, $B$, $C$ such that $A\cup B\cup C\equiv U$. Since the state  of $U$ is pure we have $\tmi=0$.  We want instead to look at all the possible values of $\tmi$, for all the possible inequivalent choices of $A$, $B$, $C$ where the total number of qubits $k$ in $A\cup B\cup C$ takes value in $\{3,4,\cdots,N-1\}$, i.e., subsystems obtained by tracing out at least one and at most $N-3$ qubits of the entire system.

We follow the following canonical algorithm for computing the tripartite mutual information.  For a given value of $k$ we consider all the possible ways to partition the qubits into three subsystems (up to qubit permutations), this is given by the list of possible decompositions of $k$ into three integers. Fixing $k$ and the  type of tripartition chosen, we compute all the possible values of $\tmi$ considering the full set of qubits permutations, and retain the maximal value. The choice is inspired  by the fact that its sign clearly tells us whether there is a violation of the monogamy of mutual information for at least one permutation of the qubits.  To wit,
\begin{equation}
\begin{split}
&\tmi(k_1|k_2|k_3) = \max_{\text{perms}} \; \tmi\left(a_{\alpha_1}\cdots a_{\alpha_{k_1}} | 
b_{\beta_1}\cdots b_{\beta_{k_2}}
|c_{\gamma_1}\cdots a_{\gamma_{k_3}} \right) 
\\
& \text{for} \;\; A = \{a_{\alpha_i}\} , \;
B = \{b_{\beta_i}\},  \; C = \{c_{\gamma_i} \} \,,
% & \text{for} \;\; A = \{a_{\alpha_i}, i = 1\cdots ,k_1\} , 
% B = \{b_{\beta_i}, i =1,\cdots,k_2\}, C = \{c_{\gamma_i} , i = 1,\cdots k_3 \} 
 \\   
 &
  k_1+k_2+k_3 = k \in \{3,4,\cdots, N-3\} \,.
\end{split}
\label{eq:I3def}
\end{equation}
Since the global state is pure, we will find certain equivalences among values of $\tmi$ for different kinds of partitions $\{k_1,k_2,k_3\}$ and different values of $k$.

We start with a system of six qubits, where $k$ takes values in $\{3,4,5\}$. The possible kinds of partitions for different values of $k$ are easily listed:
\begin{equation}
\begin{split}
k=3\qquad &\tmi(1|1|1) \\
k=4\qquad &\tmi(1|1|2) \\
k=5\qquad &\tmi(1|1|3)\,,\hspace{0.5cm}\tmi(1|2|2)
\end{split}
\label{eq:6qks}
\end{equation}
Moreover, as promised it is simple to check the following equivalence relations:
\begin{equation}
\begin{split}
\tmi(1|1|1)&\equiv \tmi(1|1|3) \\
\tmi(1|1|2)&\equiv \tmi(1|2|2)
\end{split}
\label{eq:6qI3classes}
\end{equation}
We remind the reader that expressions like $I3(1|1|1)$ here represent the set of values of $\tmi$ for a particular tripartition and all the possible choices of the qubits.

Fig.~\ref{subfig:6qI3kinds} shows the maximal value of I3 for these two classes of equivalent partitioninings, one can notice that $I3(1|1|1)$ is the quantity that get closer to the violation of monogamy. For such a partitioning we then compare the maximal value of $\tmi$ to the value of the averaged specific robustness and the tangle, see Fig.~\ref{subfig:6qI3tangle}. 

% 8 QUBITS I3
%
\begin{figure}[tb]
\centering
\begin{subfigure}{0.49\textwidth}
\includegraphics[width=\textwidth]{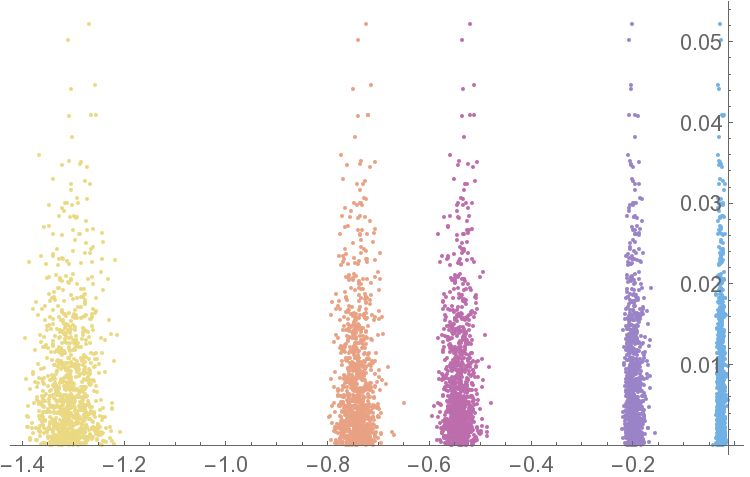}
\put(-4,0){\makebox(0,0){{\scriptsize $\tmi$}}}
\put(-5,144){\makebox(0,0){{\scriptsize $\tau_8$}}}
\put(-35,120){\makebox(0,0){{\tiny$\tmi(1|1|2)$}}}
\put (-35,110) {\makebox(0,0){
	\begin{tikzpicture}
	\draw[->] (0,0)--(0pt,-10pt);
	\end{tikzpicture}
}}
\put(-114,120){\makebox(0,0){{\tiny $\tmi(1|2|2)$}}}
\put (-114,110) {\makebox(0,0){
	\begin{tikzpicture}
	\draw[->] (0,0)--(0,-10pt);
	\end{tikzpicture}
}}
\put(-200,120){\makebox(0,0){{\tiny $\tmi(2|2|2)$}}}
\put (-200,110) {\makebox(0,0){
	\begin{tikzpicture}
	\draw[->] (0,0)--(0,-10pt);
	\end{tikzpicture}
}}
\caption{}
\label{subfig:8qI3kinds}
\end{subfigure}
\hfill
\begin{subfigure}{0.49\textwidth}
\includegraphics[width=\textwidth]{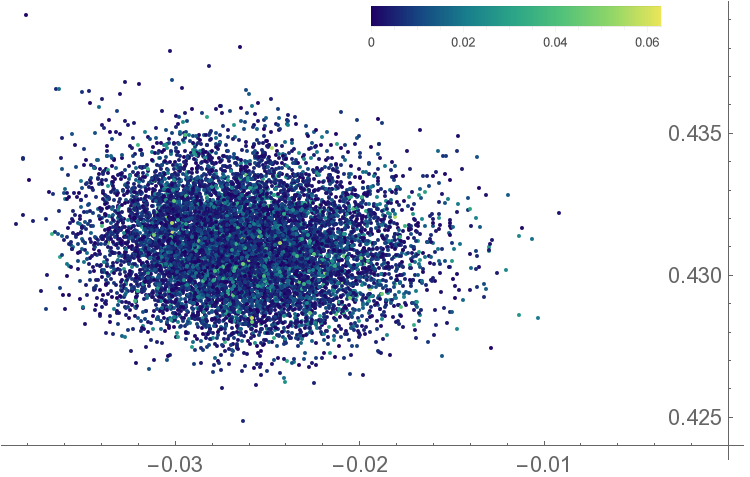}
\put(-4,0){\makebox(0,0){{\scriptsize $\tmi$}}}
\put(-5,145){\makebox(0,0){{\tiny $\overline{\mathcal{R}}$}}}
\put(-63,140){\makebox(0,0){{\tiny $\tau_8$}}}
\caption{}
\label{subfig:8qI3tangle}
\end{subfigure}
\caption{a) Maximal value of $\tmi$ for the five inequivalent kinds of partitionings of \ref{eq:8qI3classes}, from right to left. The values of $\tmi$ are compared to the tangle, $1000$ states per class, b) Max value of $I3(1|1|1)$ compared to tangle and the average ratio (color map), $10000$ states.}
\label{fig:8qI3}
\end{figure}
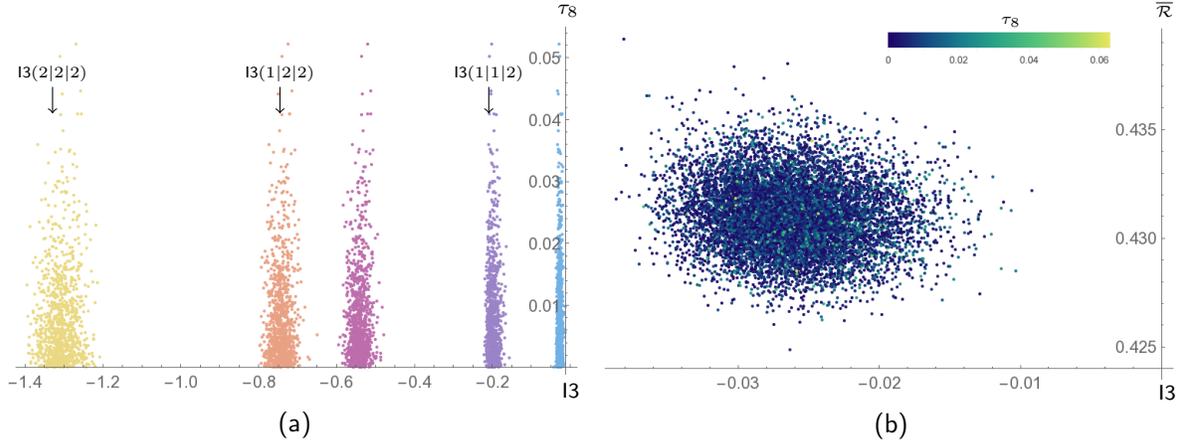
%\afterpage{\clearpage}

We can repeat this analysis for the eight qubits case. The list of all the equivalence classes of $\tmi$ for different values of $k$ is:
\begin{equation}
\begin{split}
I3(1|1|1)&\equiv I3(1|1|5) \\
I3(1|1|2)&\equiv I3(1|1|4)\equiv I3(1|2|4) \\
I3(1|1|3)&\equiv I3(1|3|3) \\
I3(1|2|2)&\equiv I3(1|2|3)\equiv I3(2|2|3) \\
I3(2|2|2)
\end{split}
\label{eq:8qI3classes}
\end{equation}
Fig.~\ref{subfig:8qI3kinds} shows the results for the five inequivalent values of $\tmi$. As for the six qubit case, we also report the relation with the average ratio and the tangle, see Fig.~\ref{subfig:8qI3tangle}. 

As expected from the results for four qubits, the monogamy of mutual information appears to be generically satisfied.  Thus the monogamy of mutual information in these systems is not particularly restrictive vis a vis applications to holographic considerations. We expect the result to be true for larger systems as well.

 From the eight qubits case another suggestive pattern appears to emerge. The values of $\tmi$ that get closer to zero correspond to the case where single qubits are involved, eg., $I3(1|1|1)$. On the other hand tripartitions into larger subsystems seem to produce the lowest values of $\tmi$, eg., $I3(2|2|2)$. 
The result seems to suggest that mutual information is essentially monogamous for large regions and that in search of a violation one should look at regions of the smallest possible size.

%~~~~~~~~~~~~~~~~~~~~~~~~~~~~~~~~~~~~~~~~~~~~~~~
\section{Discussion: Lessons for holography}
\label{sec:discuss}
%~~~~~~~~~~~~~~~~~~~~~~~~~~~~~~~~~~~~~~~~~~~~~~

The main message of our discussion has been to demonstrate  how several measures of quantum entanglement (more generally correlations) can be used to investigate the \textit{structure of entanglement} of pure states of non-interacting qubits. 
Specifically, we examined the properties of this entanglement that are captured by  (a) the monogamy of negativity, (b) the specific robustness, (c) tripartite mutual information, and (d) the tangles. Our primary interest was to use these measures to 
delineate the distribution of entanglement inside a given pure state of $N$-qubits. For the most part we resorted to random sampling from the space of states, though in specific circumstances (e.g., 3 and 4 qubit systems) we did make use of available classification schemes. 

Even if the systems under consideration were a vast oversimplification of continuum QFTs, we believe that they have rather useful message to impart in the context of holography. This should in part be attributed to the non-trivial structure entanglement inherent in them, coupled with their eminent tractability. Let us therefore try to abstract some general lessons for holographic systems.

Consider a state of a  holographic QFT (assuming $c_\text{eff} \gg 1$) which is dual to a classical bulk geometry. This state can be pure or mixed (eg., the thermal density matrix  which is dual to a black hole).  We pick a spatial region $\regA$ on the background geometry where the QFT resides, to make up our subsystem. The holographic entanglement entropy prescriptions of 
\cite{Ryu:2006bv,Ryu:2006ef,Hubeny:2007xt} associate the area of an extremal surface to the von Neumann entropy $S_\regA$.

There are two interpretation of $S_\regA$, which are sometimes called \textit{objective} and \textit{subjective} in the literature. The objective interpretation, which is perhaps more physical, relies on the intuition of entropy as measure of disorder of the system. The subjective point of view, on the other hand, is typical of information theory. If the entropy of $\regA$ is thermal there is still no correlation among degrees of freedom inside $\regA$, but the system contains complete information about some other system, i.e., its purification. It is in this second case that the von Neumann entropy can be interpreted as a measure of entanglement between the two parties, namely entanglement entropy. From the point of view of understanding the structure of quantum entanglement, the subjective viewpoint is more appropriate. We will therefore focus on pure states of some extended system;  e.g., instead of the thermal density matrix we pick the thermofield double state \cite{Maldacena:2001kr}. Given a subregion $\regA$, it should therefore be borne in mind that the complement $\regAc$ could include the purifying degrees of freedom.

Given this configuration, lets say that we are handed an algorithm for reconstructing the bulk geometry from the information theoretic content of the field theory. To be sure, such an algorithm does not exist to date, but it has been speculated that the picture is somewhat akin to tensor networks which encapsulate the entanglement pattern of the state \cite{Swingle:2009bg,Hartman:2013qma}. The closest one gets is the error correction model discussed recently in \cite{Pastawski:2015qua} (see \cite{Almheiri:2014lwa} for the genesis of this set of ideas).   In this context, the extremal surface and spacetime regions associated with it, such as the entanglement wedge are distinguished, in that they capture the long range correlations of the degrees of freedom  contained in the subregion $\regA$ of interest and its complement.

Per se tensor networks or other models are but a tool to characterize the structure of correlations of a state. We want to analyse the entanglement pattern from a more operational perspective. Given a global pure state of a system $U$ consider $N$-parties ${\cal O}_i$, each of which has access only to some subregion $\regA_i$ of $U$ (in general $\cup_{i=1}^N\, \regA_i \subseteq U$) and suppose that all the parties are allowed to perform operations on their subsystems and are also allowed to communicate through classical channels.\footnote{ This is a cartoon of the typical quantum information set-up where different parties are allowed to perform LOCC. We ignore any causal constraint (note that actually these regions live on a time-slice of the theory). This is somewhat akin to the discussions of \cite{Czech:2014tva} who attempt to give an operational definition to the concept of differential entropy.} The  entanglement amongst these subregions is the resource that ${\cal O}_i$'s can use to implement typical tasks that would not be achievable if they were restricted to classical correlations. A common intuitive procedure for example would be the distillation of Bell pairs (for two parties) or GHZ$_N$ states (for $N>2$) which can be stored and later used for other purposes. Entanglement measures, both bipartite and multipartite, are designed to quantify this resource and characterize its properties. As mentioned earlier, in the case of two parties the logarithmic negativity provides an upper bound to distillable entanglement.\footnote{ As  explained hitherto we used instead the negativity because of its interpretation for pure states, but one can map from one quantity to the other.} Unfortunately this being hard to compute in all but the simplest cases, we resorted to qubit experiments to gain some intuition; likewise to  our knowledge there is no clean measure of multipartite entanglement defined for continuous systems. Effectively, we are making the approximation $\regA_i = \text{span}\{\ket{0}, \ket{1}\}$, which is dramatic but useful truncation.\footnote{ It would be interesting to upgrade our explorations where we replace a single qubit by a composite system of many qubits; ramping up the internal dimension could allow exploration of free vector or matrix like models. We thank Don Marolf for a discussion on this issue. }

Alternatively, imagine an external agent that knowns the detailed properties of the state and wants to disrupt the entanglement for a particular bipartition. As far as only bipartitions of the entire state are concerned, the minimal amount of noise that she has to inject into the system is captured by the negativity. A question we want to ask is to what extent is the dual geometry stable against these operations. Similarly one may also ask what is the consequence of the distillation procedure on the state. It would be nice to have a model where one could test the effect of different protocols explicitly. In its absence, we limited our analysis to the minimal sufficient condition for the disruption of the state, namely the violation of monogamy of mutual information. 

The first lesson one learns from our study of $\tmi$ in qubit systems is that the monogamy constraint for mutual information, which is known to be satisfied by holographic states \cite{Hayden:2011ag}, is not a particularly restrictive condition. Statistically, random states of four qubits tend in general to satisfy monogamy;  this only strengthens as the number of qubits increases. We discussed how for a given number of qubits, the possible values of $\tmi$ for different choices of the three subregions are related. When large regions are involved, $\tmi$ becomes more and more negative and, statistically, the matching of the monogamy restriction becomes even more favoured. This in particular suggests that in search of a violation of monogamy one should look at the smallest possible regions. This would for instance suggest that in multiboundary wormhole spacetimes of \cite{Brill:1995jv,Brill:1998pr}, which was analyzed in  \cite{Balasubramanian:2014hda}, one should retain the domain of outer communication of the smallest set of black holes. 

Being a linear combination of mutual informations, $\tmi$ mixes quantum and classical correlations. We compared its behaviour to the multipartite entanglement captured by the tangle. Our results suggest that states with high values of quantum multipartite entanglement have a higher probability to violate monogamy of mutual information. For the GHZ$_4$ state this is a known result, which was used in  \cite{Gharibyan:2013aha} to argue that a $4$-boundary state built from many copies of GHZ$_4$ cannot be dual to a smooth classical geometry (see also \cite{Balasubramanian:2014hda,Susskind:2014yaa}). Our result extends this argument to generic states with a high multipartite entanglement. We discussed  a possible interpretation of multipartite entanglement that relies on the residual entanglement which is left after one takes into account all the (mixed) bipartite and tripartite entanglement in the state. A small value of $4$-partite entanglement then corresponds to strong correlation among internal parties. This may suggest that special states with high internal correlation are then more suitable for holography. Nevertheless it is immediate to check that even the W$_4$ state violates monogamy. More generally, the states of $4$-qubits with the strongest internal correlations are precisely those belonging to the special classes $\qsl{2}$-$\qsl{6}$. Under the effect of SLOCC operations a state in a class can evolve to a new state that violates monogamy. The  one exception 
are states in class $\qsl{5}$ which always respect $\tmi <0$. This class of $4$-qubit states is in a natural sense most suitable to be ``holographic''. In any event it is tempting to conjecture that states with geometric duals in holography  have strong bipartite (or tripartite) correlations.

For larger qubit systems, there isn't a statistically significant correlation between the $N$-partite entanglement and the tripartite mutual information. In fact, in our analysis it appears that increasing the number of qubits is sufficient to ensure that the $\tmi <0$. However, absent a classification, we have been forced to examine generic states in these systems, so it would be interesting to further analyze if there is a special sub-class of states with a specific pattern that mimics the class $\qsl{5}$ of $4$-qubits.

We then argued that for pure states with a given bipartitioning, even if the amount of entanglement is measured by the entropy, other measures can provide further resolution of the entanglement structure. Negativity is good measure of quantum entanglement (which can be computed in the continuum) -- it allows for some resolution of  the pattern of entanglement inherent in the state. We showed that for $4$-qubits states the internal pattern of negativity allows for a partial resolution of the classes. 

In particular, we looked at the ratio between these two quantities for bipartitions of a pure state, which we called \textit{specific robustness}. We proposed that this captures the minimal amount of noise necessary to disentangle Bell pairs in the bipartition. We further demonstrated how it is related to the distribution of internal entanglement. 
When a state is highly entangled (vis a vis $S$), the specific robustness is highly constrained. On the other hand when $S$ is small 
the behaviour depends on the specific details of the state. In general one could say that large values of  specific robustness  correspond to nematic type order: the local entanglement for the bipartitioning across some entangling surface is large, but the entanglement entropy for the reduced density matrices themselves is small. 

On a slightly different note,  holographic states are known to satisfy other interesting constraints on the distribution of internal correlations. There are situations where the Araki-Lieb inequality (AL) can be saturated  to leading order in $c_\text{eff}$, leading to the entanglement plateaux phenomena \cite{Hubeny:2013gta}. The prototypical example is a thermofield double state (a pure state in ${\cal H}\otimes {\cal H}$), where for a subsystem $\regA \in  {\cal H}$ one finds $S_\regA = S_{{\cal H}\backslash\regA} + S_\text{thermal}$. As described there (and further explained in \cite{Headrick:2013zda,Rangamani:2014ywa}), one can visualize this as saying that the degrees of freedom in $\regA$ can be decomposed in two groups; one that carries entanglement across the entangling surface in ${\cal H}$ and the other carries the thermal correlations built into the thermofield state.\footnote{ Geometrically this is realized when the extremal surface associated with a region $\regA$ splits up into a disjoint union of a small surface that is anchored on $\regAc = {\cal H}\backslash \regA$ and the bifurcation surface of a black hole, see \cite{Hubeny:2013gta} for illustrative examples.} This implies a factorization of the global state into two components.

We explained how the AL can be interpreted as a constraint on the internal pattern of entanglement of the state and related its saturation to the disentangling theorem for the negativity (cf., \cite{Rangamani:2014ywa}). When the conditions for the disentangling theorem are matched then the degrees of freedom satisfy the factorization mentioned above. Curiously, it was possible to see such behaviour in special states of even small numbers of qubits. Our analysis demonstrates conclusively that the conditions for the disentangling theorem  are not only sufficient, but also necessary for the saturation of Araki-Lieb, strengthening thus the entropic results of \cite{Zhang:2012fp}.

All in all, qubit systems appear to provide an excellent playground for understanding the general properties of entanglement that one might hope to understand in holographic contexts. While our analysis has been restricted to the simplest of possible scenarios, the rich structure seen in the qubit states, leads us to believe that one could extract general lessons from examining them closely. It would be interesting to build in minimal dynamics and or consider networks of qubits as in graph states or tensor type networks \cite{Pastawski:2015qua}, to gain more insight into the interplay of entanglement and geometry.

%~~~~~~~~~~~~~~~~~~~~~~~~~~~~~~~~~~~~~~~~~~~~~~
\acknowledgments 
%~~~~~~~~~~~~~~~~~~~~~~~~~~~~~~~~~~~~~~~~~~~~~~

It is a pleasure to thank  Jyotirmoy Bhattacharya, Matthew Headrick, Felix Haehl, Veronika Hubeny, Wei Li and Henry Maxfield for extremely useful discussions. We would  in particular like to thank Eric Perlmutter for collaboration during the early stages of the project.

M.~Rangamani would like to thank KITP, Santa Barbara for hospitality during the concluding stages of the project, where his work was supported in part by the National Science Foundation under Grant No. NSF PHY11-25915. This work was supported in part by FQXi  grant "Measures of Holographic Information" (FQXi-RFP3-1334), by the STFC Consolidated Grant ST/L000407/1 and by the European Research Council under the European Union's Seventh Framework Programme (FP7/2007-2013), ERC Consolidator Grant Agreement ERC-2013-CoG-615443: SPiN (Symmetry Principles in Nature).

\newpage

\appendix
%~~~~~~~~~~~~~~~~~~~~~~~~~~~~~~~~~~~~~~~~~~~~~~~
\section{Four qubit states: Detailed analysis of SLOCC classes}
\label{sec:appendix}
%~~~~~~~~~~~~~~~~~~~~~~~~~~~~~~~~~~~~~~~~~~~~~~
In this appendix we collect the results for the SLOCC classes of 4-qubit states. These complement the discussion of \S\ref{subsec:4qclasses} in that they encompass the partitions of qubits we did not consider in the main text.

\subsection*{$\qsl{1}$}

\begin{figure}[H]
\centering
\begin{subfigure}{0.49\textwidth}
\includegraphics[width=\textwidth]{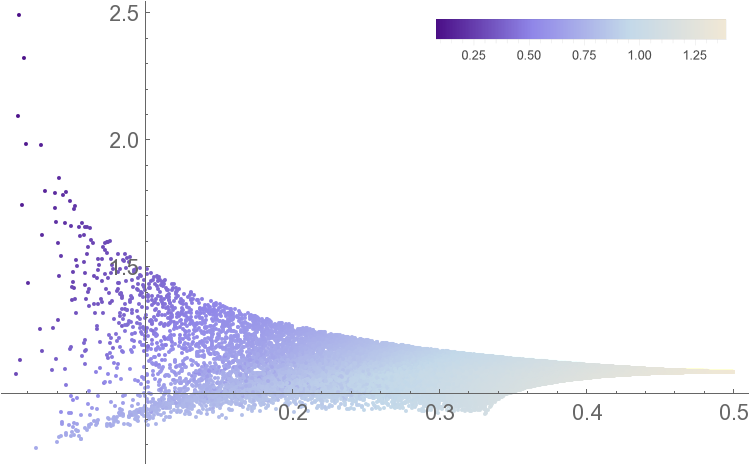}
\put(-177,139){\makebox(0,0){{\tiny $\mathcal{R}_{ac|bd}$}}}
\put(-9,3){\makebox(0,0){{\tiny $\overline{\Neg^\Sigma_{1|2}}$}}}
\put(-47,133){\makebox(0,0){{\tiny $S$}}}
\subcaption{The result for the cut $\mathcal{R}_{ad|bc}$ is equivalent.}
\label{}
\end{subfigure}
\hfill
\begin{subfigure}{0.49\textwidth}
\includegraphics[width=\textwidth]{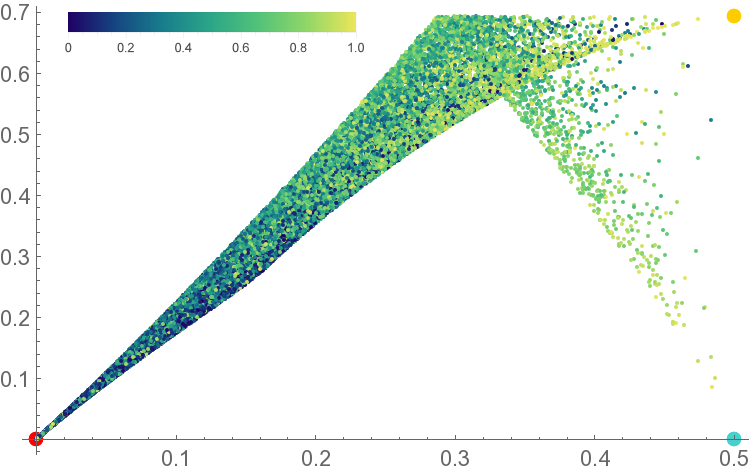}
\put(-12,-3){\makebox(0,0){{\tiny $\Delta\Neg_{ABC}$}}}
\put(-204.5,140){\makebox(0,0){{\tiny $\Delta S_{AB}$}}}
\put(-154,136){\makebox(0,0){{\tiny $\tau_4$}}}
\put (-13.5,15) {\makebox(0,0){
	\begin{tikzpicture}
	\draw[->] (0,0)--(8.5pt,-5.26pt);
	\end{tikzpicture}
}}
\put(-22,21){\makebox(0,0){{\tiny{$\Phi$}}}}
\put (-201,21) {\makebox(0,0){
	\begin{tikzpicture}
	\draw[->] (0,0)--(-3.71pt,-9.28pt);
	\end{tikzpicture}
}}
\put(-197,30){\makebox(0,0){{\tiny{$\Xi$}}}}
\put (-15,135) {\makebox(0,0){
	\begin{tikzpicture}
	\draw[->] (0,0)--(9.8pt,-1.96pt);
	\end{tikzpicture}
}}
\put(-29,139){\makebox(0,0){{\tiny{GHZ}}}}
\caption{$A=\{a\}$, $B=\{bc\}$, $C=\{d\}$}
\label{subfig:AL1b}
\end{subfigure}
\caption{(a) Specific robustness for different choices of $\Sigma$. (b) Saturation of AL for alternative permutation of qubits, see Tab.~\ref{tab:AL1} for the full list of possible cases.}
\label{}
\end{figure}
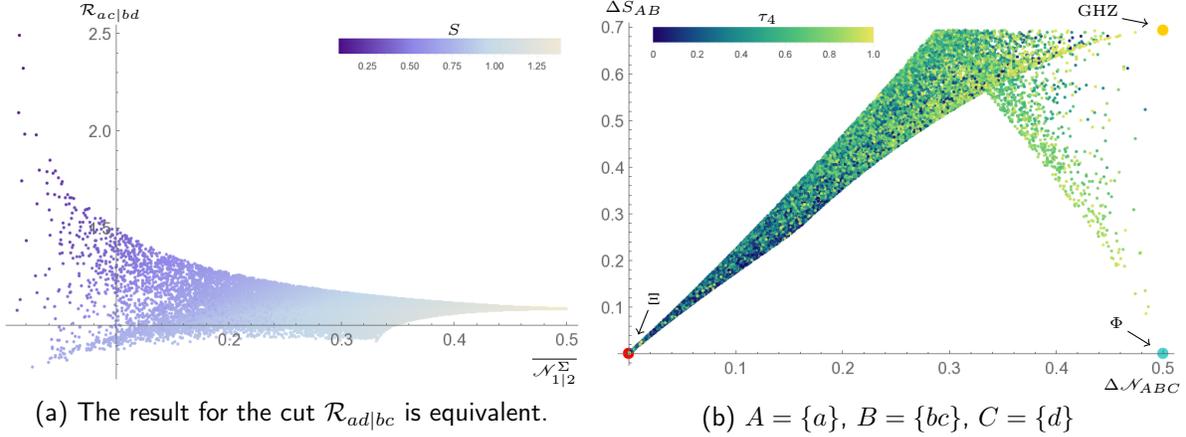

\begin{table}[H]
\centering
\begin{tabular}{l|l}
$A=\{b\}\,,\; B=\{cd\}\,,\; C=\{a\}$ & $A=\{a\}\,,\; B=\{bc\}\,,\; C=\{d\}$\\
$A=\{a\}\,,\; B=\{cd\}\,,\; C=\{b\}$ & $A=\{a\}\,,\; B=\{bd\}\,,\; C=\{c\}$\\
$A=\{d\}\,,\; B=\{ab\}\,,\; C=\{c\}$ & $A=\{b\}\,,\; B=\{ac\}\,,\; C=\{d\}$\\
$A=\{c\}\,,\; B=\{ab\}\,,\; C=\{d\}$ & $A=\{b\}\,,\; B=\{ad\}\,,\; C=\{c\}$\\
  &   $A=\{d\}\,,\; B=\{ac\}\,,\; C=\{b\}$\\
  &   $A=\{d\}\,,\; B=\{bc\}\,,\; C=\{a\}$\\
  &   $A=\{c\}\,,\; B=\{ad\}\,,\; C=\{b\}$\\
  &   $A=\{c\}\,,\; B=\{bd\}\,,\; C=\{a\}$\\
\end{tabular}
\caption{Possible permutations of qubits for the disentangling theorem of the negativity and the saturation of AL inequality. The left column shows the choices which give the result shown in the main text, the right column corresponds to Fig.~\ref{subfig:AL1b}.}
\label{tab:AL1}
\end{table}

\newpage

\subsection*{$\qsl{2}$}

\begin{figure}[H]
\centering
\begin{subfigure}{0.49\textwidth}
\includegraphics[width=\textwidth]{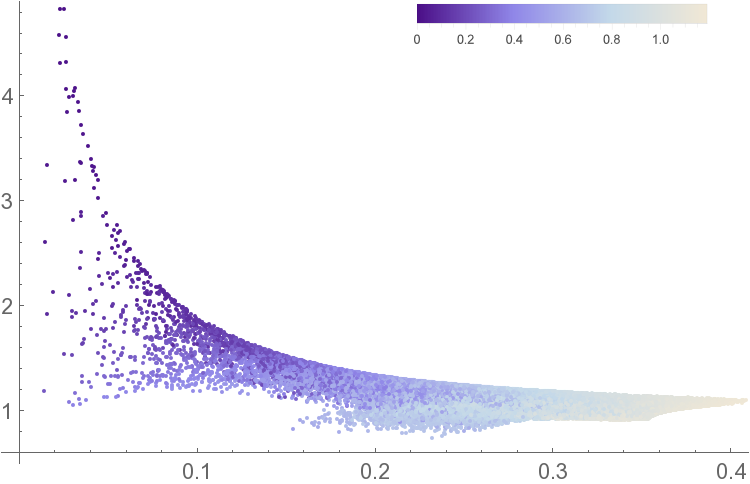}
\put(-212,143){\makebox(0,0){{\tiny $\mathcal{R}_{ac|bd}$}}}
\put(-8,-8){\makebox(0,0){{\tiny $\overline{\Neg^\Sigma_{1|2}}$}}}
\put(-54,142){\makebox(0,0){{\tiny $S$}}}
\subcaption{The result for the cut $\mathcal{R}_{ad|bc}$ is equivalent. The plot is truncated, few states with small values of $\Neg$ and ratio up to $\mathcal{R}\sim 7.2$ are not shown ($\mathcal{R}\sim 4.8$ for the other cut).}
\label{}
\end{subfigure}
\hfill
\begin{subfigure}{0.49\textwidth}
\includegraphics[width=\textwidth]{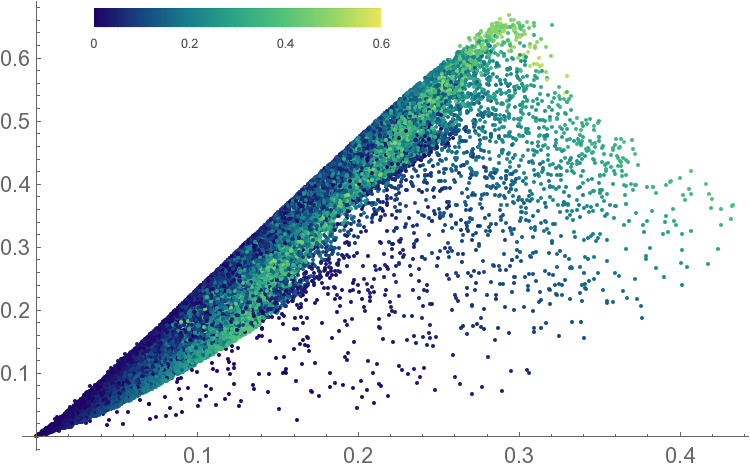}
\put(-12,-6){\makebox(0,0){{\tiny $\Delta\Neg_{ABC}$}}}
\put(-204.5,140){\makebox(0,0){{\tiny $\Delta S_{AB}$}}}
\put(-148,137){\makebox(0,0){{\tiny $\tau_4$}}}
\caption{$A=\{a\}$, $B=\{bc\}$, $C=\{d\}$}
\label{subfig:AL2b}
\end{subfigure}
\caption{(a) Specific robustness for different choices of $\Sigma$. (b) Saturation of AL for alternative permutation of qubits, see Tab.~\ref{tab:AL2} for the full list of possible cases.}
\label{}
\end{figure}

\begin{table}[H]
\centering
\begin{tabular}{l|l}
$A=\{a\}\,,\; B=\{cd\}\,,\; C=\{b\}$ & $A=\{a\}\,,\; B=\{bc\}\,,\; C=\{d\}$\\
$A=\{b\}\,,\; B=\{cd\}\,,\; C=\{a\}$ & $A=\{a\}\,,\; B=\{bd\}\,,\; C=\{c\}$\\
$A=\{d\}\,,\; B=\{ab\}\,,\; C=\{c\}$ & $A=\{b\}\,,\; B=\{ac\}\,,\; C=\{d\}$\\
$A=\{c\}\,,\; B=\{ab\}\,,\; C=\{d\}$ & $A=\{b\}\,,\; B=\{ad\}\,,\; C=\{c\}$\\
  &   $A=\{d\}\,,\; B=\{ac\}\,,\; C=\{b\}$\\
  &   $A=\{d\}\,,\; B=\{bc\}\,,\; C=\{a\}$\\
  &   $A=\{c\}\,,\; B=\{ad\}\,,\; C=\{b\}$\\
  &   $A=\{c\}\,,\; B=\{bd\}\,,\; C=\{a\}$\\
\end{tabular}
\caption{Possible permutations of qubits for the disentangling theorem of the negativity and the saturation of AL inequality. The left column shows the choices which give the result shown in the main text, the right column corresponds to Fig.~\ref{subfig:AL2b}.}
\label{tab:AL2}
\end{table}

\newpage

\subsection*{$\qsl{3}$}

\begin{figure}[H]
\centering
\begin{subfigure}{0.49\textwidth}
\includegraphics[width=\textwidth]{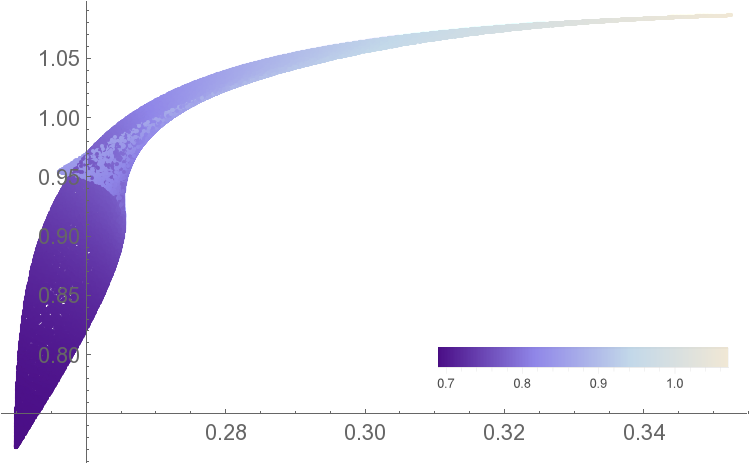}
\put(-191.5,139){\makebox(0,0){{\tiny $\mathcal{R}_{ab|cd}$}}}
\put(-8,5){\makebox(0,0){{\tiny $\overline{\Neg^\Sigma_{1|2}}$}}}
\put(-47,38){\makebox(0,0){{\tiny $S$}}}
\subcaption{The result for the cut $\mathcal{R}_{ad|bc}$ is equivalent.}
\label{subfig:R3a}
\end{subfigure}
\hfill
\begin{subfigure}{0.49\textwidth}
\includegraphics[width=\textwidth]{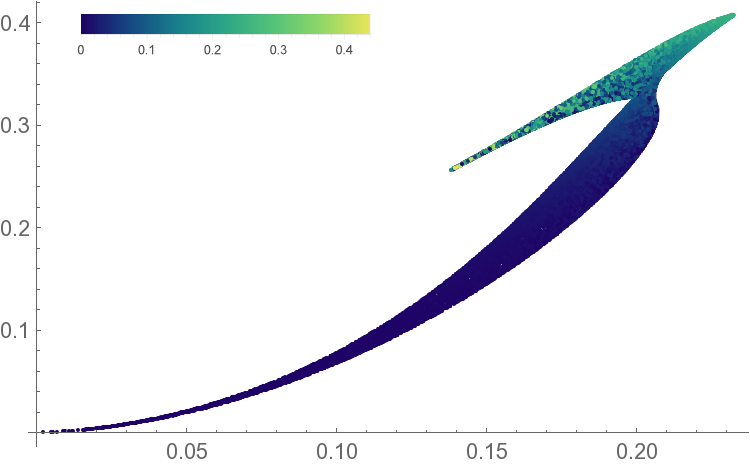}
\put(-12,-6){\makebox(0,0){{\tiny $\Delta\Neg_{ABC}$}}}
\put(-204.5,140){\makebox(0,0){{\tiny $\Delta S_{AB}$}}}
\put(-151,134){\makebox(0,0){{\tiny $\tau_4$}}}
\caption{$A=\{a\}$, $B=\{bc\}$, $C=\{d\}$}
\label{subfig:AL3b}
\end{subfigure}

\begin{subfigure}{0.49\textwidth}
\includegraphics[width=\textwidth]{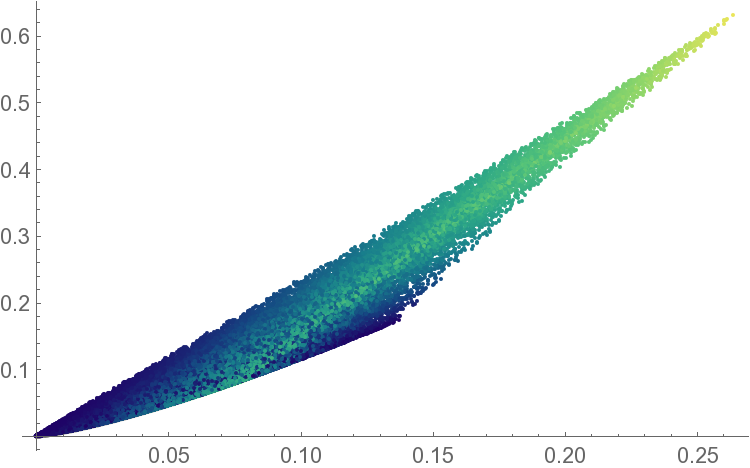}
\put(-12,-6){\makebox(0,0){{\tiny $\Delta\Neg_{ABC}$}}}
\put(-204.5,140){\makebox(0,0){{\tiny $\Delta S_{AB}$}}}
\caption{$A=\{a\}$, $B=\{bd\}$, $C=\{c\}$}
\label{subfig:AL3c}
\end{subfigure}
\hfill
\begin{subfigure}{0.49\textwidth}
\includegraphics[width=\textwidth]{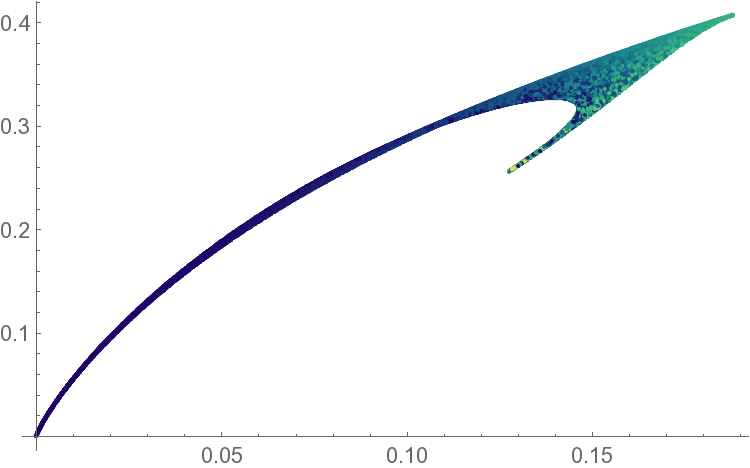}
\put(-12,-6){\makebox(0,0){{\tiny $\Delta\Neg_{ABC}$}}}
\put(-204.5,140){\makebox(0,0){{\tiny $\Delta S_{AB}$}}}
\caption{$A=\{b\}$, $B=\{ad\}$, $C=\{c\}$}
\label{subfig:AL3d}
\end{subfigure}
\caption{(a) Specific robustness for different choices of $\Sigma$. (b)(c)(d) Saturation of AL for alternative permutations of qubits, see Tab.~\ref{tab:AL3} for the full list of possible cases.}
\label{fig:AL3app}
\end{figure}

\begin{table}[H]
\centering
\begin{tabular}{l|l}
$A=\{b\}\,,\; B=\{ac\}\,,\; C=\{d\}$ & $A=\{a\}\,,\; B=\{bc\}\,,\; C=\{d\}$\\
$A=\{d\}\,,\; B=\{ac\}\,,\; C=\{b\}$ & $A=\{a\}\,,\; B=\{cd\}\,,\; C=\{b\}$\\
  &   $A=\{c\}\,,\; B=\{ab\}\,,\; C=\{d\}$\\
  &   $A=\{c\}\,,\; B=\{ad\}\,,\; C=\{b\}$\\
\hline
\hline
$A=\{a\}\,,\; B=\{bd\}\,,\; C=\{c\}$ & $A=\{b\}\,,\; B=\{ad\}\,,\; C=\{c\}$\\
$A=\{c\}\,,\; B=\{bd\}\,,\; C=\{a\}$ & $A=\{b\}\,,\; B=\{cd\}\,,\; C=\{a\}$\\
  &   $A=\{d\}\,,\; B=\{ab\}\,,\; C=\{c\}$\\
  &   $A=\{d\}\,,\; B=\{bc\}\,,\; C=\{a\}$\\
\end{tabular}
\caption{Possible permutations of qubits for the disentangling theorem of the negativity and the saturation of AL inequality. The top left cases give the result shown in the main text. Top right, Fig.~\ref{subfig:AL3b}. Bottom left, Fig.~\ref{subfig:AL3c}. Bottom right, Fig.~\ref{subfig:AL3d}.}
\label{tab:AL3}
\end{table}

\newpage

\subsection*{$\qsl{4}$}

\begin{figure}[H]
\centering
\begin{subfigure}{0.49\textwidth}
\includegraphics[width=\textwidth]{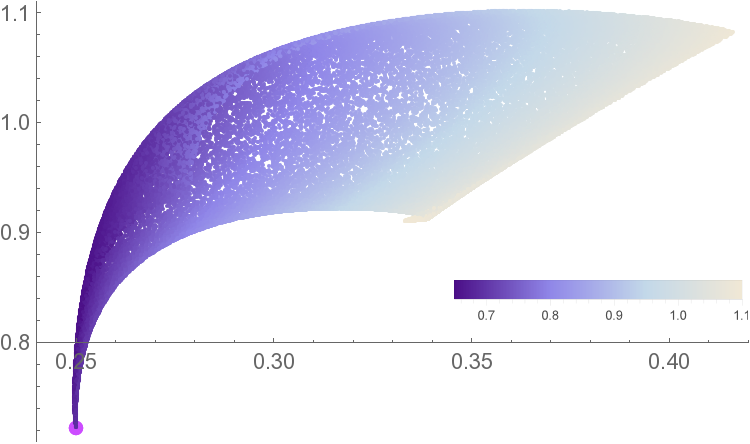}
\put(-206.5,134){\makebox(0,0){{\tiny $\mathcal{R}_{ab|cd}$}}}
\put(-8,12){\makebox(0,0){{\tiny $\overline{\Neg^\Sigma_{1|2}}$}}}
\put(-43,51){\makebox(0,0){{\tiny $S$}}}
\put (-182,4) {\makebox(0,0){
	\begin{tikzpicture}
	\draw[->] (0,0)--(-10pt,0);
	\end{tikzpicture}
}}
\put(-171,4){\makebox(0,0){{\tiny{W}}}}
\subcaption{The result for the cut $\mathcal{R}_{ad|bc}$ is equivalent to the one shown in the main text.}
\label{subfig:R4a}
\end{subfigure}
\hfill
\begin{subfigure}{0.49\textwidth}
\includegraphics[width=\textwidth]{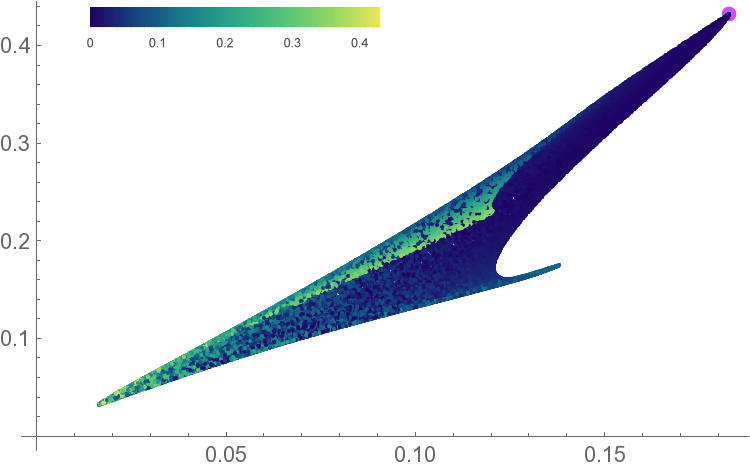}
\put(-12,-6){\makebox(0,0){{\tiny $\Delta\Neg_{ABC}$}}}
\put(-204.5,140){\makebox(0,0){{\tiny $\Delta S_{AB}$}}}
\put(-149,137){\makebox(0,0){{\tiny $\tau_4$}}}
\put (-16,131) {\makebox(0,0){
	\begin{tikzpicture}
	\draw[->] (0,0)--(10pt,0);
	\end{tikzpicture}
}}
\put(-27,131){\makebox(0,0){{\tiny{W}}}}
\caption{$A=\{a\}$, $B=\{cd\}$, $C=\{b\}$}
\label{subfig:AL4b}
\end{subfigure}
\caption{(a) Specific robustness for different choices of $\Sigma$. (b) Saturation of AL for alternative permutation of qubits, see Tab.~\ref{tab:AL4} for the full list of possible cases.}
\label{fig:AL4app}
\end{figure}

\begin{table}[H]
\centering
\begin{tabular}{l|l}
$A=\{a\}\,,\; B=\{bc\}\,,\; C=\{d\}$ & $A=\{a\}\,,\; B=\{cd\}\,,\; C=\{b\}$\\
$A=\{a\}\,,\; B=\{bd\}\,,\; C=\{c\}$ & $A=\{b\}\,,\; B=\{cd\}\,,\; C=\{a\}$\\
$A=\{b\}\,,\; B=\{ac\}\,,\; C=\{d\}$ & $A=\{d\}\,,\; B=\{ab\}\,,\; C=\{c\}$\\
$A=\{b\}\,,\; B=\{ad\}\,,\; C=\{c\}$ & $A=\{c\}\,,\; B=\{ab\}\,,\; C=\{d\}$\\
$A=\{d\}\,,\; B=\{ac\}\,,\; C=\{b\}$  &  \\
$A=\{d\}\,,\; B=\{bc\}\,,\; C=\{a\}$  &  \\
$A=\{c\}\,,\; B=\{ad\}\,,\; C=\{b\}$  &  \\
$A=\{c\}\,,\; B=\{bd\}\,,\; C=\{a\}$  &  \\
\end{tabular}
\caption{Possible permutations of qubits for the disentangling theorem of the negativity and the saturation of AL inequality. The left column shows the choices which give the result shown in the main text, the right column corresponds to Fig.~\ref{subfig:AL4b}.}
\label{tab:AL4}
\end{table}

\newpage

\subsection*{$\qsl{5}$}

\begin{figure}[H]
\centering
\begin{subfigure}{0.49\textwidth}
\includegraphics[width=\textwidth]{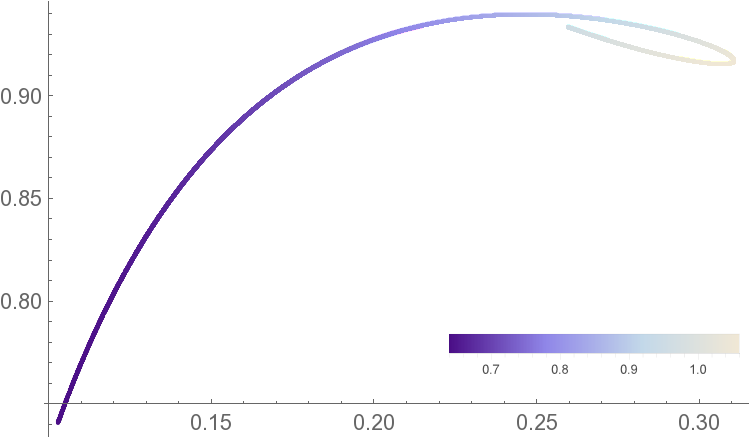}
\put(-203,133){\makebox(0,0){{\tiny $\mathcal{R}_{ac|bd}$}}}
\put(-8,-5){\makebox(0,0){{\tiny $\overline{\Neg^\Sigma_{1|2}}$}}}
\put(-42,34){\makebox(0,0){{\tiny $S$}}}
\subcaption{The result for the cut $\mathcal{R}_{ad|bc}$ is equivalent to the one shown in the main text.}
\label{}
\end{subfigure}

\begin{subfigure}{0.49\textwidth}
\includegraphics[width=\textwidth]{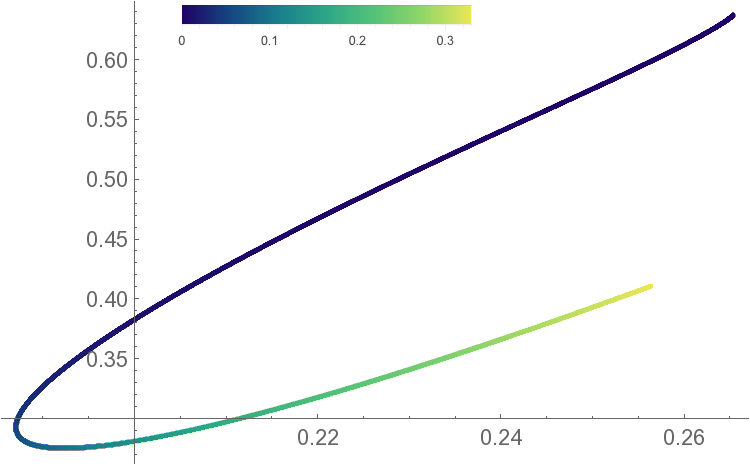}
\put(-12,-3){\makebox(0,0){{\tiny $\Delta\Neg_{ABC}$}}}
\put(-179,140){\makebox(0,0){{\tiny $\Delta S_{AB}$}}}
\put(-124,137){\makebox(0,0){{\tiny $\tau_4$}}}
\caption{$A=\{a\}$, $B=\{bd\}$, $C=\{c\}$}
\label{subfig:AL5b}
\end{subfigure}
\hfill
\begin{subfigure}{0.49\textwidth}
\includegraphics[width=\textwidth]{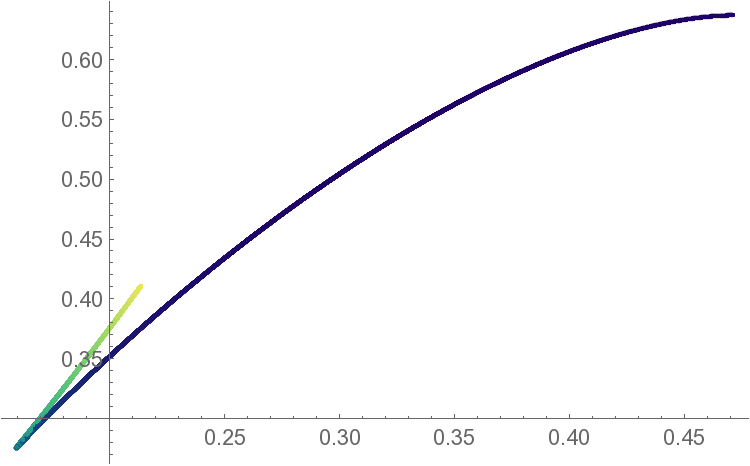}
\put(-12,-3){\makebox(0,0){{\tiny $\Delta\Neg_{ABC}$}}}
\put(-187,140){\makebox(0,0){{\tiny $\Delta S_{AB}$}}}
\caption{$A=\{b\}$, $B=\{ac\}$, $C=\{d\}$}
\label{subfig:AL5c}
\end{subfigure}
\caption{(a) Specific robustness for different choices of $\Sigma$. (b)(c) Saturation of AL for alternative permutations of qubits, see Tab.~\ref{tab:AL5} for the full list of possible cases.}
\label{}
\end{figure}

\begin{table}[H]
\centering
\begin{tabular}{l|l|l}
$A=\{a\}\,,\; B=\{bc\}\,,\; C=\{d\}$ & $A=\{a\}\,,\; B=\{bd\}\,,\; C=\{c\}$ & $A=\{b\}\,,\; B=\{ac\}\,,\; C=\{d\}$\\
$A=\{a\}\,,\; B=\{cd\}\,,\; C=\{b\}$ & $A=\{c\}\,,\; B=\{bd\}\,,\; C=\{a\}$ & $A=\{d\}\,,\; B=\{ac\}\,,\; C=\{b\}$\\
$A=\{b\}\,,\; B=\{ad\}\,,\; C=\{c\}$ &  & \\
$A=\{b\}\,,\; B=\{cd\}\,,\; C=\{a\}$ &  & \\
$A=\{d\}\,,\; B=\{ab\}\,,\; C=\{c\}$ &  & \\
$A=\{d\}\,,\; B=\{bc\}\,,\; C=\{a\}$  &  &  \\
$A=\{c\}\,,\; B=\{ab\}\,,\; C=\{d\}$  &  &  \\
$A=\{c\}\,,\; B=\{ad\}\,,\; C=\{b\}$  &  &  \\
\end{tabular}
\caption{Possible permutations of qubits for the disentangling theorem of the negativity and the saturation of AL inequality. The left column shows the choices which give the result shown in the main text. The center column corresponds to Fig.~\ref{subfig:AL5b}, the right one to Fig.~\ref{subfig:AL5c}.}
\label{tab:AL5}
\end{table}

\newpage

\subsection*{$\qsl{6}$}

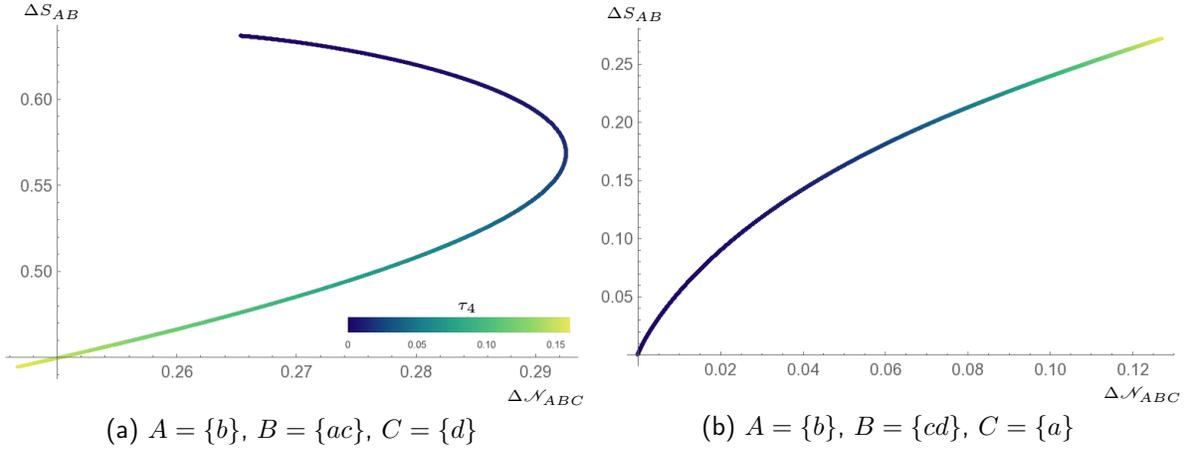
\begin{figure}[H]
\centering
\begin{subfigure}{0.49\textwidth}
\includegraphics[width=\textwidth]{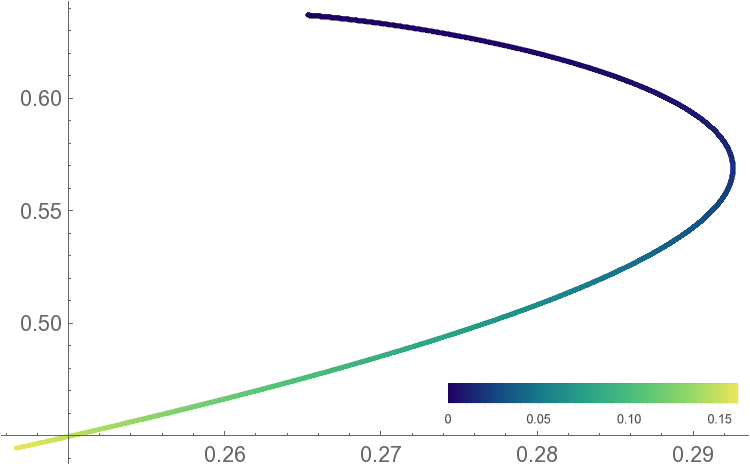}
\put(-12,-6){\makebox(0,0){{\tiny $\Delta\Neg_{ABC}$}}}
\put(-199,140){\makebox(0,0){{\tiny $\Delta S_{AB}$}}}
\put(-42,28){\makebox(0,0){{\tiny $\tau_4$}}}
\caption{$A=\{b\}$, $B=\{ac\}$, $C=\{d\}$}
\label{subfig:AL6b}
\end{subfigure}
\hfill
\begin{subfigure}{0.49\textwidth}
\includegraphics[width=\textwidth]{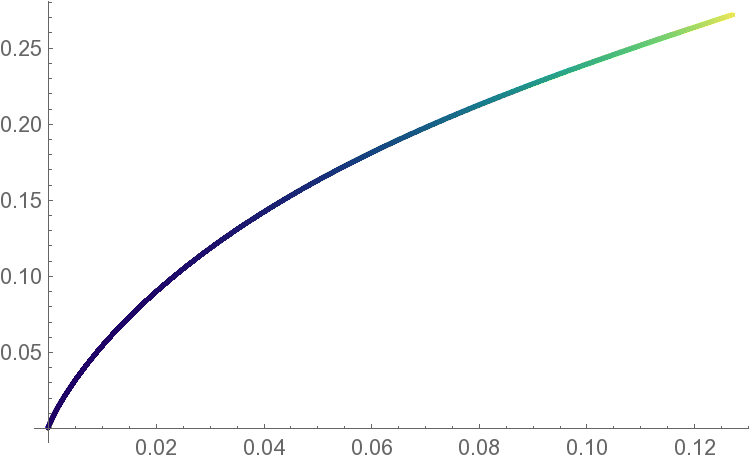}
\put(-12,-6){\makebox(0,0){{\tiny $\Delta\Neg_{ABC}$}}}
\put(-204,138){\makebox(0,0){{\tiny $\Delta S_{AB}$}}}
\caption{$A=\{b\}$, $B=\{cd\}$, $C=\{a\}$}
\label{subfig:AL6c}
\end{subfigure}
\caption{Saturation of AL for alternative permutations of qubits, see Tab.~\ref{tab:AL6} for the full list of possible cases.}
\label{}
\end{figure}

\begin{table}[H]
\centering
\begin{tabular}{l|l|l}
$A=\{a\}\,,\; B=\{bc\}\,,\; C=\{d\}$ & $A=\{b\}\,,\; B=\{ac\}\,,\; C=\{d\}$ & $A=\{b\}\,,\; B=\{cd\}\,,\; C=\{a\}$\\
$A=\{a\}\,,\; B=\{bd\}\,,\; C=\{c\}$ & $A=\{b\}\,,\; B=\{ad\}\,,\; C=\{c\}$ & $A=\{d\}\,,\; B=\{bc\}\,,\; C=\{a\}$\\
$A=\{a\}\,,\; B=\{cd\}\,,\; C=\{b\}$ & $A=\{d\}\,,\; B=\{ab\}\,,\; C=\{c\}$ & $A=\{c\}\,,\; B=\{bd\}\,,\; C=\{a\}$\\
 & $A=\{d\}\,,\; B=\{ac\}\,,\; C=\{b\}$ & \\
 & $A=\{c\}\,,\; B=\{ab\}\,,\; C=\{d\}$ & \\
 & $A=\{c\}\,,\; B=\{ad\}\,,\; C=\{b\}$ & \\
\end{tabular}
\caption{Possible permutations of qubits for the disentangling theorem of the negativity and the saturation of AL inequality. The left column shows the choices which give the result shown in the main text. The center column corresponds to Fig.~\ref{subfig:AL6b}, the right one to Fig.~\ref{subfig:AL6c}.}
\label{tab:AL6}
\end{table}

\newpage

%%%%%%%%%%%%%%%%%%%%%%%%%%%%%%%%%%%%%%%%%%%
%  \bibliographystyle{JHEP}
% \bibliography{negativity}

\begin{thebibliography}{10}

\bibitem{Ryu:2006bv}
S.~Ryu and T.~Takayanagi, {\it {Holographic derivation of entanglement entropy
  from AdS/CFT}},  {\em Phys.Rev.Lett.} {\bf 96} (2006) 181602,
  [\href{http://xxx.lanl.gov/abs/hep-th/0603001}{{\tt hep-th/0603001}}].

\bibitem{Ryu:2006ef}
S.~Ryu and T.~Takayanagi, {\it {Aspects of Holographic Entanglement Entropy}},
  {\em JHEP} {\bf 0608} (2006) 045,
  [\href{http://xxx.lanl.gov/abs/hep-th/0605073}{{\tt hep-th/0605073}}].

\bibitem{Hubeny:2007xt}
V.~E. Hubeny, M.~Rangamani, and T.~Takayanagi, {\it {A Covariant holographic
  entanglement entropy proposal}},  {\em JHEP} {\bf 0707} (2007) 062,
  [\href{http://xxx.lanl.gov/abs/0705.0016}{{\tt arXiv:0705.0016}}].

\bibitem{Swingle:2009bg}
B.~Swingle, {\it {Entanglement Renormalization and Holography}},  {\em
  Phys.Rev.} {\bf D86} (2012) 065007,
  [\href{http://xxx.lanl.gov/abs/0905.1317}{{\tt arXiv:0905.1317}}].

\bibitem{VanRaamsdonk:2009ar}
M.~Van~Raamsdonk, {\it {Comments on quantum gravity and entanglement}},
  \href{http://xxx.lanl.gov/abs/0907.2939}{{\tt arXiv:0907.2939}}.

\bibitem{VanRaamsdonk:2010pw}
M.~Van~Raamsdonk, {\it {Building up spacetime with quantum entanglement}},
  {\em Gen.Rel.Grav.} {\bf 42} (2010) 2323--2329,
  [\href{http://xxx.lanl.gov/abs/1005.3035}{{\tt arXiv:1005.3035}}].

\bibitem{Maldacena:2013xja}
J.~Maldacena and L.~Susskind, {\it {Cool horizons for entangled black holes}},
  {\em Fortsch.Phys.} {\bf 61} (2013) 781--811,
  [\href{http://xxx.lanl.gov/abs/1306.0533}{{\tt arXiv:1306.0533}}].

\bibitem{Brill:1995jv}
D.~R. Brill, {\it {Multi - black hole geometries in (2+1)-dimensional
  gravity}},  {\em Phys.Rev.} {\bf D53} (1996) 4133--4176,
  [\href{http://xxx.lanl.gov/abs/gr-qc/9511022}{{\tt gr-qc/9511022}}].

\bibitem{Aminneborg:1997pz}
S.~Aminneborg, I.~Bengtsson, D.~Brill, S.~Holst, and P.~Peldan, {\it {Black
  holes and wormholes in (2+1)-dimensions}},  {\em Class.Quant.Grav.} {\bf 15}
  (1998) 627--644, [\href{http://xxx.lanl.gov/abs/gr-qc/9707036}{{\tt
  gr-qc/9707036}}].

\bibitem{Brill:1998pr}
D.~Brill, {\it {Black holes and wormholes in (2+1)-dimensions}},  {\em
  Lect.Notes Phys.} {\bf 537} (2000) 143,
  [\href{http://xxx.lanl.gov/abs/gr-qc/9904083}{{\tt gr-qc/9904083}}].

\bibitem{Gharibyan:2013aha}
H.~Gharibyan and R.~F. Penna, {\it {Are entangled particles connected by
  wormholes? Evidence for the ER=EPR conjecture from entropy inequalities}},
  {\em Phys.Rev.} {\bf D89} (2014), no.~6 066001,
  [\href{http://xxx.lanl.gov/abs/1308.0289}{{\tt arXiv:1308.0289}}].

\bibitem{Balasubramanian:2014hda}
V.~Balasubramanian, P.~Hayden, A.~Maloney, D.~Marolf, and S.~F. Ross, {\it
  {Multiboundary Wormholes and Holographic Entanglement}},  {\em
  Class.Quant.Grav.} {\bf 31} (2014) 185015,
  [\href{http://xxx.lanl.gov/abs/1406.2663}{{\tt arXiv:1406.2663}}].

\bibitem{Hayden:2011ag}
P.~Hayden, M.~Headrick, and A.~Maloney, {\it {Holographic Mutual Information is
  Monogamous}},  {\em Phys.Rev.} {\bf D87} (2013), no.~4 046003,
  [\href{http://xxx.lanl.gov/abs/1107.2940}{{\tt arXiv:1107.2940}}].

\bibitem{Maldacena:2001kr}
J.~M. Maldacena, {\it {Eternal black holes in anti-de Sitter}},  {\em JHEP}
  {\bf 0304} (2003) 021, [\href{http://xxx.lanl.gov/abs/hep-th/0106112}{{\tt
  hep-th/0106112}}].

\bibitem{Vidal:2002zz}
G.~Vidal and R.~Werner, {\it {Computable measure of entanglement}},  {\em
  Phys.Rev.} {\bf A65} (2002) 032314.

\bibitem{Calabrese:2012ew}
P.~Calabrese, J.~Cardy, and E.~Tonni, {\it {Entanglement negativity in quantum
  field theory}},  {\em Phys.Rev.Lett.} {\bf 109} (2012) 130502,
  [\href{http://xxx.lanl.gov/abs/1206.3092}{{\tt arXiv:1206.3092}}].

\bibitem{Calabrese:2012nk}
P.~Calabrese, J.~Cardy, and E.~Tonni, {\it {Entanglement negativity in extended
  systems: A field theoretical approach}},  {\em J.Stat.Mech.} {\bf 1302}
  (2013) P02008, [\href{http://xxx.lanl.gov/abs/1210.5359}{{\tt
  arXiv:1210.5359}}].

\bibitem{Calabrese:2014yza}
P.~Calabrese, J.~Cardy, and E.~Tonni, {\it {Finite temperature entanglement
  negativity in conformal field theory}},  {\em J.Phys.} {\bf A48} (2015),
  no.~1 015006, [\href{http://xxx.lanl.gov/abs/1408.3043}{{\tt
  arXiv:1408.3043}}].

\bibitem{Kulaxizi:2014nma}
M.~Kulaxizi, A.~Parnachev, and G.~Policastro, {\it {Conformal Blocks and
  Negativity at Large Central Charge}},  {\em JHEP} {\bf 1409} (2014) 010,
  [\href{http://xxx.lanl.gov/abs/1407.0324}{{\tt arXiv:1407.0324}}].

\bibitem{Coser:2014gsa}
A.~Coser, E.~Tonni, and P.~Calabrese, {\it {Entanglement negativity after a
  global quantum quench}},  {\em J.Stat.Mech.} {\bf 1412} (2014), no.~12
  P12017, [\href{http://xxx.lanl.gov/abs/1410.0900}{{\tt arXiv:1410.0900}}].

\bibitem{Hoogeveen:2014bqa}
M.~Hoogeveen and B.~Doyon, {\it {Entanglement negativity and entropy in
  non-equilibrium conformal field theory}},
  \href{http://xxx.lanl.gov/abs/1412.7568}{{\tt arXiv:1412.7568}}.

\bibitem{Wen:2015qwa}
X.~Wen, P.-Y. Chang, and S.~Ryu, {\it {Entanglement negativity after a local
  quantum quench in conformal field theories}},
  \href{http://xxx.lanl.gov/abs/1501.0056}{{\tt arXiv:1501.0056}}.

\bibitem{Wichterich:2009fk}
H.~Wichterich, J.~Molina-Vilaplana, and S.~Bose, {\it Scaling of entanglement
  between separated blocks in spin chains at criticality},  {\em Physical
  Review A} {\bf 80} (2009), no.~1 010304.

\bibitem{Calabrese:2013uq}
P.~Calabrese, L.~Tagliacozzo, and E.~Tonni, {\it Entanglement negativity in the
  critical ising chain},  {\em Journal of Statistical Mechanics: Theory and
  Experiment} {\bf 2013} (2013), no.~05 P05002.

\bibitem{Eisler:2015kx}
V.~Eisler and Z.~Zimbor{\'a}s, {\it On the partial transpose of fermionic
  gaussian states},  {\em New Journal of Physics} {\bf 17} (2015), no.~5
  053048.

\bibitem{Coser:2015rt}
A.~Coser, E.~Tonni, and P.~Calabrese, {\it Partial transpose of two disjoint
  blocks in xy spin chains},   \href{http://arxiv.org/abs/1503.09114}{{\tt arXiv:1503.09114}}.

\bibitem{Rangamani:2014ywa}
M.~Rangamani and M.~Rota, {\it {Comments on Entanglement Negativity in
  Holographic Field Theories}},  {\em JHEP} {\bf 1410} (2014) 60,
  [\href{http://arxiv.org/abs/1406.6989}{{\tt arXiv:1406.6989}}].

\bibitem{Lewkowycz:2014jia}
A.~Lewkowycz and E.~Perlmutter, {\it {Universality in the geometric dependence
  of R{\'e}nyi entropy}},  {\em JHEP} {\bf 1501} (2015) 080,
  [\href{http://arxiv.org/abs/1407.8171}{{\tt arXiv:1407.8171}}].

\bibitem{Perlmutter:2015zr}
E.~Perlmutter, M.~Rangamani, and M.~Rota, {\it {Positivity, negativity, and
  entanglement}},    \href{http://arxiv.org/abs/1506.01679}{{\tt arXiv:1506.01679}}.

\bibitem{Eckert:2002rw}
K.~{Eckert}, J.~{Schliemann}, D.~{Bru{\ss}}, and M.~{Lewenstein}, {\it {Quantum
  Correlations in Systems of Indistinguishable Particles}},  {\em Annals of
  Physics} {\bf 299} (July, 2002) 88--127,
  [\href{http://xxx.lanl.gov/abs/quant-ph/0203060}{{\tt quant-ph/0203060}}].

\bibitem{Haque:2007gf}
M.~Haque, O.~Zozulya, and K.~Schoutens, {\it Entanglement entropy in fermionic
  laughlin states},  {\em Phys. Rev. Lett.} {\bf 98} (Feb, 2007) 060401.

\bibitem{Zozulya:2007ul}
O.~S. Zozulya, M.~Haque, K.~Schoutens, and E.~H. Rezayi, {\it Bipartite
  entanglement entropy in fractional quantum hall states},  {\em Phys. Rev. B}
  {\bf 76} (Sep, 2007) 125310.

\bibitem{Herdman:2014qr}
C.~M. Herdman, P.-N. Roy, R.~G. Melko, and A.~Del~Maestro, {\it Particle
  entanglement in continuum many-body systems via quantum monte carlo},  {\em
  Phys. Rev. B} {\bf 89} (Apr, 2014) 140501.

\bibitem{Horodecki:2009aa}
R.~Horodecki, P.~Horodecki, M.~Horodecki, and K.~Horodecki, {\it Quantum
  entanglement},  {\em Rev.Mod.Phys.} {\bf 81} (2009) 865--942,
  [\href{http://xxx.lanl.gov/abs/quant-ph/0702225}{{\tt quant-ph/0702225}}].

\bibitem{Vidal:1999aa}
G.~Vidal and R.~Tarrach, {\it Robustness of entanglement},  {\em Phys.Rev.}
  {\bf A59} (1999) 141--155,
  [\href{http://xxx.lanl.gov/abs/quant-ph/9806094}{{\tt quant-ph/9806094}}].

\bibitem{Groisman:aa}
B.~Groisman, S.~Popescu, and A.~Winter, {\it On the quantum, classical and
  total amount of correlations in a quantum state},
  \href{http://xxx.lanl.gov/abs/quant-ph/0410091}{{\tt quant-ph/0410091}}.

\bibitem{He:2014aa}
H.~He and G.~Vidal, {\it Disentangling theorem and monogamy for entanglement
  negativity},  \href{http://xxx.lanl.gov/abs/1401.5843}{{\tt
  arXiv:1401.5843}}.

\bibitem{Lin:2014hva}
J.~Lin, M.~Marcolli, H.~Ooguri, and B.~Stoica, {\it {Tomography from
  Entanglement}},  \href{http://xxx.lanl.gov/abs/1412.1879}{{\tt
  arXiv:1412.1879}}.

\bibitem{Lashkari:2014kda}
N.~Lashkari, C.~Rabideau, P.~Sabella-Garnier, and M.~Van~Raamsdonk, {\it
  {Inviolable energy conditions from entanglement inequalities}},
  \href{http://xxx.lanl.gov/abs/1412.3514}{{\tt arXiv:1412.3514}}.

\bibitem{Bhattacharya:2014vja}
J.~Bhattacharya, V.~E. Hubeny, M.~Rangamani, and T.~Takayanagi, {\it
  {Entanglement density and gravitational thermodynamics}},
  \href{http://xxx.lanl.gov/abs/1412.5472}{{\tt arXiv:1412.5472}}.

\bibitem{Guhne:2009aa}
O.~G{\"u}hne and G.~Toth, {\it Entanglement detection},  {\em Physics Reports}
  {\bf 474} (2009) 1, [\href{http://xxx.lanl.gov/abs/0811.2803}{{\tt
  arXiv:0811.2803}}].

\bibitem{Gisin:1998aa}
N.~Gisin and H.~Bechmann-Pasquinucci, {\it Bell inequality, bell states and
  maximally entangled states for n qubits},  {\em Phys.Lett.} {\bf A246} (1998)
  1--6, [\href{http://xxx.lanl.gov/abs/quant-ph/9804045}{{\tt
  quant-ph/9804045}}].

\bibitem{Coffman:2000aa}
V.~Coffman, J.~Kundu, and W.~K. Wootters, {\it Distributed entanglement},  {\em
  Phys.Rev.A} {\bf 61} (2000) 052306,
  [\href{http://xxx.lanl.gov/abs/quant-ph/9907047}{{\tt quant-ph/9907047}}].

\bibitem{Wong:aa}
A.~Wong and N.~Christensen, {\it A potential multipartide entanglement
  measure},  \href{http://xxx.lanl.gov/abs/quant-ph/0010052}{{\tt
  quant-ph/0010052}}.

\bibitem{Gour:2010aa}
G.~Gour and N.~R. Wallach, {\it All maximally entangled four qubits states},
  \href{http://xxx.lanl.gov/abs/1006.0036}{{\tt arXiv:1006.0036}}.

\bibitem{Dur:2000aa}
W.~D{\"u}r, G.~Vidal, and J.~I. Cirac, {\it Three qubits can be entangled in
  two inequivalent ways},  {\em Phys. Rev. A} {\bf 62} (2000) 062314,
  [\href{http://xxx.lanl.gov/abs/quant-ph/0005115}{{\tt quant-ph/0005115}}].

\bibitem{Vidal:2000aa}
G.~Vidal, {\it Entanglement monotones},  {\em J.Mod.Opt.} {\bf 47} (2000) 355,
  [\href{http://xxx.lanl.gov/abs/quant-ph/9807077}{{\tt quant-ph/9807077}}].

\bibitem{Hubeny:2013gta}
V.~E. Hubeny, H.~Maxfield, M.~Rangamani, and E.~Tonni, {\it {Holographic
  entanglement plateaux}},  {\em JHEP} {\bf 1308} (2013) 092,
  [\href{http://xxx.lanl.gov/abs/1306.4004}{{\tt arXiv:1306.4004}}].

\bibitem{Araki:1970ba}
H.~Araki and E.~Lieb, {\it {Entropy inequalities}},  {\em Commun.Math.Phys.}
  {\bf 18} (1970) 160--170.

\bibitem{Zhang:2012fp}
L.~Zhang and J.~Wu, {\it {On Conjectures of Classical and Quantum Correlations
  in Bipartite States}},  {\em J.Phys.} {\bf A45} (2012) 025301.

\bibitem{Headrick:2013zda}
M.~Headrick, {\it {General properties of holographic entanglement entropy}},
  {\em JHEP} {\bf 1403} (2014) 085,
  [\href{http://xxx.lanl.gov/abs/1312.6717}{{\tt arXiv:1312.6717}}].

\bibitem{Verstraete:aa}
F.~Verstraete, J.~Dehaene, B.~D. Moor, and H.~Verschelde, {\it Four qubits can
  be entangled in nine different ways},  {\em Phys. Rev. A, 65: 052112 (2002).}
  (2001) [\href{http://xxx.lanl.gov/abs/quant-ph/0109033}{{\tt
  quant-ph/0109033}}].

\bibitem{Lamata:2007aa}
L.~Lamata, J.~Leon, D.~Salgado, and E.~Solano, {\it Inductive entanglement
  classification of four qubits under slocc},  {\em Phys. Rev. A} {\bf 75}
  (2007) 022318, [\href{http://xxx.lanl.gov/abs/quant-ph/0610233}{{\tt
  quant-ph/0610233}}].

\bibitem{Hartman:2013qma}
T.~Hartman and J.~Maldacena, {\it {Time Evolution of Entanglement Entropy from
  Black Hole Interiors}},  {\em JHEP} {\bf 1305} (2013) 014,
  [\href{http://xxx.lanl.gov/abs/1303.1080}{{\tt arXiv:1303.1080}}].

\bibitem{Pastawski:2015qua}
F.~Pastawski, B.~Yoshida, D.~Harlow, and J.~Preskill, {\it {Holographic quantum
  error-correcting codes: Toy models for the bulk/boundary correspondence}},
  \href{http://xxx.lanl.gov/abs/1503.0623}{{\tt arXiv:1503.0623}}.

\bibitem{Almheiri:2014lwa}
A.~Almheiri, X.~Dong, and D.~Harlow, {\it {Bulk Locality and Quantum Error
  Correction in AdS/CFT}},  \href{http://xxx.lanl.gov/abs/1411.7041}{{\tt
  arXiv:1411.7041}}.

\bibitem{Czech:2014tva}
B.~Czech, P.~Hayden, N.~Lashkari, and B.~Swingle, {\it {The Information
  Theoretic Interpretation of the Length of a Curve}},
  \href{http://xxx.lanl.gov/abs/1410.1540}{{\tt arXiv:1410.1540}}.

\bibitem{Susskind:2014yaa}
L.~Susskind, {\it {ER=EPR, GHZ, and the Consistency of Quantum Measurements}},
  \href{http://xxx.lanl.gov/abs/1412.8483}{{\tt arXiv:1412.8483}}.

\end{thebibliography}

%%%%%%%%%%%%%%%%%%%%%%%%%%%%%%%%%%%%%%%%%%%
\providecommand{\href}[2]{#2}\begingroup\raggedright\endgroup

\end{document}